\UseRawInputEncoding 

\documentclass{jfm}
\usepackage{graphicx}
\usepackage{color}
\usepackage{amsmath}
\usepackage{subfig}

\shorttitle{Turbulence and heat transfer on a rotating, heated half soap bubble}
\shortauthor{ X. Q. He, A. D. Bragg, Y. L. Xiong, P. Fischer}

\title{Turbulence and heat transfer on a rotating, heated half soap bubble}

\author{X. Q. HE\aff{1,2},
  A. D. Bragg\aff{3},
  Y. L. Xiong\aff{1,2,3}
  \corresp{\email{xylcfd@hust.edu.cn}},
  P. Fischer\aff{4},
   }

\affiliation{
\aff{1}Department of Mechanics, Huazhong University of Science and Technology, China
\aff{2}Hubei Key Laboratory of Engineering Structural Analysis and Safety Assessment, China
\aff{3}Pratt School of Engineering, Duke University, USA
\aff{4}Institut de Math{\'e}matiques de Bordeaux (IMB), Universit{\'e} de Bordeaux, CNRS UMR 5251,  France
}

\begin{document}

\maketitle

\begin{abstract}

We use Direct Numerical Simulations to study the two-dimensional flow of a rotating, half soap bubble that is heated at its equator. The heating produces buoyancy and rotation generates a Coriolis forces in the fluid. However, due to the curved surface of the bubble, the buoyancy and Coriolis forces vary with latitude on the bubble, giving rise to rich flow behavior. We first explore the single-point properties of the flow, including the Reynolds and Nusselt numbers, mean fields, and Reynolds stresses, all as a function of latitude. For a given Rayleigh number, we observe a non-monotonic dependence on the Rossby number $Ro$, and large scale mean circulations that are strongly influenced by rotation. We then consider quantities that reveal the multiscale nature of the flow, including spectrums and spectral fluxes of kinetic and thermal energy, and enstrophy, and structure functions of velocity and temperature. The fluxes show that just a for non-buoyant two-dimensional turbulence on a flat surface, there is an upscale flux of kinetic energy at larger scales (fed by buoyancy injection of turbulent kinetic energy at smaller scales), and a downscale flux of enstrophy at smaller scales. The kinetic energy spectrum and velocity structure functions are well described by Bolgiano-Obukhov (BO) scaling at scales where the effects of rotation are weak. The temperature structure functions do not, however, satisfy BO scaling in general, due to strong intermittency in the temperature field.

\end{abstract}


\section{Introduction}
Thermal convection is ubiquitous in natural and engineered systems \citep{StevensClercx81,LohseXia29,AhlersGrossmann32}.
For example, convection plays a dominant role in many solar energy flat-plate collectors \citep{DasRoy116}, and heat transfer from the core to the exterior of stellar structure is of paramount importance \citep{RSLONG2020}.
In such flows, the fluid motion is driven by buoyancy forces that arise from density variations due to thermal gradients, and the flows can be classified based on the relative directions of the imposed temperature gradient and gravity \citep{DasRoy116}. Some studies have considered flows where the temperature gradient is perpendicular to gravity \citep{HussamSheard121,BasakRoy120,BasakRoy118,BasakRoy119}, but perhaps the most commonly studied situation is where the temperature gradient is parallel to gravity, such as in Rayleigh-B\'enard convection (RBC). However, in many real world circumstances, the imposed temperature gradient is neither parallel nor perpendicular to gravity \citep{Bejan117}. Such situations can occur for thermal convection inside enclosures with irregular geometries, which have been reviewed in detail recently by \citet{DasRoy116}. Among the many different possible geometries, thermal convection on a spherical surface is an interesting model for studying astrophysical and geophysical flows \citep{Kellay2017}, and this is the subject of the present work.

Two key parameters that determine the behavior of a (non-rotating) thermally convective flow are the Rayleigh number, $Ra$, and the Prandtl number $Pr$. Given $Ra, Pr$, two emergent parameters in the flow that are of utmost importance are the Reynolds number, $Re$, and the Nusselt number $Nu$, with $Nu$ characterizing the heat transport properties of the flow.

Rayleigh-B\'enard convection (RBC) is an important model for the study of the turbulent thermal convection, and can be implemented conveniently in experiments or numerical simulations \citep{StevensClercx81,LohseXia29,AhlersGrossmann32}.
Research on Rayleigh-B\'enard convection may be broadly described in terms of two aspects: the small-scale and large scale dynamics.
The study of the small-scale properties has tended to focus on the scaling of velocity and temperature structure functions, and intermittent behavior in the flow.
A recent review of the small-scale properties can be found in the work of \citet{LohseXia29}.
Studies on the large scale properties of RBC tend to focus on exploring the dependence of the emergent properties $Nu$ and $Re$ on the control parameters $Ra$ and $Pr$, as well as the properties of thermal plumes and large-scale circulations (LSC). A detailed discussion of theoretical and experimental progress on RBC can be found in \citet{AhlersGrossmann32}.

Another important problem concerns rotating Rayleigh-B\'enard convection (RRBC) \citep{StevensClercx81}, wherein the Rayleigh-B\'enard system rotates about the direction parallel to gravity (the vertical direction). In this case, the other important control parameter is the Rossby number, $Ro$. When $Ro\gg 1$ the effect of rotation on the flow is weak, while it is strong for $Ro\ll 1$. When the dynamics are considered in the rotating frame of reference, the effect of rotation is seen through the addition of the Coriolis force, which may counteract the effect of buoyancy in the flow and stabilize the system \citep{RajaeiAlards74,HornSchmid72}. This stabilizing effect was demonstrated analytically by \citet{Chandrasekhar110} using linear-stability analysis. Note that for an incompressible flow, the centrifugal force may be considered to be absorbed into the pressure field that constrains the velocity field to be divergence-free.

Many studies have explored the properties of RRBC, exploring the role of rotation on the heat transport and flow structures, e.g. \cite{FavierGuervilly73,RajaeiAlards74,JoshiRajaei71,HornSchmid72,ZhongSterl70,HornShishkina80,GuervillyHughes76,KingStellmach78,KunnenClercx79,StevensClercx81,KingStellmach75,PharasiKannan87,WeissAhlers88,WeissAhlers89,KunnenStevens90,KunnenGeurts64,ZhongAhlers65,ScheelMutyaba77,GroomsJulien85,KunnenClercx91,KingStellmach95,ZhongStevens96,OrestaStringano97,VorobieffEcke94,BodenschatzPesch67,Sakai98,JulienLegg68,JulienLegg93,ZhongEcke69,BoubnovGolitsyn101,Rossby92}. Collectively, these studies have shown that as $Ro$ is decreased, RRBC goes through three successive regimes, which are named regime \uppercase\expandafter{\romannumeral1}, \uppercase\expandafter{\romannumeral2} and \uppercase\expandafter{\romannumeral3} respectively \citep{RajaeiAlards74,JoshiRajaei71,HornSchmid72}.

In regime \uppercase\expandafter{\romannumeral1}, $Ro$ is large enough for the Coriolis force to be negligible compared to buoyancy, and hence rotation has little effect on the flow \citep{KingStellmach78}. In this regime, LSC are the prominent flow structures, and $Nu$ does not vary with $Ro$ \citep{WeissAhlers88,WeissAhlers89}. In regime \uppercase\expandafter{\romannumeral 2}, $Ro$ is such that the system depends on the interaction between the Coriolis and buoyancy forces.
Moreover, small vertical plumes parallel to the rotation axis take the place of LSC as the dominant flow structures, while heat transport is enhanced with decreasing $Ro$ \citep{ZhongSterl70,HornShishkina80,GuervillyHughes76,KingStellmach75,WeissAhlers88,WeissAhlers89,KunnenGeurts64,ScheelMutyaba77,VorobieffEcke94,Sakai98,BoubnovGolitsyn101}. In this regime, variations of the flow in the vertical direction are suppressed due to the well-known Taylor-Proudman effect. The vertical plumes also convey hot fluid from the Ekman layer to the cold bulk of the fluid, a phenomena is refer to ``Ekman pumping''. This pumping is what causes $Nu$ to increase with decreasing $Ro$ in regime \uppercase\expandafter{\romannumeral 2}. In regime \uppercase\expandafter{\romannumeral3}, $Ro$ is small enough such that the Coriolis force dominates the system and $Nu$ plunges as $Ro$ is further decreased \citep{ZhongAhlers65,KingStellmach95,ZhongStevens96,KunnenClercx102,KunnenClercx103,VorobieffEcke94,LiuEcke105,JulienLegg68,ZhongEcke69,Rossby92}.
In this regime, turbulence is quenched and the efficiency of heat transport is greatly diminished.

Recently, an alternative system has been explored to understand turbulent motion and heat transfer on a curved surface, namely Kellay's soap bubble \citep{Kellay2017}, which has yielded interesting results on the behavior of turbulent convection, as well as a model for hurricane tracking \citep{MeuelCoudert11,MeuelXiong14,SeychellesAmarouchene2008,SeychellesIngremeau9}. In Kellay's experiment, a half soap bubble (hemisphere) is heated from its equator, producing buoyancy forces and convection, resulting in a quasi two-dimensional flow on a hemispherical surface. Unlike standard RBC, the local buoyancy force in this flow varies with location not only due to variations in the local fluid temperature, but also because the component of gravity acting on the flow varies with latitude on the bubble. In \citet{BruneauFischer19}, Direct Numerical Simulations (DNS) of the soap bubble computed $Re$ and $Nu$ as a function of $Ra$ and found scaling behavior that is remarkably similar to that in standard RBC. In \cite{MeuelCoudert11} the system was further explored by subjecting the soap bubble to rotation. Unlike RRBC, in the rotating soap bubble flow, the effect of the Coriolis force varies with location due to geometrical reasons, being zero at the equator of the bubble. Significant effects of the rotation were found on the structure functions of velocity and temperature in \cite{MeuelCoudert11}.

The study of \cite{MeuelCoudert11} only rather briefly explored the properties of the flow on the rotating soap bubble, and there is much to understand about this flow, and the various ways in which its properties are similar and dissimilar to RRBC. Such a detailed study is the goal of this present paper. The outline of the paper is as follows. In section \ref{GEDNS}, the equations of motion for the system, their properties and non-dimensional parameters are discussed, as well as the DNS used to solve the equations. In section \ref{SPI} we explore the single-point properties of the flow, including the behavior of $Re, Nu$, the mean flow and Reynolds stresses. In \ref{SDI} we explore the properties of the flow at different scales, considering fluxes of kinetic energy, enstrophy and entropy. Structure functions of the velocity and temperature field are also explored, along with a detailed consideration of their scaling behavior. Finally, in section \ref{conc} we draw conclusions to the work and discuss future directions for study.

%
%
\section{Governing Equations and Direct Numerical Simulations}\label{GEDNS}

\subsection{Governing equations and control parameters}

We consider the flow of a half soap bubble of radius $R$ that is heated from its equator, mimicking the experimental setup of \cite{Kellay2017}. The bubble geometry is such that its thickness is negligible compared to its radius, and the flow in the radial direction has little effect on the heat and mass transfer of the system.
Therefore, the system may be modeled as a two-dimensional flow on a hemispherical surface of radius $R$, with boundary conditions applied at the equator. Furthermore, we consider a system where the bubble rotates at a rate $\Omega\equiv\|\boldsymbol{\Omega}\|$ about its North Pole.

The bubble may be described in terms of the three-dimensional Cartesian coordinate system, with coordinates $(x,y,z)$ and unitary basis vectors $\boldsymbol{e_x},\boldsymbol{e_y},\boldsymbol{e_z}$. With respect to this, the bubble under consideration has its equator on the $(x,y,z=0)$ plane, and rotates about $\boldsymbol{e_z}$, with $\boldsymbol{e_z\cdot \Omega}=\Omega$. Given the curved surface of the bubble, it is also convenient to use a geographical coordinate system with coordinates $(r,\theta,\phi)$, and basis vectors  $\boldsymbol{e_{r}}(\theta,\phi),\boldsymbol{e_{\theta}}(\theta,\phi),\boldsymbol{e_{\phi}}(\theta,\phi)$, where $\boldsymbol{e_{r}}(\theta,\phi)\times\boldsymbol{e_{\theta}}(\theta,\phi)=\boldsymbol{e_{\phi}}(\theta,\phi)$. The latitudinal coordinate $\theta\in[0,\pi]$ increases from 0 at the equator, and the longitudinal coordinate is $\phi\in[0,2\pi)$. In these coordinates, the two-dimensional flow is on the surface $(r=R,\theta,\phi)$, and there is no flow in the direction $\boldsymbol{e_{r}}(\theta,\phi)$.

The two-dimensional flow on the hemisphere is governed by the incompressible Navier-Stokes equations with the Oberbeck-Boussinesq approximation
\begin{align}
	\frac{D\boldsymbol{U}}{Dt}
	&= -\frac{1}{\rho}\bnabla p+\nu\triangle\boldsymbol{U} - \beta T \boldsymbol{g} - 2\boldsymbol{\Omega}\times\boldsymbol{U} - F\boldsymbol{U},\label{NSE}\\
	\bnabla\bcdot\boldsymbol{U}&=0,\\
	\frac{DT}{Dt} &= \alpha\triangle T - ST,\label{TE}
\end{align}
where $\boldsymbol{U}$ is the velocity field, $p$ is the pressure field (that accounts for the centrifugal forces through incompressibility), $T$ is the temperature field, and $\rho$ is the constant reference density. We consider the case where gravity and rotation are aligned with $\boldsymbol{e_z}$, and $\boldsymbol{e_z\cdot g}=-g$.

Boundary conditions are applied on the hemisphere equator, with no-slip for the velocity, and fixed value for the temperature field. Buoyancy driven flow on the hemisphere is therefore fundamentally different to the classical Rayleigh-B\'enard systems for which there is a boundary through which heat flows out. A consequence of this is that when solving \eqref{NSE}--\eqref{TE} without the terms involving $S$ and $F$, energy would accumulate inside the bubble, and the system may become numerically unstable. In contrast, in Kellay's heated soap bubble experiment \citep{Kellay2017}, part of the energy is lost through exchange with the cold air outside and inside the bubble. The terms in \eqref{NSE} and \eqref{TE} involving $S$ and $F$ are supposed to represent this energy loss to the surrounding air. This is analogous to the way in which DNS of two-dimensional turbulence often uses a friction term to prevent accumulation of energy at the large scales of the flow \citep{Boffetta12}. We will return momentarily to discuss the specification of $S$ and $F$. Concerning the initial conditions for the system, the initial temperature of the bubble is the same as the surrounding air, and the velocity is initially zero everywhere.

%

In order to define the various control parameters in the system, we take the characteristic length to be the radius of the bubble $R$, and the temperature difference
$\delta T$ to be the difference in temperature between the equator and the cold air surrounding the bubble. Using these, we define the Raleigh number $Ra$, Rossby number $Ro$, and Prandtl number $Pr$
 \begin{align}
	Ra &\equiv \frac{g\beta R^3 \delta T}{\nu \alpha},\\
	Ro &\equiv \frac{\sqrt{g/R}}{2\Omega},\\
	\Pran &\equiv \frac{\nu}{\alpha}.
\end{align}
On the bubble, the velocity field may be represented as
\begin{equation}
\boldsymbol{U} = U_{\theta}\boldsymbol{e_{\theta}}+U_{\phi}\boldsymbol{e_{\phi}},
\end{equation}
where $U_\theta\equiv\boldsymbol{e_{\theta}\cdot U}$, $U_\phi\equiv\boldsymbol{e_{\phi}\cdot U}$, while for the radial direction, $U_r\equiv\boldsymbol{e_r \cdot U}=0$ since the flow is two dimensional. We also define the fluctuations $\boldsymbol{U}'\equiv\boldsymbol{U}-\langle \boldsymbol{U}\rangle$ with components $U_\theta'\equiv U_\theta-\langle U_\theta\rangle, U_\phi'\equiv U_\phi-\langle U_\phi\rangle$. Using these, we define the other two crucial parameters in the system, the Reynolds number $Re$ and the Nusselt number $Nu$
\begin{align}
	Re &\equiv \frac{\sqrt{2 E_{turb}} R}{\nu},\\
	Nu &\equiv \frac{ \langle U_\theta' T \rangle - \partial_\theta\langle T\rangle}{ (\beta \delta T/R)},
\end{align}
where $E_{turb}\equiv(1/2)\langle U_\theta' U_\theta' +U_\phi' U_\phi'\rangle$ is the flow Turbulent Kinetic Energy (TKE). Since $Re$ and $Nu$ depend on the properties of the flow they are emergent quantities and depend implicitly on $Ra, Ro, \Pran$. Furthermore, given the statistical symmetries of the system, the flow parameters $Re, Nu$ do not depend on the longitudinal coordinate $\phi$, but they do depend on the latitudinal coordinate $\theta$.

The flow under consideration is driven by buoyancy, with heating at the equator. As such, the fluid will convect away from the equator, and the intensity of the turbulence will increase with increasing $Ra$. Furthermore, $- \beta T \boldsymbol{g} = \beta Tg_\theta \boldsymbol{e_\theta}$, where $ g_\theta=g\boldsymbol{e_z\cdot e_\theta}$, hence irrespective of spatial variations in $T$, buoyancy forces will vary from being maximum at the equator where $\boldsymbol{e_z\cdot e_\theta}=1$, to minimum at the North Pole where $\boldsymbol{e_z\cdot e_\theta}=0$. This geometrical variation, caused by the curved surface of the bubble, makes flow of the heated bubble distinct from RBC for which such geometrical variation of the buoyancy force is absent.

The Coriolis term can significantly affect the flow when $Re\leq O(1)$, although it makes no direct contribution to the TKE since $(\boldsymbol{\Omega}\times\boldsymbol{U})\boldsymbol{\cdot U}=0$. For the system under consideration, the Coriolis force may be expressed as
\begin{align}
-2\boldsymbol{\Omega}\times\boldsymbol{U}= -2\Omega U_{\theta}\boldsymbol{e_z}\times\boldsymbol{e_{\theta}} -2\Omega U_{\phi}\boldsymbol{e_z}\times\boldsymbol{e_{\phi}},
\end{align}
and at the equator, $\boldsymbol{e_z}\times\boldsymbol{e_{\theta}}=\boldsymbol{0}$ and $\boldsymbol{e_z}\times\boldsymbol{e_{\phi}}=\boldsymbol{e_r}$. Since there is no flow in the radial direction, then at the equator the Coriolis force makes no contribution to the two-dimensional flow on the hemisphere, but its effect becomes increasingly large as $\theta$ increases. As a result, irrespective of spatial variations in $U_{\theta},U_{\phi}$, the Coriolis force varies on the surface of the bubble. This geometrical variation again makes flow of the rotating soap bubble quite different from standard RRBC.

Taking the curl of the steady, inviscid, linearized form of \eqref{NSE}, with $S=F=0$ and $\boldsymbol{\Omega}$ a constant, we obtain $(\boldsymbol{\Omega\cdot\nabla})\boldsymbol{U}=-\beta\boldsymbol{\nabla}\times(T\boldsymbol{g})$. In the limit $Ro\to 0$, this recovers the Taylor-Proudman theorem $(\boldsymbol{\Omega\cdot\nabla})\boldsymbol{U}=\boldsymbol{0}$, describing the constancy of $\boldsymbol{U}$ in the direction of $\boldsymbol{\Omega}$, associated with Taylor-Proudman columns in the flow. In coordinate form, $(\boldsymbol{\Omega\cdot\nabla})\boldsymbol{U}=\boldsymbol{0}$ may be written on the bubble as
\begin{align}
\partial_z(\boldsymbol{e_{\phi}} U_\phi +\boldsymbol{e_{\theta}} U_\theta) =R\cos\theta\partial_\theta (\boldsymbol{e_{\phi}} U_\phi +\boldsymbol{e_{\theta}} U_\theta)=\boldsymbol{0}. \label{TPsc}
\end{align}
Since this is true for arbitrary $\theta$, then it implies constancy of the flow in the latitudinal direction. For the fully non-linear system \eqref{NSE}, if $Ro$ is sufficiently small, the Taylor-Proudman behavior may be still observed at the large scales of the flow where nonlinearity is weakest. This may be seen by introducing a scale-dependent Rossby number $Ro_\ell\equiv 1/(\tau_\ell\Omega)$, that compares the eddy turnover time at scale $\ell$, namely $\tau_\ell$, to the period of rotation, $1/\Omega$. Since $\tau_\ell$ increases with increasing $\ell$, then $Ro_\ell$ decreases with increasing $\ell$. At scales where $Ro_\ell\ll 1$, the Taylor-Proudman behavior may still be observed in the full system described by \eqref{NSE}, whereas the effect of rotation will be sub-leading at all scales where $Ro_\ell>1$.

Near the equator, the no slip condition on the soap bubble generates strong viscous effects, and the Taylor-Proudman theorem does not apply. Instead, one may observe a regime where the Coriolis term balances viscous forces in the flow, at scales where $Ro_\ell\ll 1$. This balance can give rise to Eckman transport, producing momentum transport into, or out of the boundary layer near the equator

The other key feature influencing the bubble flow is its two-dimensionality. This geometry prohibits both vortex stretching and strain self-amplification that are fundamental to the energy cascade in three-dimensional turbulence \citep{carbone20,johnson20}. This prohibition gives rise to an additional inviscid constant of motion in two-dimensional compared with three dimensional turbulence, namely the inviscid conservation of enstrophy, and this leads to an inverse energy cascade in two-dimensional turbulence \citep{Boffetta12}. DNS for flow on a non-rotating bubble surface also observed an inverse energy cascade \citep{BruneauFischer19}, similar to two-dimensional turbulence on a flat surface.

\subsection{Details of Direct Numerical Simulations}

Following \citet{BruneauFischer19}, we solve \eqref{NSE}--\eqref{TE} using the stereographic coordinate system for numerical simplicity. The governing equations in the stereographic coordinate system are discretized using a finite difference method on a uniform staggered grid. The discrete values of the pressure and temperature are located at the center of each cell, and those of the velocity components are located at the middle of the sides. The unsteady term is discretized by the second-order Gear scheme, and the nonlinear term is handling by the third-order Murman-like scheme. The linear terms of the governing equations are treated implicitly, while the non-linear terms are treated explicitly. The pressure and velocity are directly solved using the Cramer method in a full coupled form, then the temperature equation is solved using the Conjugate gradient method. Further details on the code used and numerical methods may be found in \citet{BruneauFischer19}.

\begin{table}
  \begin{center}
\def~{\hphantom{0}}
  \begin{tabular}{cccccccc}
      Run					&$Ra$  						      & $Pr$   			&$1/Ro$ 			  																			&$S$				    	& $F$       	&Time       	&Resolution\\
      $A1$					&$3\times10^9$ 			& $7$ 			   &$0$																						&$0.06$	  		  		& $0.06$	   &$1100$     	&$512\times512$\\
      $A2$					&$3\times10^9$ 			& $7$ 			   &$0$																						&$0.06$	  		  		& $0.06$	   &$1100$     	&$1024\times1024$\\
      $A3$					&$3\times10^9$ 			& $7$ 			   &$0$																						&$0.06$	  		  		& $0.06$	   &$1100$     	&$1536\times1536$\\
      $A4$					&$3\times10^9$ 			& $7$ 			   &$0$																						&$0.06$	  		  		& $0.06$	   &$500$     	&$2048\times2048$\\
      $B1$					&$3\times10^6$   			& $7$ 			 	&$0, 0.01, 0.05, 0.1, 0.5, 1.0, 1.5, 2, 5, 10$ 	   		&$0.06$					& $0.06$		&$100-1100$		  	&$1024\times1024$\\
      $B2$					&$3\times10^6$ 			& $7$ 			   &$0, 0.01, 0.05, 0.1, 0.5, 1.0, 1.5, 2, 5, 10$				&$0.08$	  		  		& $0.08$	   &$100$     	&$1024\times1024$\\
      $B3$					&$3\times10^6$ 			& $7$ 			   &$0, 0.01, 0.05, 0.1, 0.5, 1.0, 1.5, 2, 5, 10$				&$0.1$	  		  			& $0.1$	   &$100$     	&$1024\times1024$\\
      $C1$				&$3\times10^7$ 			& $7$ 			   &$0, 0.01, 0.05, 0.1, 0.5, 1.0, 1.5, 2, 5, 10$				&$0.06$	  		  		& $0.06$	   &$100-1100$     	&$1024\times1024$\\
      $C2$				&$3\times10^8$ 			& $7$ 			   &$0, 0.01, 0.05, 0.1, 0.5, 1.0, 1.5, 2, 5, 10$				&$0.06$	  		  		& $0.06$	   &$100-1100$     	&$1024\times1024$\\
      $C3$				&$3\times10^9$ 			& $7$ 			   &$0, 0.01, 0.05, 0.1, 0.5, 1.0, 1.5, 2, 5, 10$				&$0.06$	  		  		& $0.06$	   &$100-1100$     	&$1024\times1024$\\
  \end{tabular}
  \caption{The parameters for the DNS cases. Time corresponds to the number of time units for which the DNS was run, and is expressed in non-dimensional form using $\sqrt{(R/g)}$.
  }
  \label{tbl:RunList}
  \end{center}
\end{table}
The details of the different DNS and the associated parameters are given in Table \ref{tbl:RunList}.
There are three different classes of runs, A, B and C. In class $A$, $Ra$ and $Ro$ are fixed while the numerical grid resolution varies from $512\times512$ to $2048\times2048$. From this we determined that $1024\times 1024$ provides the optimum resolution for convergence and minimal computational cost for the parameter ranges we consider, as was also found in \citep{BruneauFischer19}. The appropriate resolution is mostly constrained by the thermal boundary layer thickness, since the thermal boundary layer is thinner than the velocity boundary layer for $Pr>1$. Previous studies indicate that the maximum $Ra$ for which the thermal boundary layer could be resolved by $1024\times 1024$ for $Pr=7$ is $Ra=3 \times 10^9$ \citep{BruneauFischer19}. Although they did not consider rotation, those findings also applies to our study since rotation reduces the kinetic energy dissipation rate, so that the grid resolution requirements are most stringent for the $1/Ro=0$ case. It is also to be noted that the stereographic projection method used in our numerical simulations is also beneficial to the grid resolution because the uniform grid in the projected plane corresponds in spherical coordinates to a smaller cell size near the equator than near the polar zone.

The class $B$ runs consider three different values of $S$ and $F$ in order to evaluate the impact of $S$ and $F$ on the flow statistics. We found that $S=F=0.06$ gave the optimum choice, as was also found in \citet{BruneauFischer19} for the non-rotating bubble DNS. The class $C$ runs then use the optimum grid resolution and values of $S,F$ found from runs A and B, but now with varying $Ro$ and $Ra$ in order to explore the role of rotation on the flow properties.

The results shown in the following sections correspond to the statistically stationary regime of the flow. In this regime, the ensemble average $\langle\cdot\rangle$ is approximated using a time average, and an average over $\phi$, the latter being appropriate since the system is statistically invariant with respect to $\phi$. The statistics depend only on the latitudinal coordinate $\theta$, and due to symmetry, we plot the results only over the range $\theta\in[0,\pi/2]$.

\section{Results \& Discussion: single-point information}\label{SPI}

\begin{figure}
  \centering
  \includegraphics[width = 6.5cm]{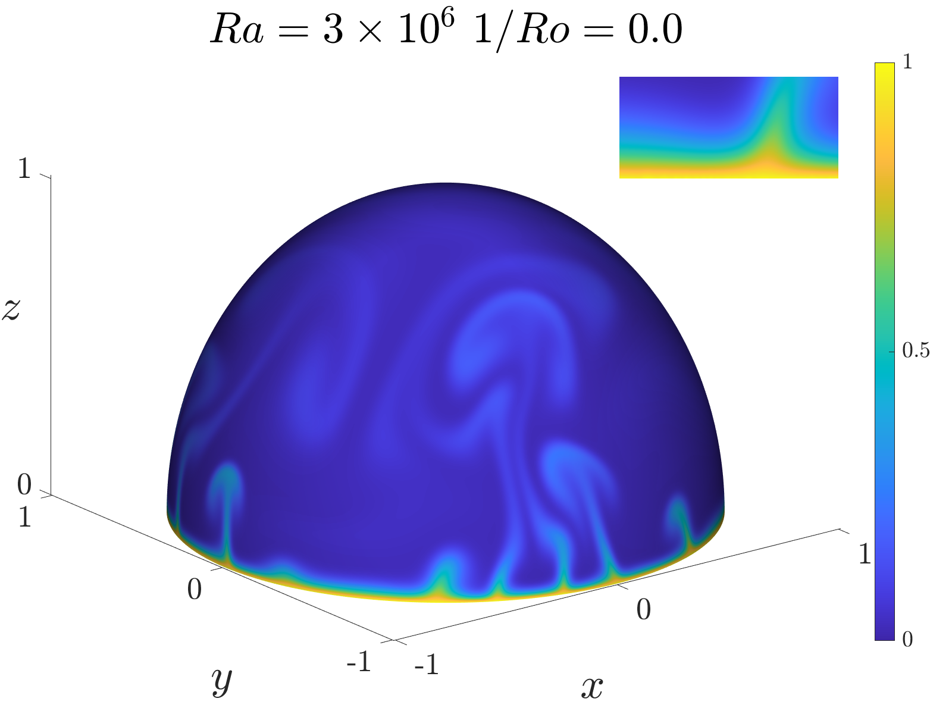}
  \includegraphics[width = 6.5cm]{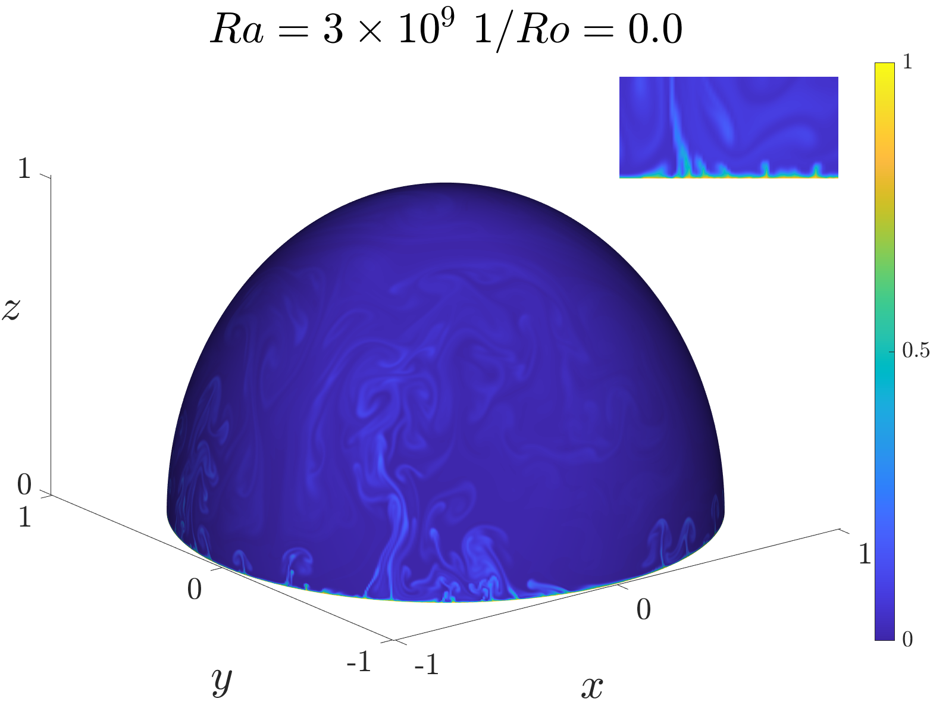}
    \includegraphics[width = 6.5cm]{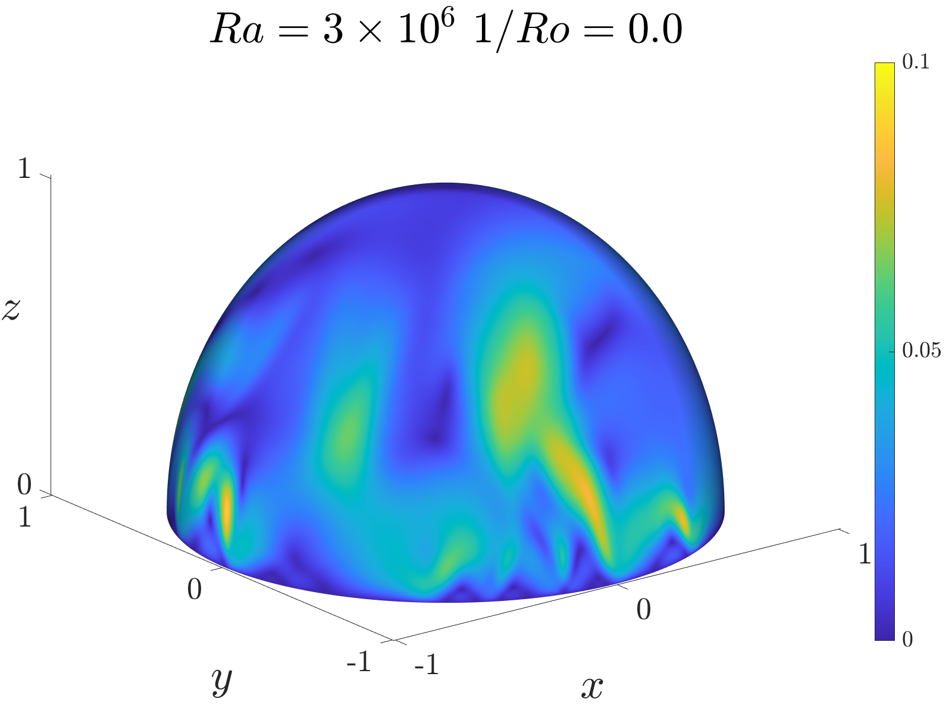}
  \includegraphics[width = 6.5cm]{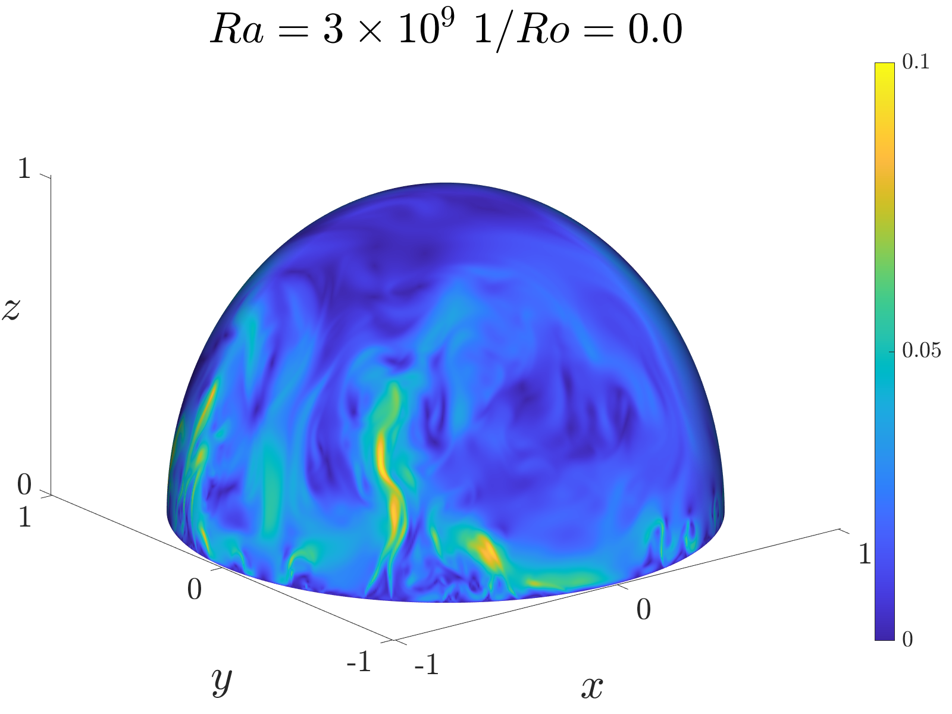}
    \includegraphics[width = 6.5cm]{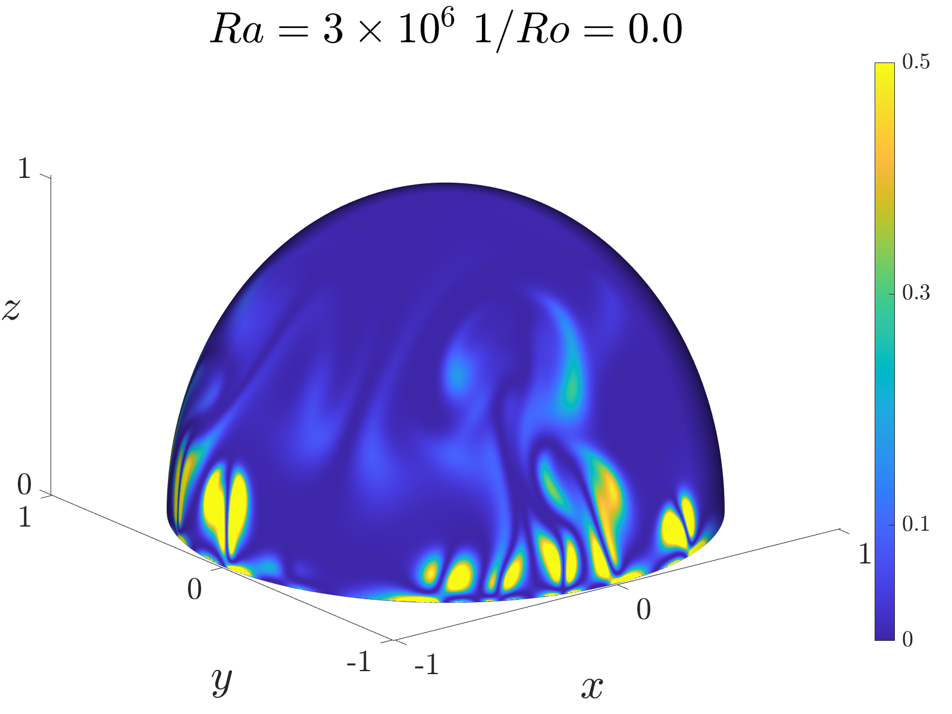}
  \includegraphics[width = 6.5cm]{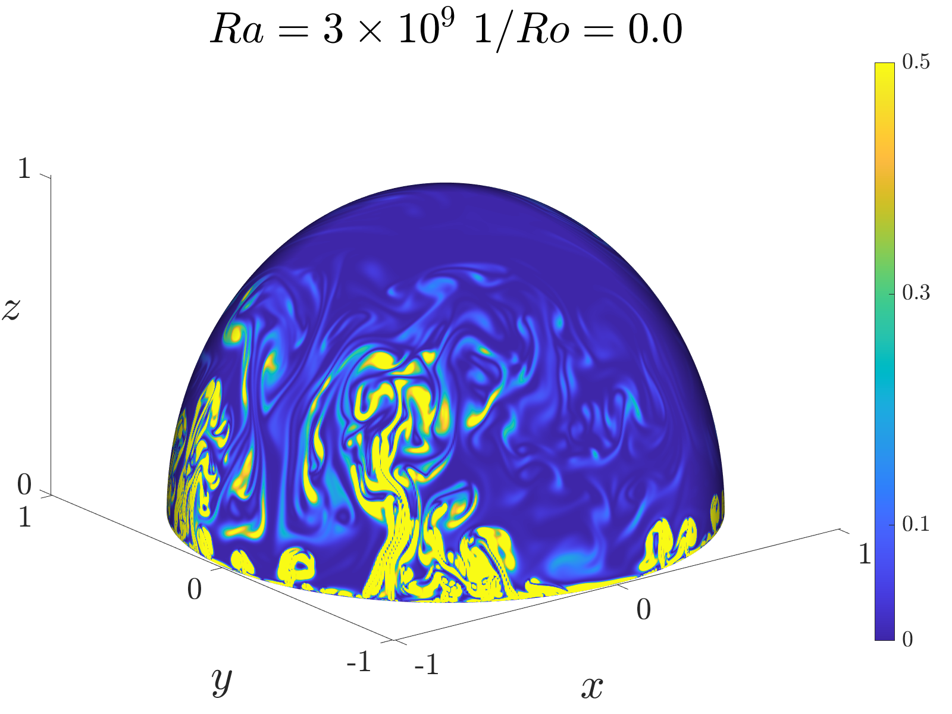}
  \caption{Instantaneous temperature (first row), TKE (second row), and enstrophy (third row) for $1/Ro=0$. Left column shows results for $Ra=3\times 10^{6}$, and right column shows results for $Ra=3\times 10^{9}$. Insets to temperature visualization highlights a section of the thermal boundary layer.  }
  \label{fig: VaryRa}
\end{figure}
\begin{figure}
  \centering
  \includegraphics[width = 6.5cm]{figs/TRa3e9R0}
  \includegraphics[width = 6.5cm]{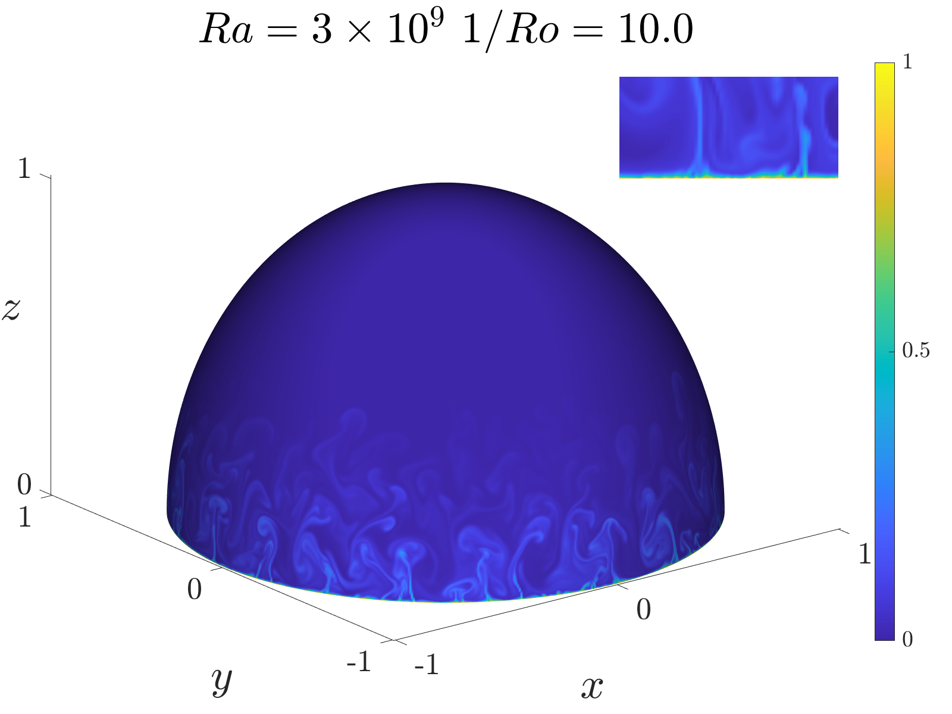}
    \includegraphics[width = 6.5cm]{figs/KRa3e9R0}
  \includegraphics[width = 6.5cm]{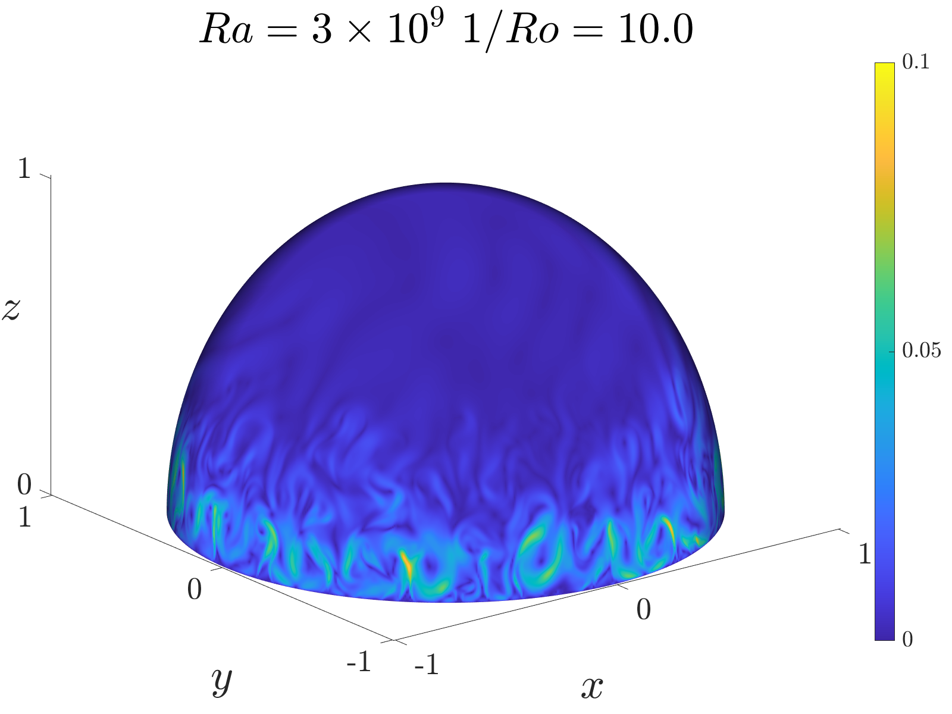}
    \includegraphics[width = 6.5cm]{figs/enstrophy3e9}
  \includegraphics[width = 6.5cm]{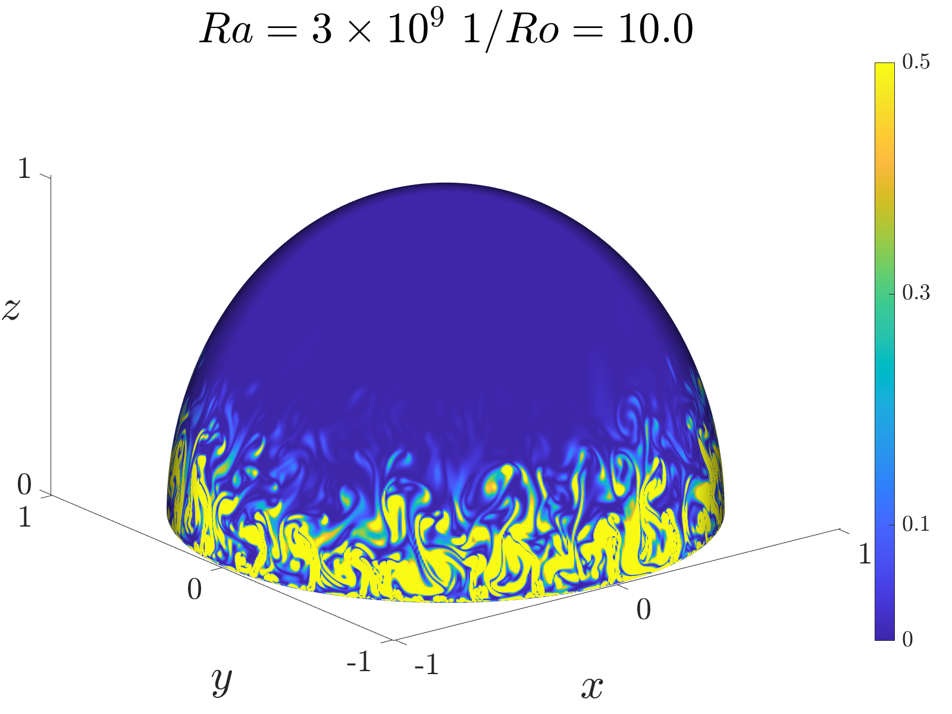}
  \caption{Instantaneous temperature (first row), TKE (second row), and enstrophy (third row) for $Ra=3\times 10^{9}$. Left column shows results for $1/Ro=0$, and right column shows results for $1/Ro=10$. Insets to temperature visualization highlights the thermal boundary layer.}
  \label{fig: VaryRo}
\end{figure}

We begin with a visual, qualitative comparison of the effects of $Ro$ and $Ra$ on the instantaneous properties of the flow. In figure \ref{fig: VaryRa} we plot the temperature, TKE, and enstrophy fields for $1/Ro=0$ and for $Ra=3\times10^6$ and $Ra=3\times10^9$ to see the effect of varying $Ra$, while in figure \ref{fig: VaryRo} we plot the same quantities for $Ra=3\times10^9$ and for $1/Ro=0$ and $1/Ro=10$ to see the effect of varying $Ro$. We note that the plumes and corresponding vortices in these visualizations are qualitatively very similar to those observed in experiments of a soap bubble \citep{MeuelCoudert11,MeuelXiong14}, and comparing the plumes for $Ra=3\times10^6$ and $Ra=3\times10^9$ in figure \ref{fig: VaryRa} we observe that the plumes become smaller and more convoluted in shape with the increase of $Ra$. This is due to the enhanced turbulence intensity as $Ra$ is increased, leading to stronger mixing in the flow. Associated with this is that the thermal boundary layer thickness reduces significantly in going from $Ra=3\times10^6$ to $Ra=3\times10^9$. Concerning the TKE and enstrophy, we find that as $Ra$ is increased, smaller scale structures in the flow emerge, with strong enstrophy occurring at higher latitudes.

The results in figure \ref{fig: VaryRo} show that for fixed $Ra$, as $1/Ro$ is increased the turbulent activity in the flow becomes restricted to lower latitudes where buoyancy is still strong enough to overcome the suppressing influence of the Coriolis force. The insets that highlight the thermal boundary layer indicate that the boundary layer and its thickness are only weakly affected by rotation. This is likely due to the fact that the Coriolis force is most active at the largest scales of the system, and plays a weaker role at the small scales of the flow, such as those that characterize the thin boundary layer at $Ra=3\times 10^9$.

\subsection{The behavior of the Reynolds and Nusselt numbers}
We now turn to quantitatively analyze the statistics of the flow, beginning with an examination of the behavior of the Reynolds $Re$ and Nusselt $Nu$ numbers in the flow. We remind the reader that based on their definitions, these are emergent properties of the flow, that depend upon the control variables $Ra, Ro$ and on the coordinate $\theta$.

\begin{figure}
  \centering
  \includegraphics[width = 6.5cm]{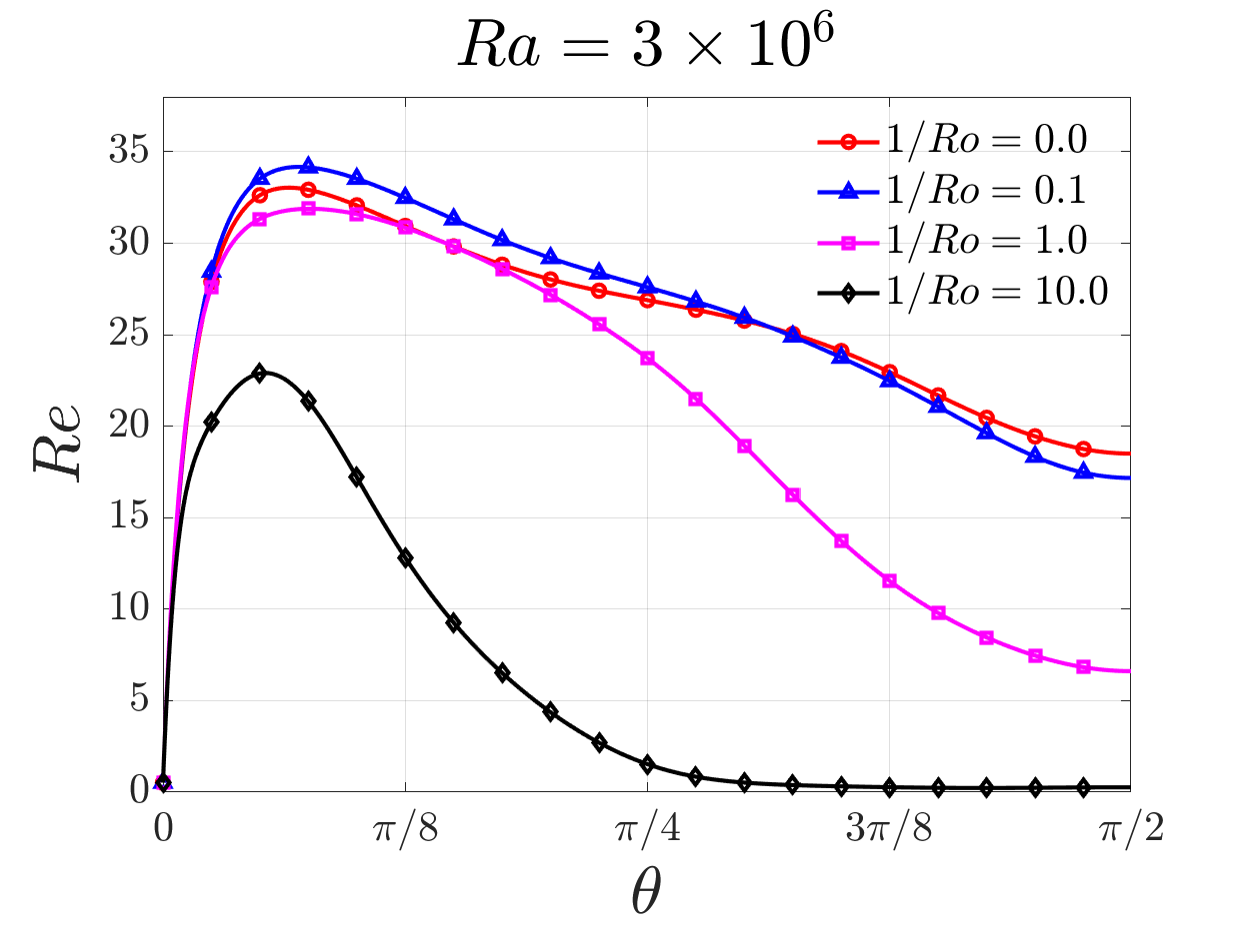}
    \includegraphics[width = 6.5cm]{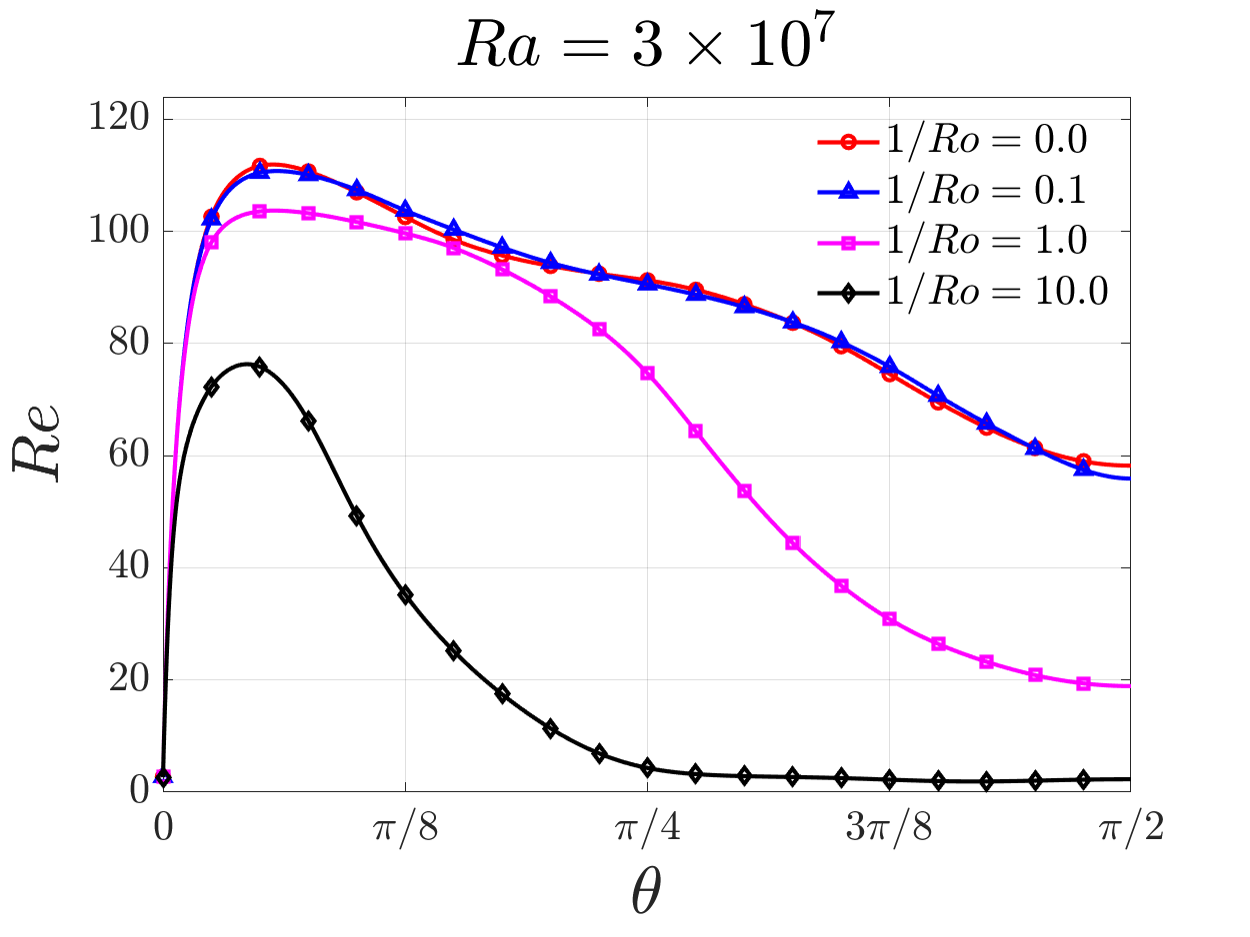}
  \includegraphics[width = 6.5cm]{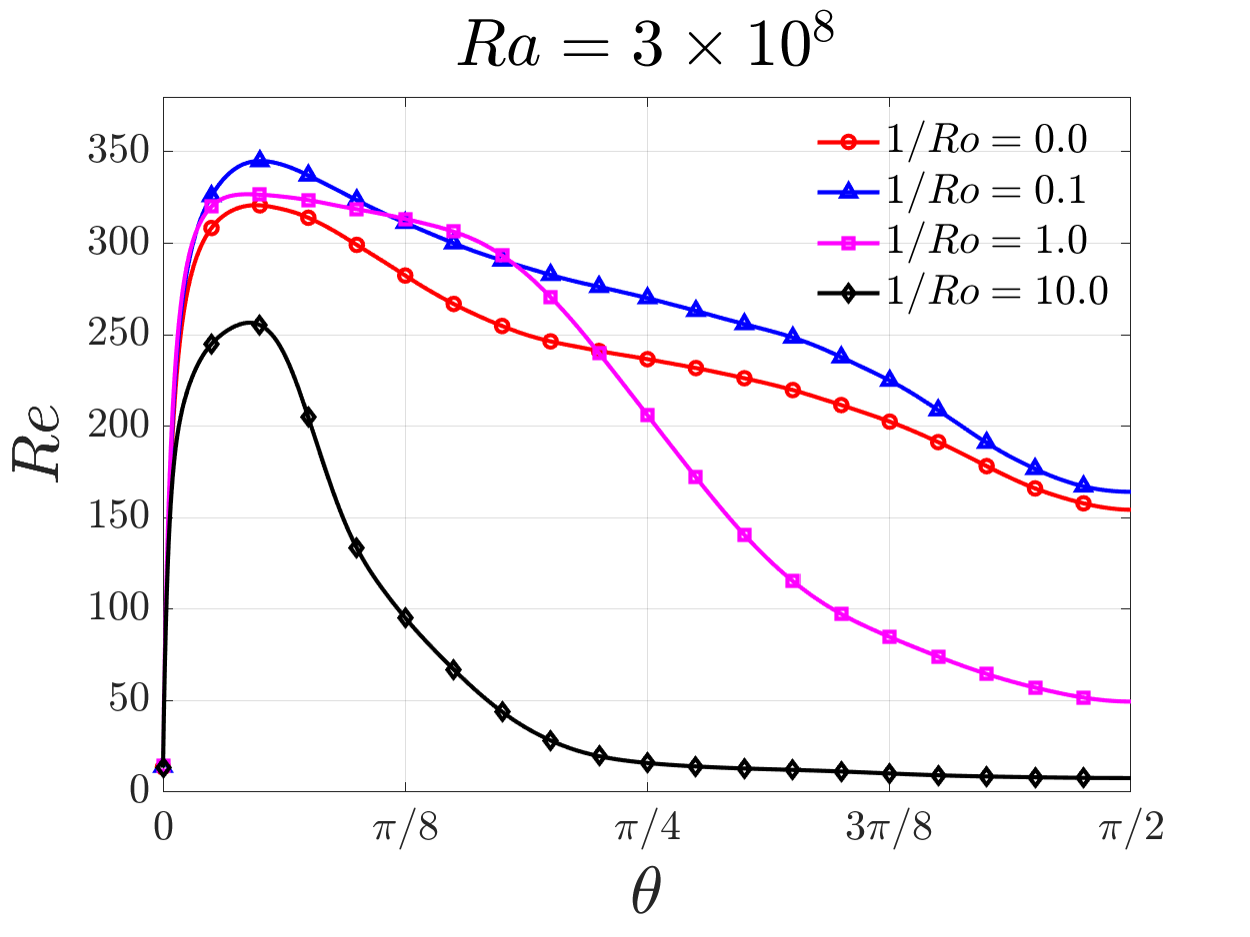}
      \includegraphics[width = 6.5cm]{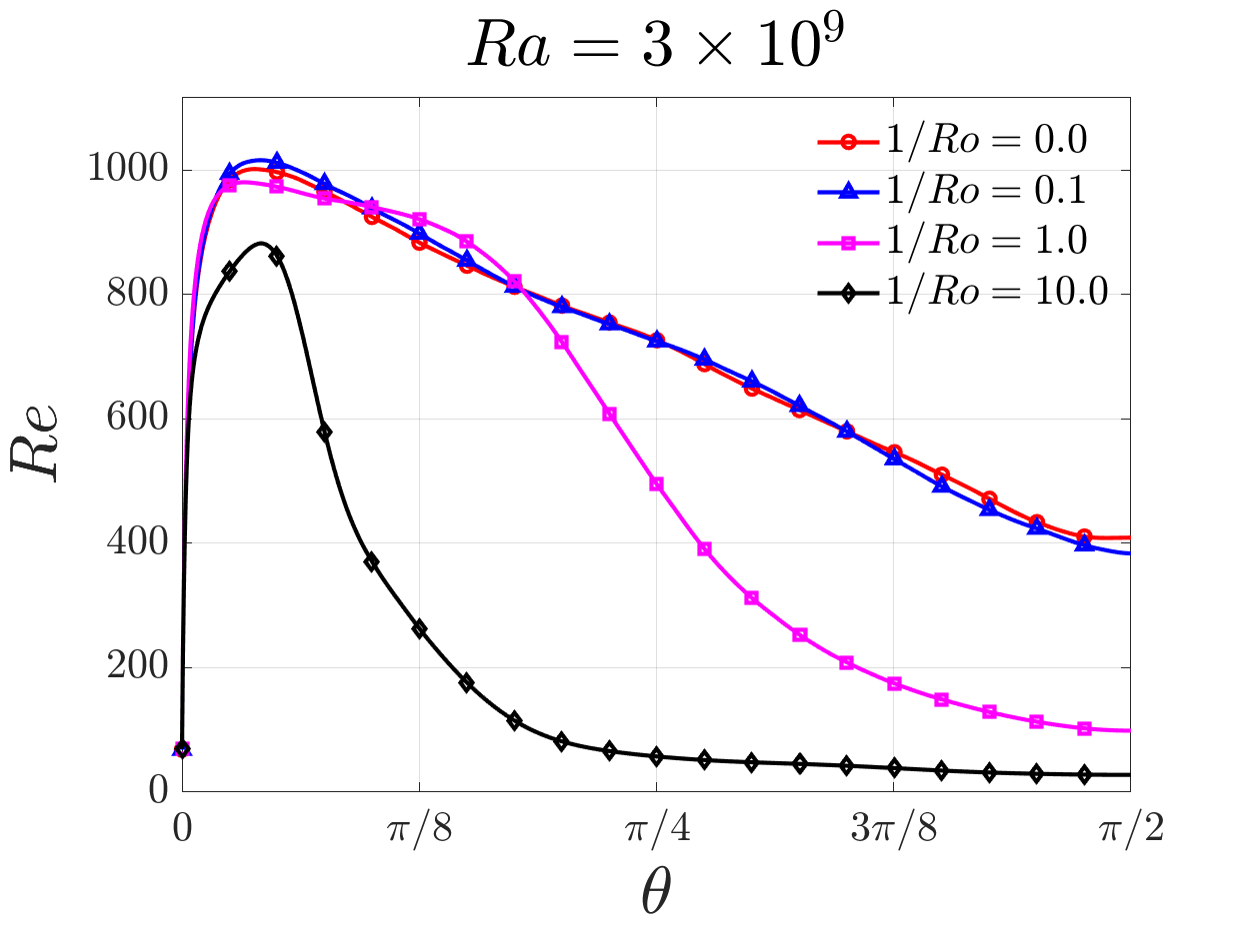}
  \caption{Variation of $Re$ with $\theta$ for varying $1/Ro$ and $Ra$. }
  \label{fig:rero1}
\end{figure}

In the figure \ref{fig:rero1} the variation of $Re$ for different $Ra$ and $1/Ro$ is shown. In each case, $Re$ reaches a maximum at some intermediate $\theta$, being small near the equator due to the no-slip condition, and small near the North Pole where buoyancy forces are weak since the heating is at the equator. The effect of $Ro$ on $Re$ is somewhat subtle, leading to an increase for some latitudes, and a decrease for others. However, once $1/Ro\geq 1$ the suppression of turbulence at higher latitudes becomes evident. This suppression occurs because as $\theta$ increases, the buoyancy force decreases, and the Coriolis force becomes dominant. The Coriolis force does not generate TKE, and the Taylor-Proudman effect (combined with the fact that $\langle U_\theta\rangle=0$ for this flow) inhibits transport in the $\theta$ direction. As a result, TKE is not able to be transported to the top of the bubble. However, as $Ra$ is increased for fixed $Ro$, the convection becomes stronger and the Reynolds number increases at high latitudes.
\begin{figure}
  \centering
  \includegraphics[width = 6.5cm]{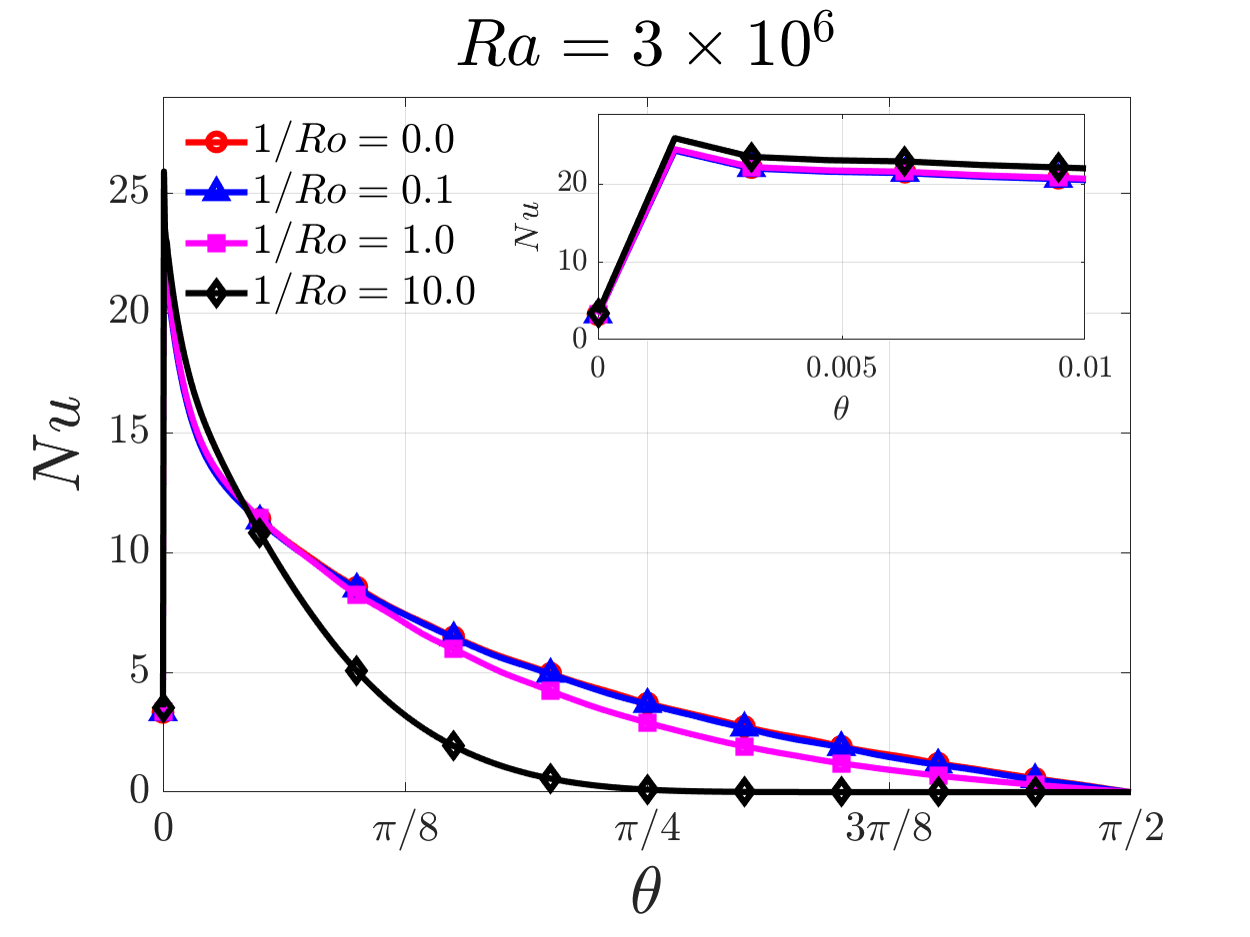}
    \includegraphics[width = 6.5cm]{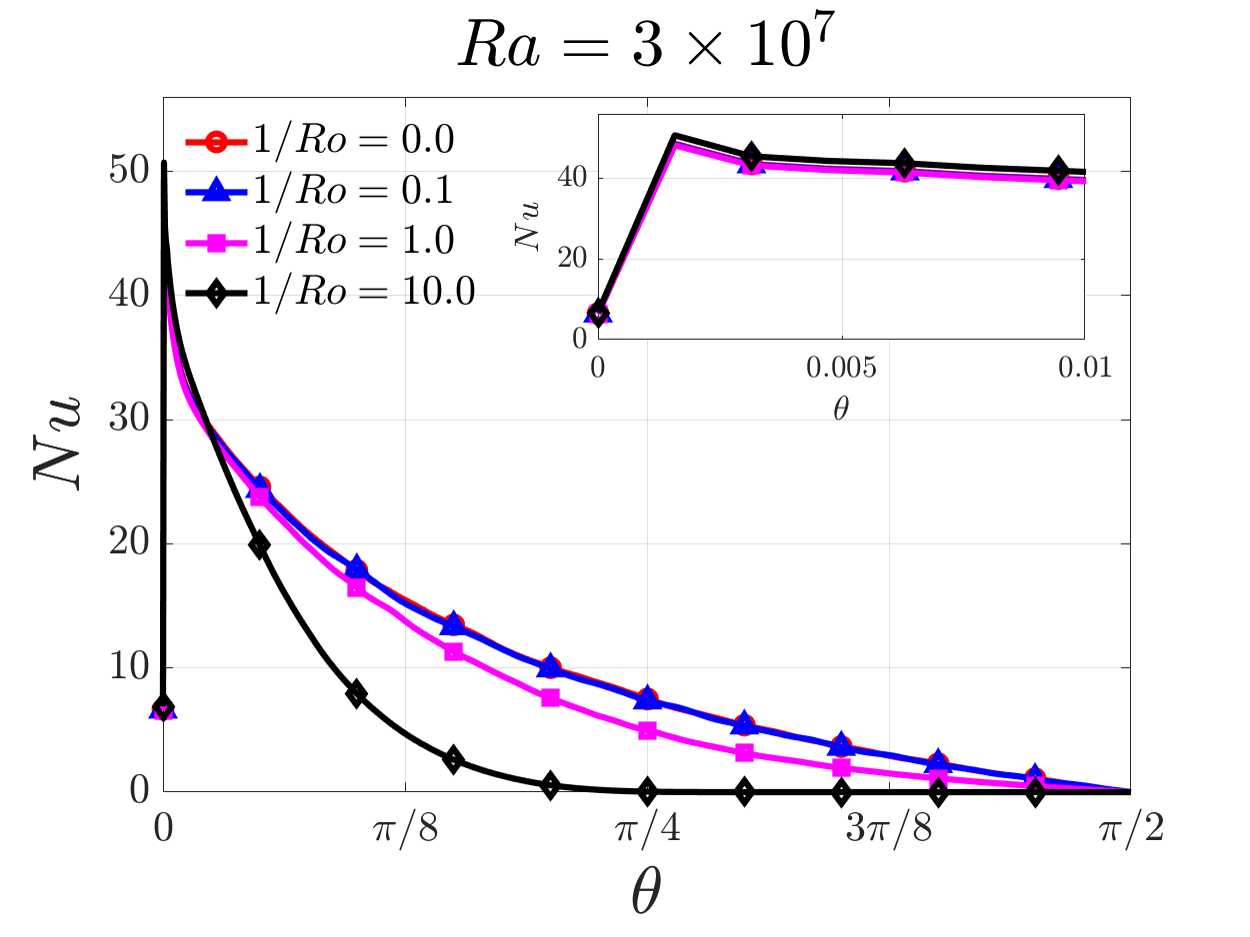}
      \includegraphics[width = 6.5cm]{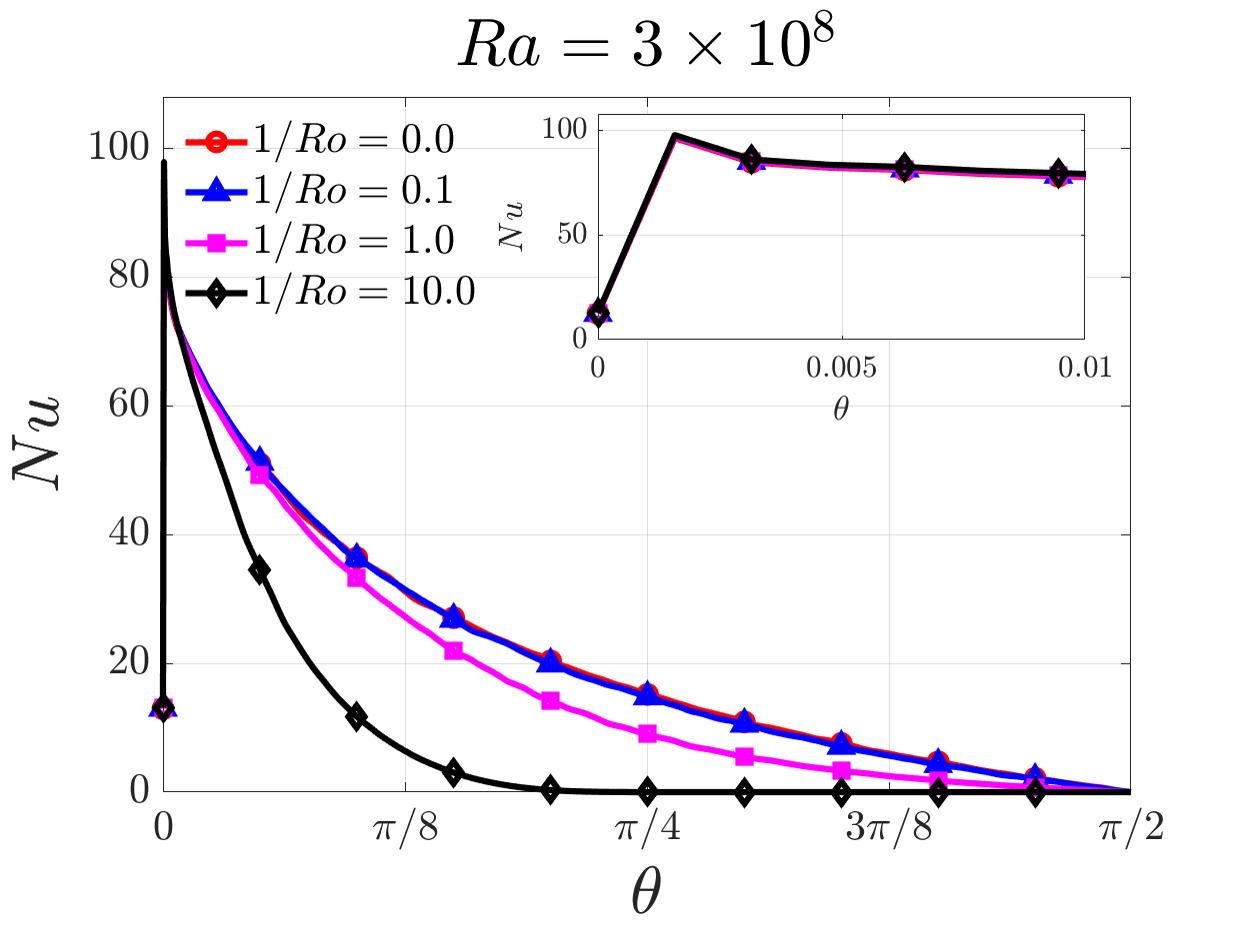}
        \includegraphics[width = 6.5cm]{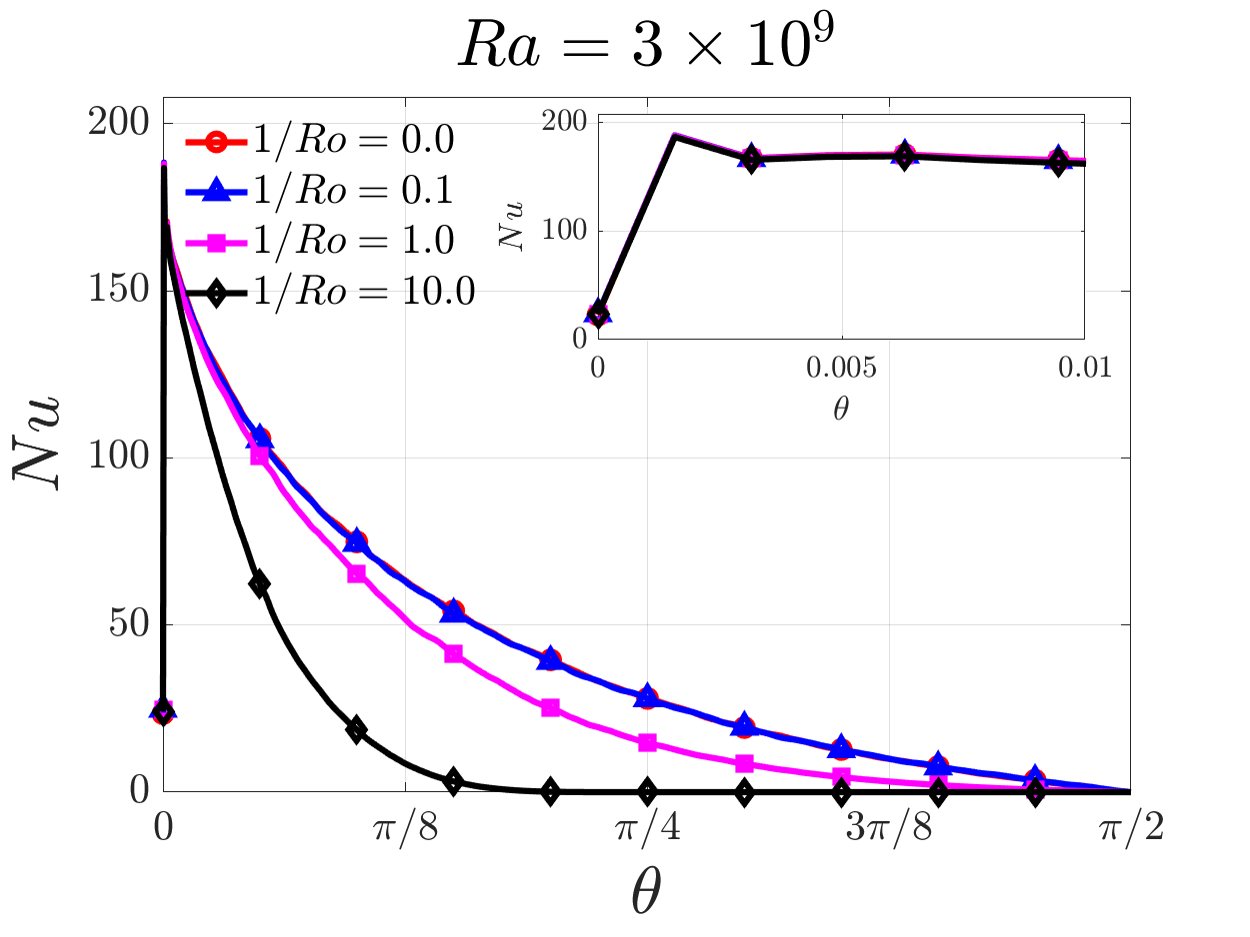}
  \caption{Variation of $Nu$ with $\theta$ for varying $1/Ro$ and $Ra$. Insets highlight behavior for small $\theta$.}
  \label{fig:nura1}
\end{figure}


In figure \ref{fig:nura1} we consider the Nusselt number $Nu$ for different $Ra$ and $1/Ro$. For the lowest $Ra$ cases, there is evidence that at low latitudes, $Nu$ reaches slightly higher values as $1/Ro$ is increased, which may be due to Eckman transport in the boundary layer. This is similar to the enhancement of $Nu$ due to rotation for intermediate $1/Ro$ that is observed in standard RRBC, i.e. regime \uppercase\expandafter{\romannumeral 2} discussed in the introduction. At higher latitudes, $Nu$ is dramatically suppressed with increasing $1/Ro$. This is due to the suppression of turbulence in the flow due to the Coriolis force that becomes increasingly important as $\theta$ increases, and the fact that buoyancy becomes weaker as $\theta$ increases. In this sense, the behavior of the flow we are considering is quite different at high latitudes from RRBC.

\subsection{The mean flows}

\begin{figure}
  \centering
  \includegraphics[width = 6.5cm]{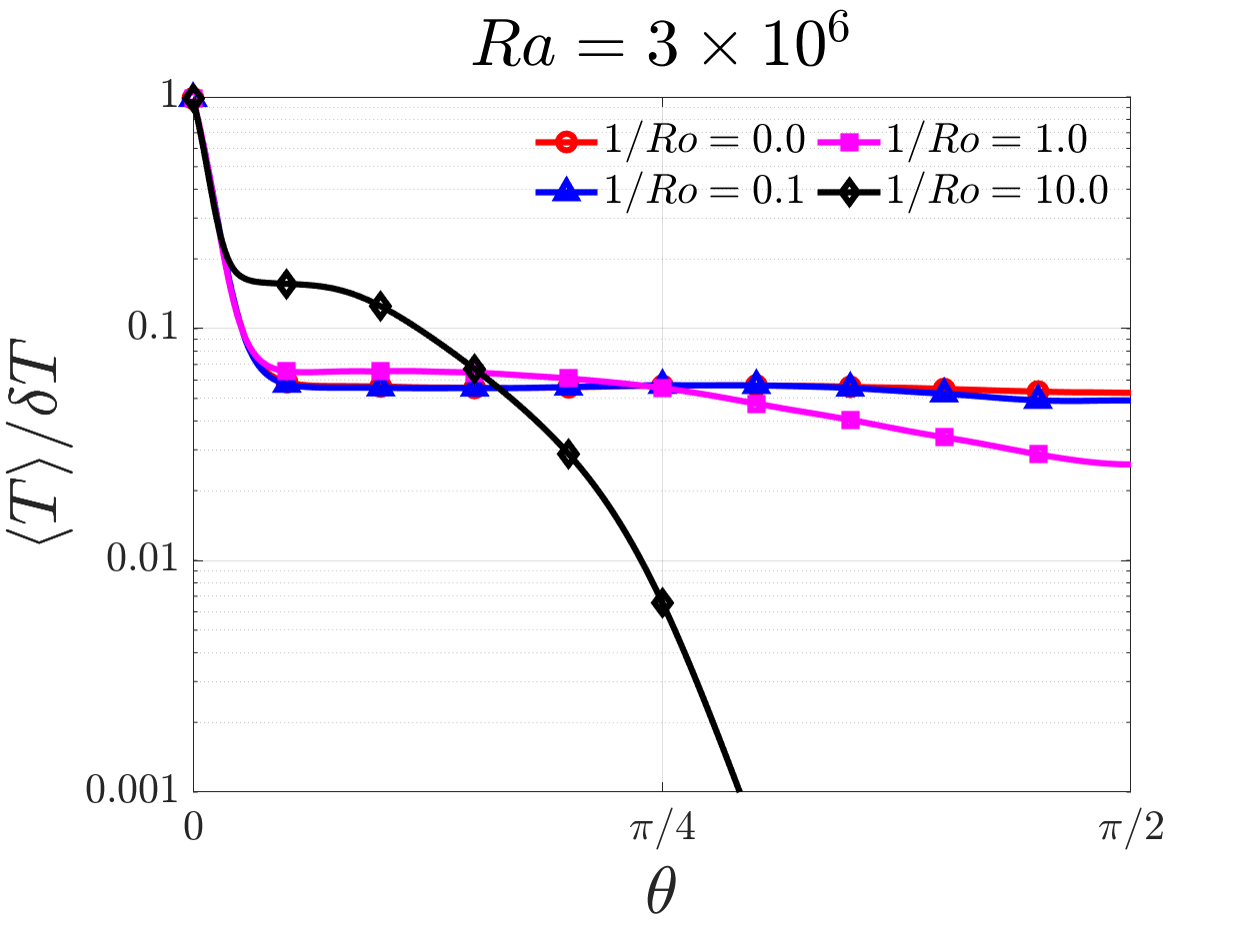}
  \includegraphics[width = 6.5cm]{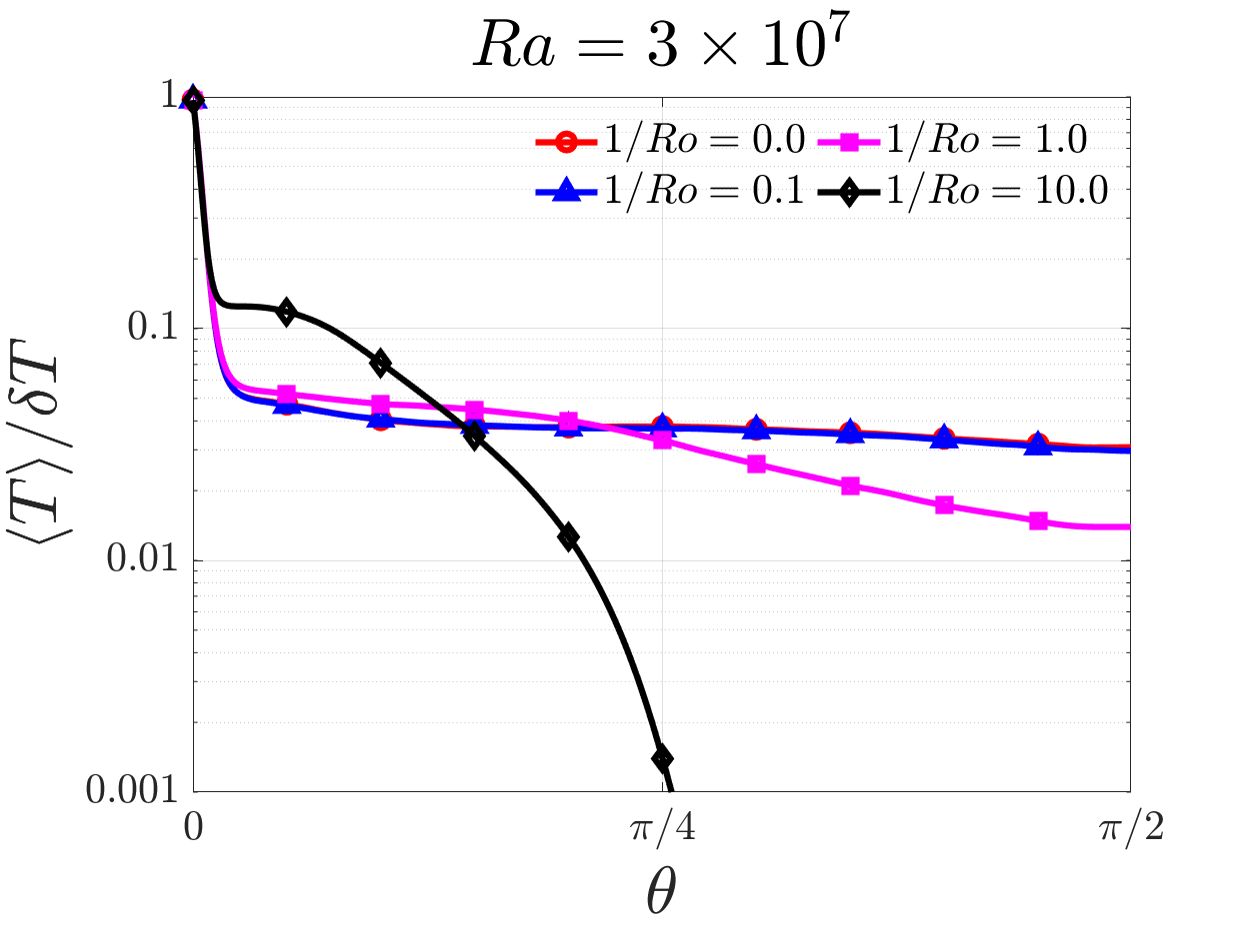}
  \includegraphics[width = 6.5cm]{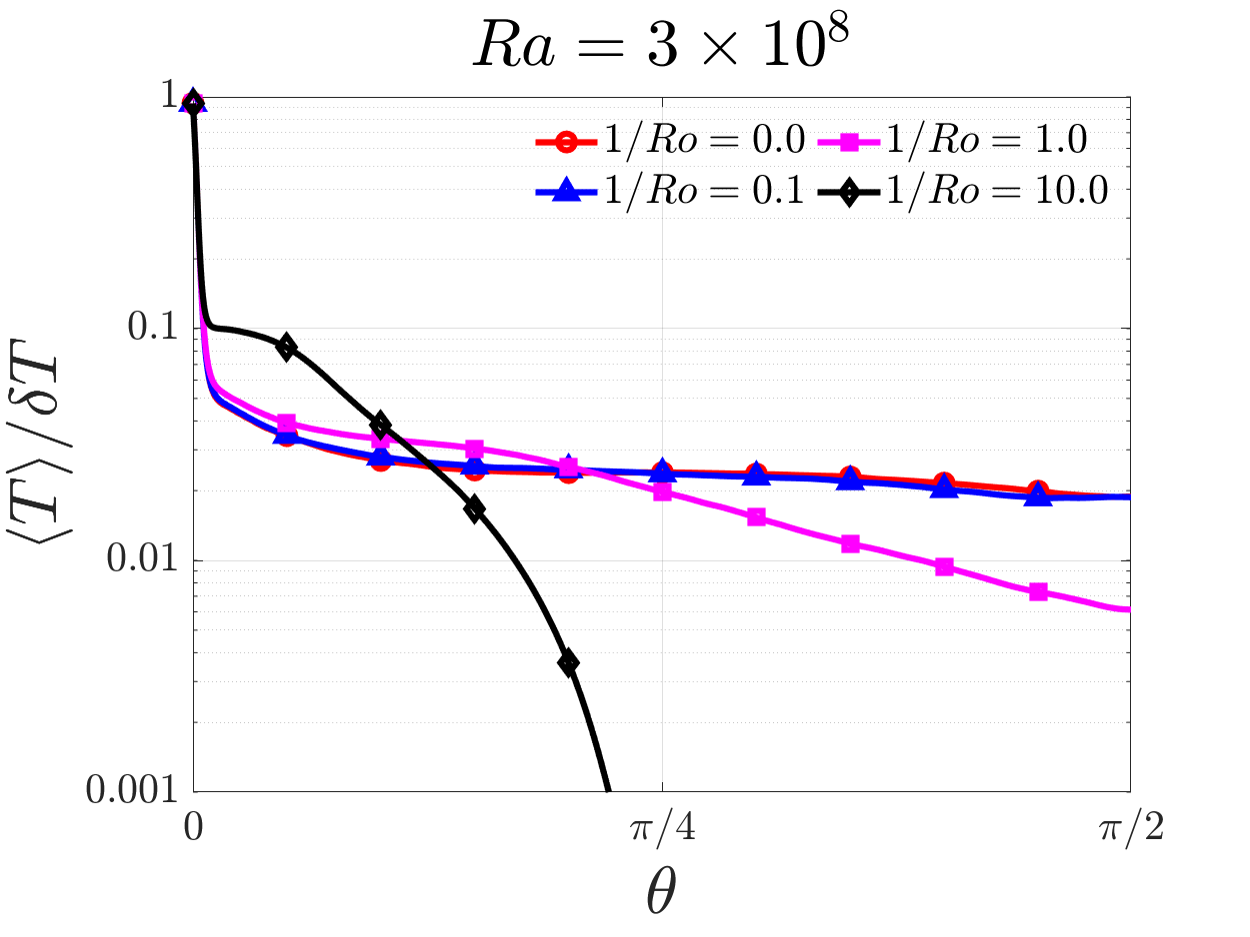}
  \includegraphics[width = 6.5cm]{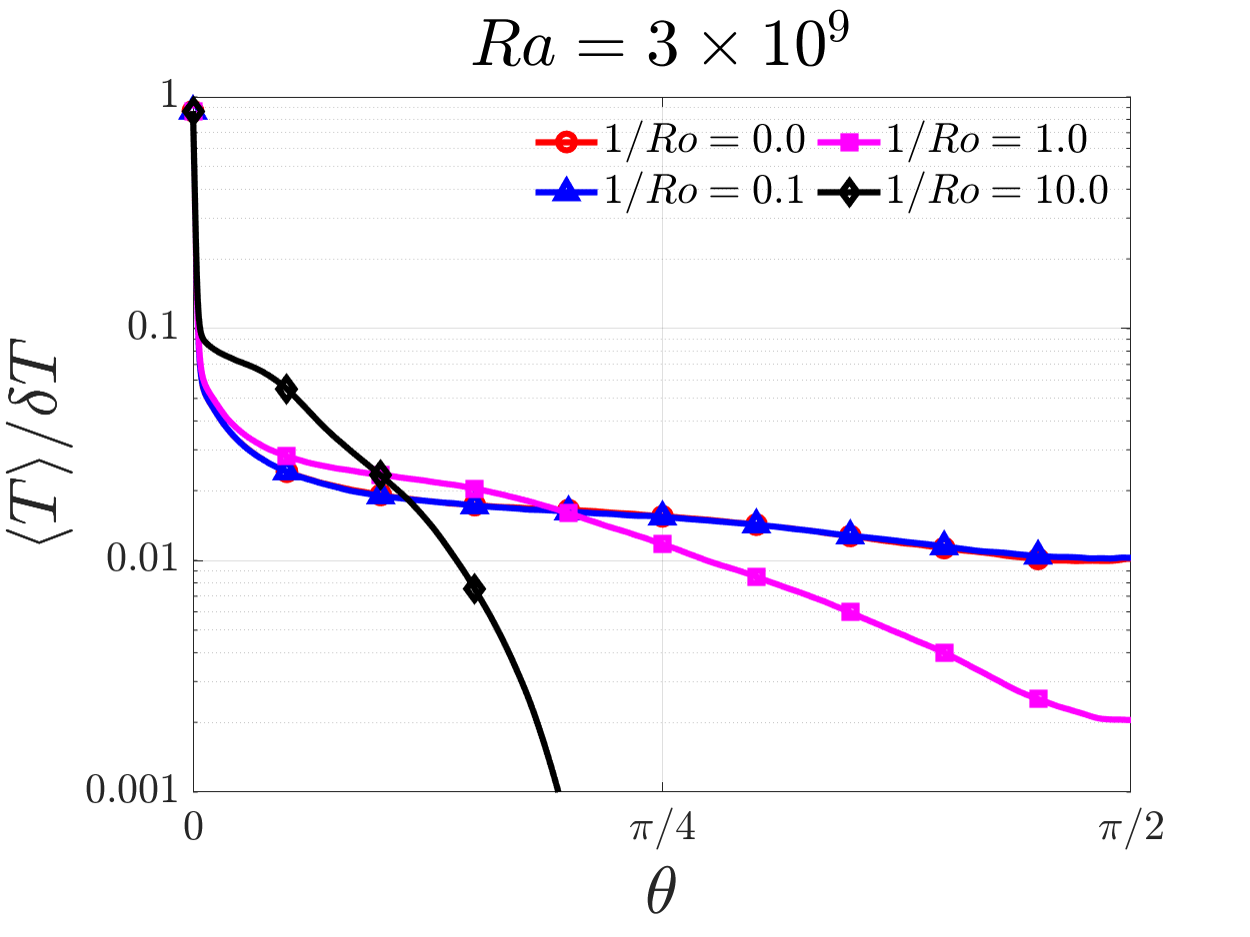}
  \caption{The averaged temperature profiles $\langle T\rangle$, normalized by $\delta T$, as a function of latitude, and for different $Ra, 1/Ro$. }
  \label{fig:tmp1}
\end{figure}
In figure \ref{fig:tmp1} we plot the normalized mean temperature $\langle T\rangle/\delta T$, for different $Ra$ and $1/Ro$. When $1/Ro<1$, $\langle T\rangle/\delta T$ is weakly affected by rotation since the role of the Coriolis force is subleading in this regime. For $1/Ro\geq 1$, the Coriolis force and the associated Taylor-Proudman effect inhibits thermal transport at high latitudes, causing $\langle T\rangle/\delta T$ to reduce significantly as $1/Ro$ is increased. However, since in this regime heat transfer towards higher latitudes is significantly reduced, heat accumulates at lower latitudes. This explains the increase in $\langle T\rangle/\delta T$ that can be seen in figure \ref{fig:tmp1} for $1/Ro\geq1$ at lower latitudes.
\begin{figure}
  \centering
  \includegraphics[width = 6.5cm]{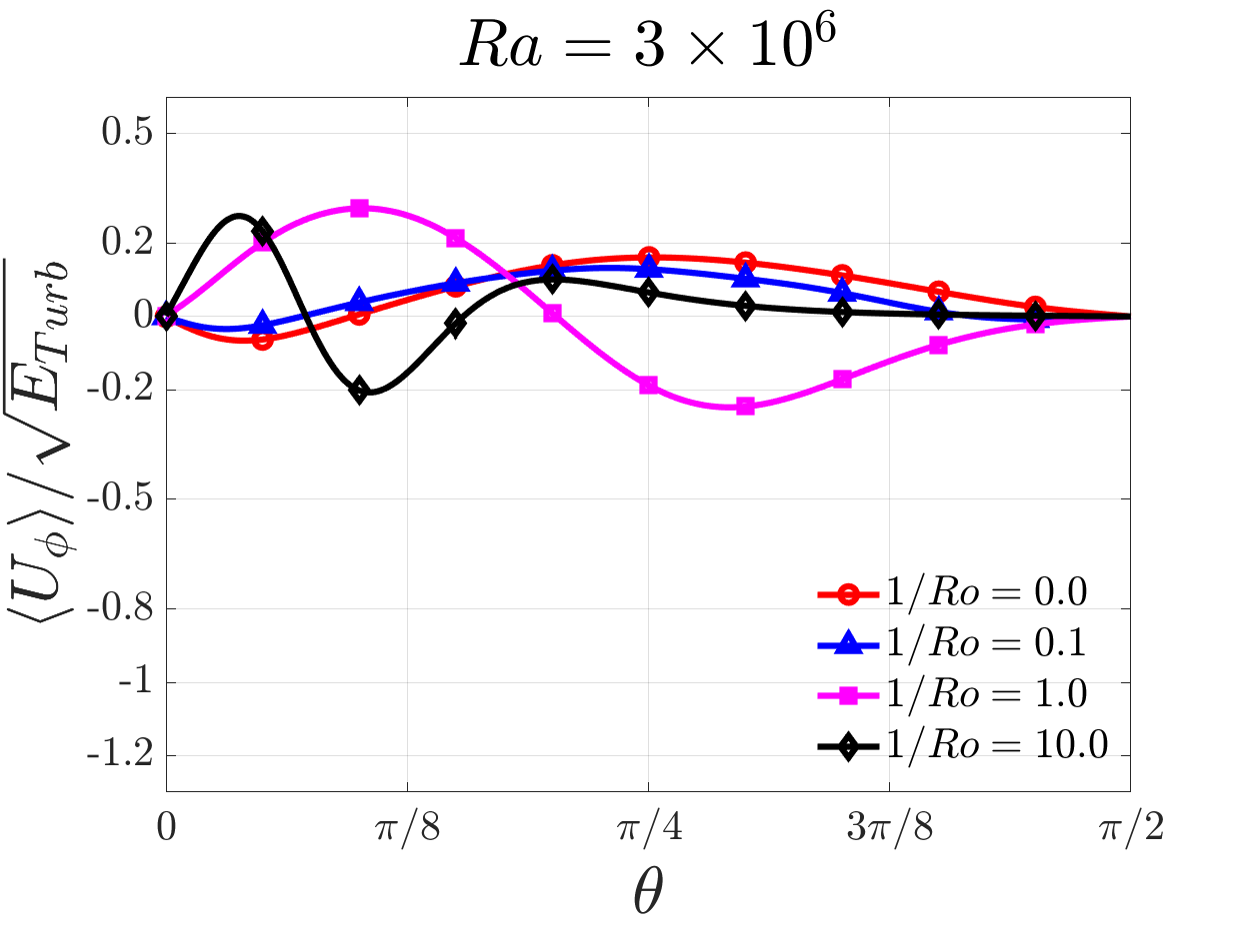}
  \includegraphics[width = 6.5cm]{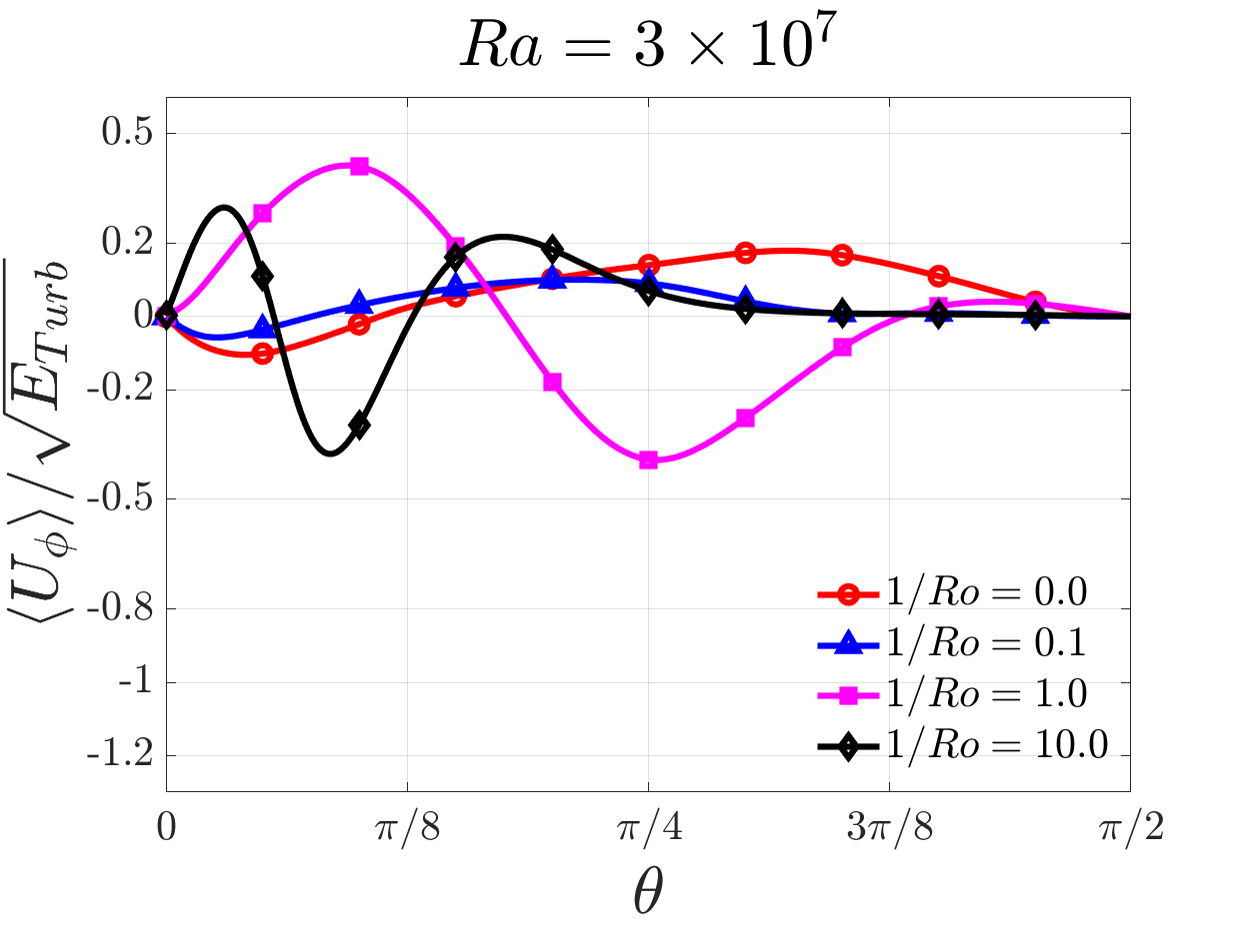}
  \includegraphics[width = 6.5cm]{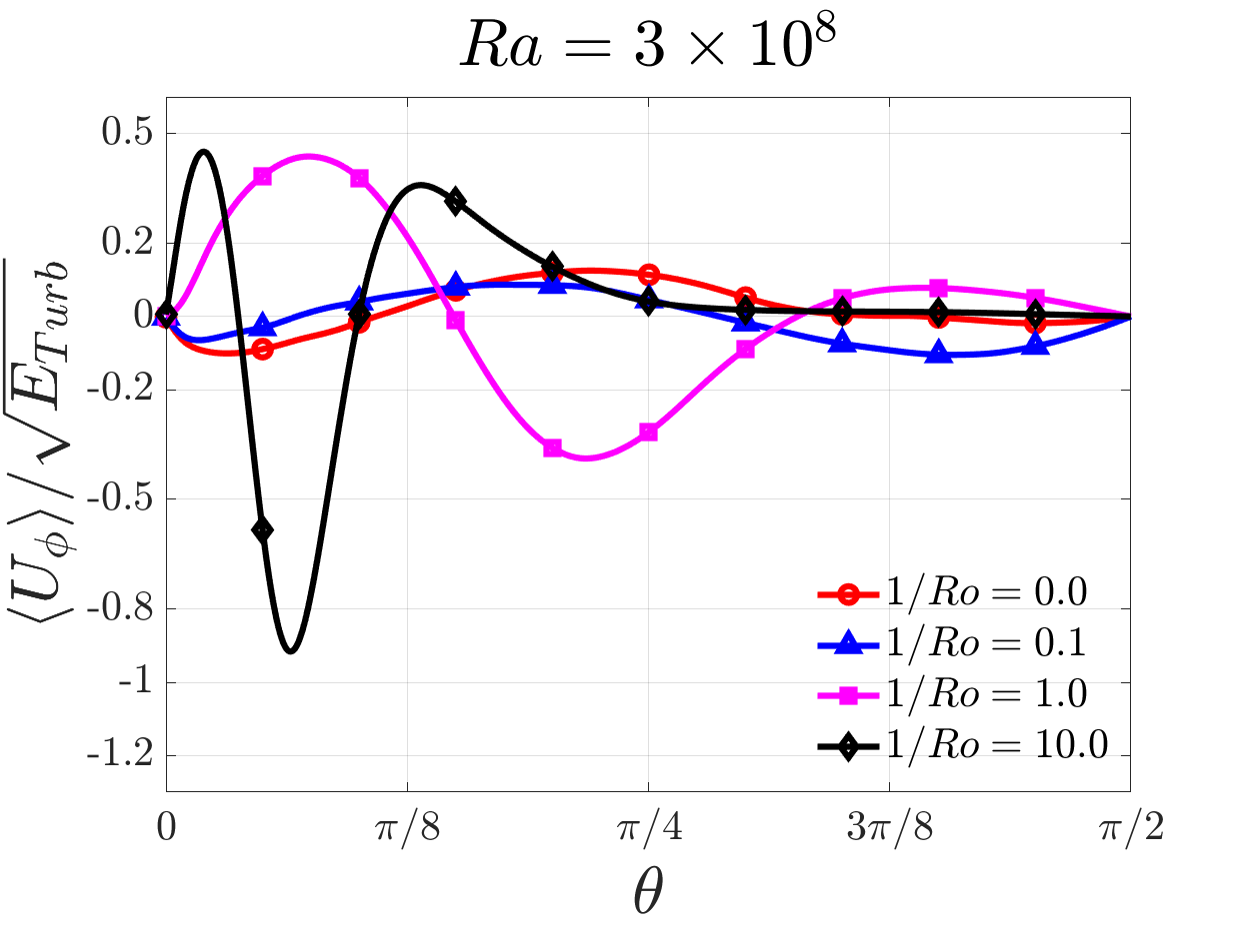}
  \includegraphics[width = 6.5cm]{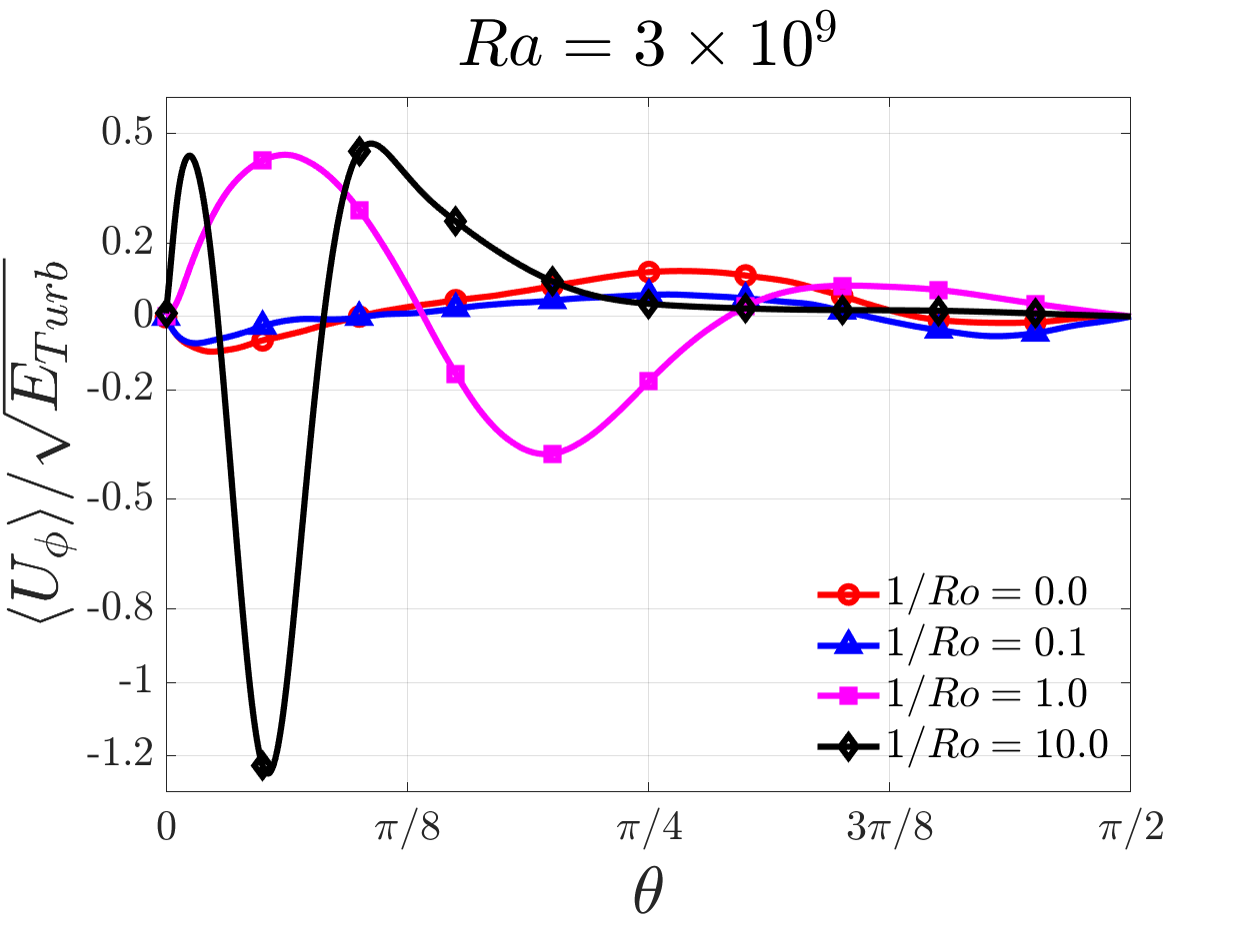}
  \caption{The mean longitudinal velocity $\langle U_{\phi}\rangle$, normalized by $\sqrt{E_{turb}}$, as a function of latitude, and for different $Ra, 1/Ro$. }
  \label{fig:uPhivTheta}
\end{figure}

In figure \ref{fig:uPhivTheta} we show results for $\langle U_{\phi}\rangle$ normalized by $\sqrt{E_{turb}}$, for different $Ra$ and $1/Ro$. In RBC, large-scale circulation (LSC) are known to play an important role in the flow. Our results in figure \ref{fig:uPhivTheta} for $1/Ro=0$ indicate that LSC are also present in our flow, with the sign of $\langle U_{\phi}\rangle$ varying with $\theta$. In particular,  $\langle U_{\phi}\rangle$ is negative at low latitudes, then becomes positive, and for the highest $Ra$ cases, becomes negative again near the North Pole. It should be noted that LSC are known to persist for very long periods of time compared with the integral timescale, and it is therefore possible that our mean-flow results have not fully converged in a statistical sense.

The effect of $1/Ro$ on $\langle U_{\phi}\rangle$ appears to be complicated. At higher latitudes, as $1/Ro$ is increased, $\langle U_{\phi}\rangle$ is suppressed due to dominance of the Coriolis force over the buoyancy force that is responsible for generating the LSC in the first place. At lower latitudes where buoyancy is still strong enough to create LSC, the behavior of $\langle U_{\phi}\rangle$ is profoundly affected by rotation when $1/Ro\geq 1$. In particular, the sign of $\langle U_{\phi}\rangle$ at low latitudes changes and becomes positive, but its sign changes multiple times as $\theta$ is increased. This may be due to inertial waves produced by rotation interacting nonlinearly with the buoyancy mechanisms that generate LSC, but given that the flow is highly nonlinear and turbulent it is difficult to know, and a more detailed investigation of this interesting behavior is left for future work.

\subsection{Temperature fluctuations}
\begin{figure}
  \centering
  \includegraphics[width = 6.5cm]{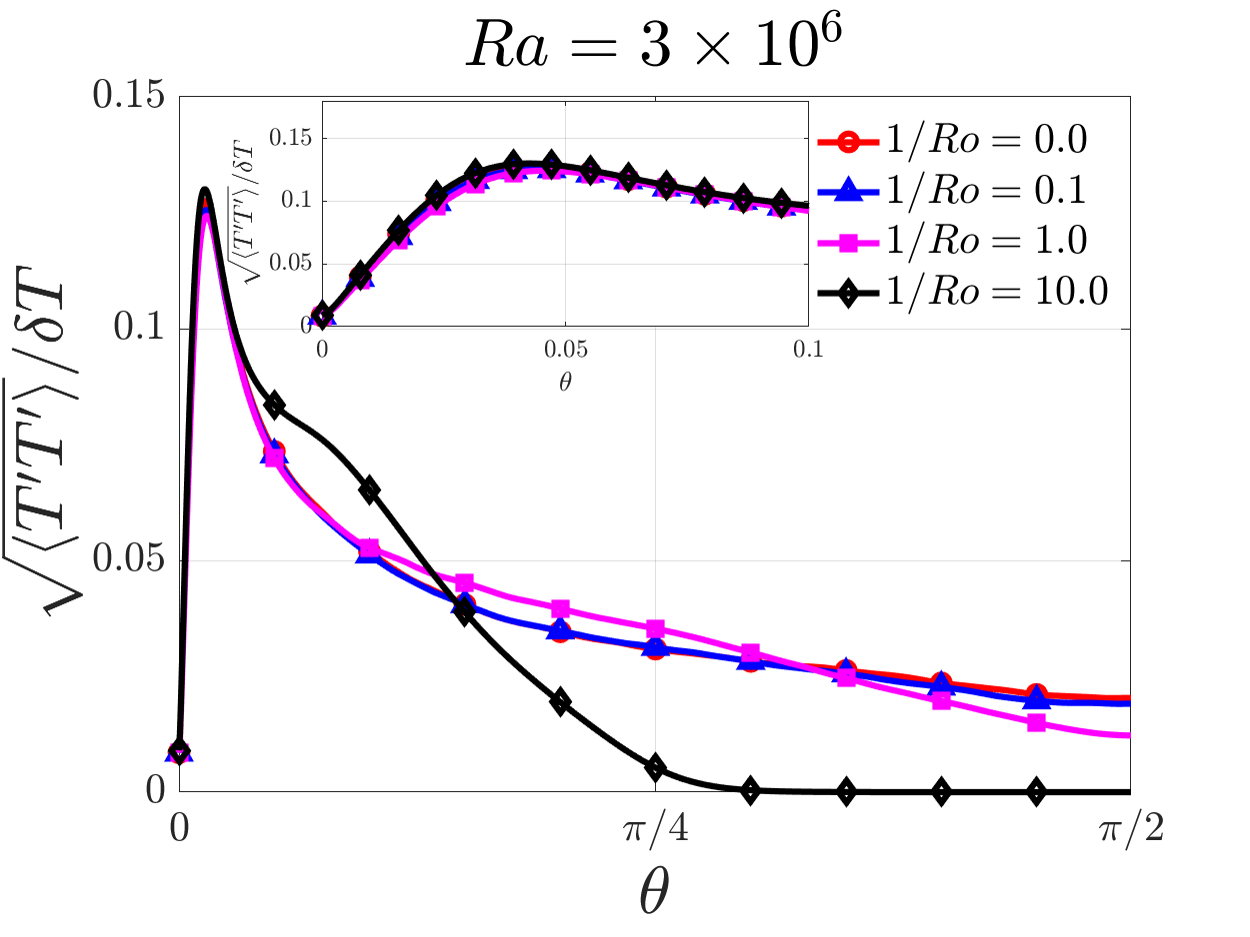}
  \includegraphics[width = 6.5cm]{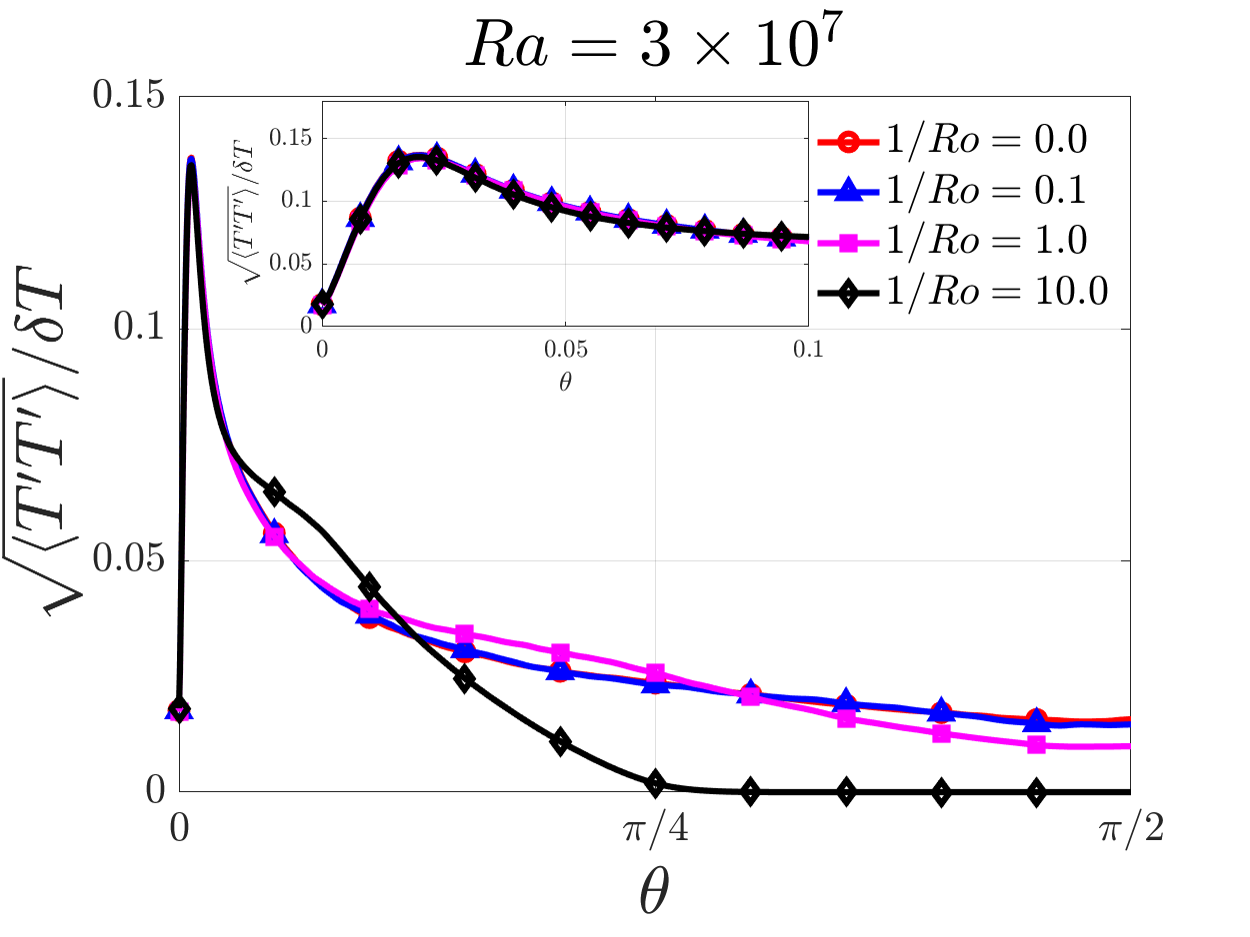}
  \includegraphics[width = 6.5cm]{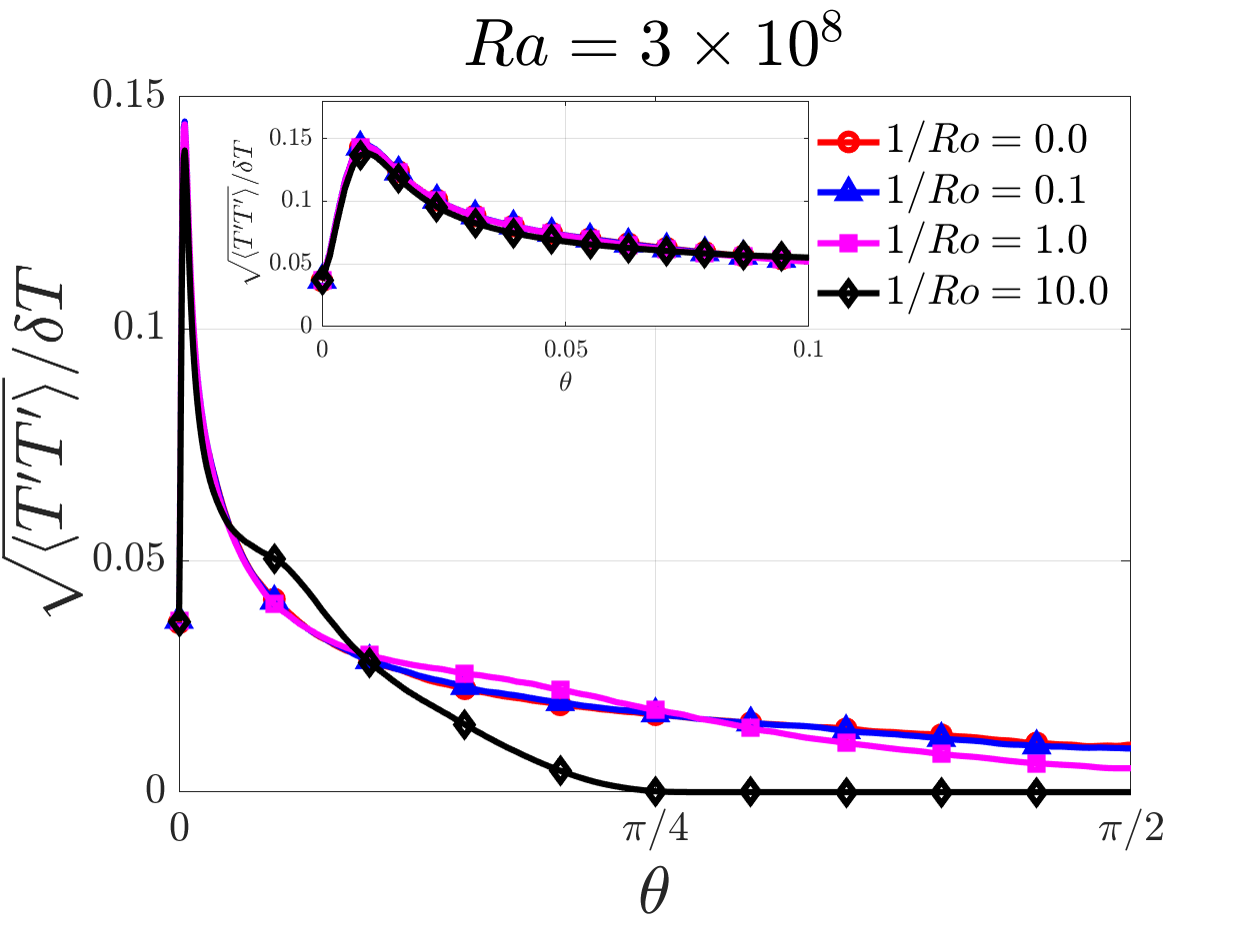}
  \includegraphics[width = 6.5cm]{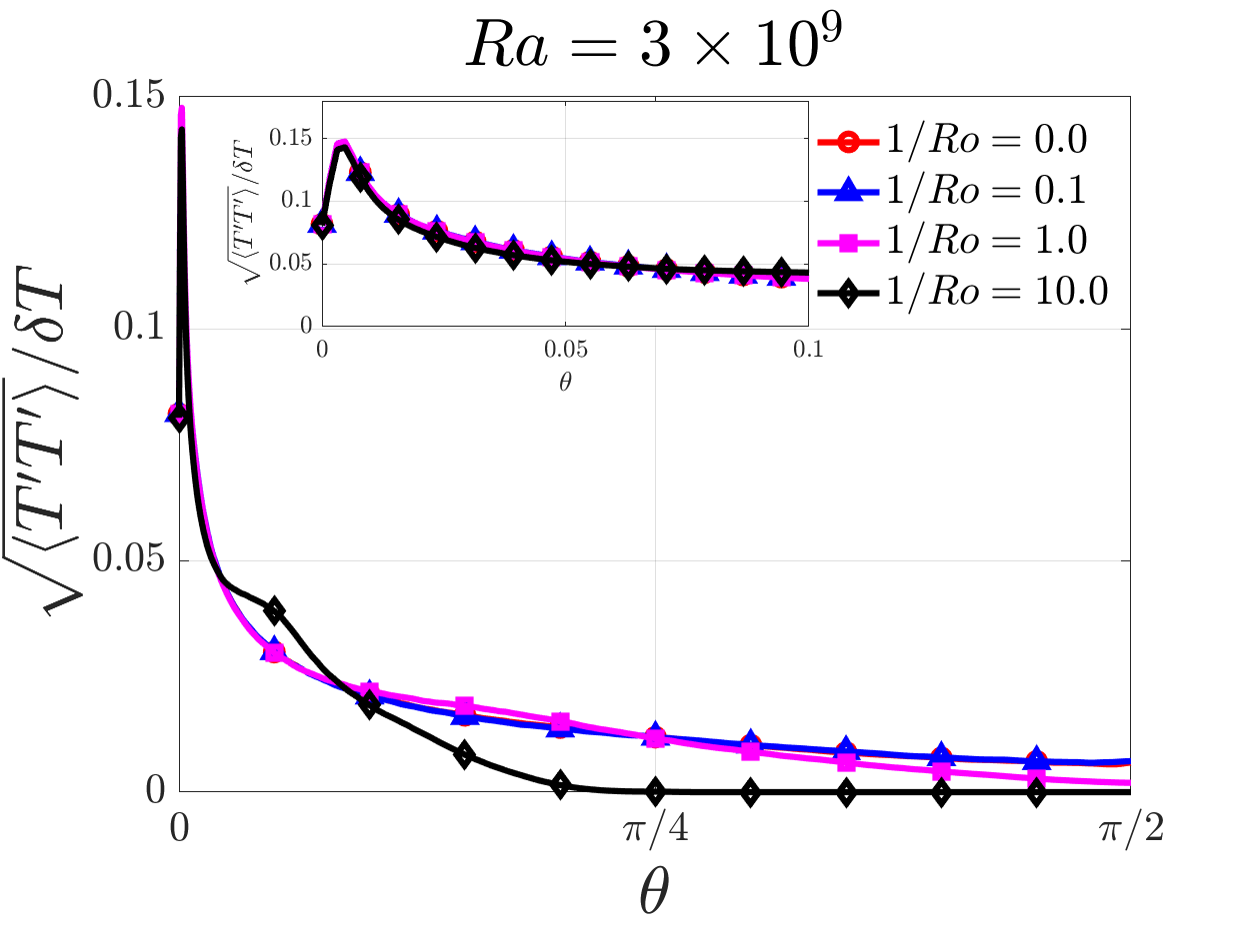}
  \caption{The rms fluctuating temperature $\sqrt{\langle T'T'\rangle}$, normalized by $\delta T$, as a function of latitude, and for different $Ra, 1/Ro$ }
  \label{fig:Trms}
\end{figure}

In figure \ref{fig:Trms} we plot the results for the rms fluctuating temperature $\sqrt{\langle T'T'\rangle}$, normalized by $\delta T$, as a function of latitude, and for different $Ra, 1/Ro$. For $1/Ro=0$, as $Ra$ is increased, the main effect is to simply shrink the thermal boundary layer, with the temperature fluctuations at high latitudes becoming weaker as $Ra$ is increased. As $1/Ro$ is increased, at higher latitudes the thermal fluctuations are significantly suppressed as the Coriolis force inhibits the transport of thermal fluctuations away from the boundary layer. However, in the vicinity of $\theta=\pi/16$ (the precise region probably depends on $Ra$), we see that increasing $1/Ro$ actually increases the thermal fluctuations. This effect may be due to something similar to Eckman suction wherein hot fluid is sucked out from the boundary layer due to the transport produced by viscous and Coriolis forces on the fluid. For $1/Ro$, the enhancement in this region reduces with increasing $Ra$, however, it is possible that the enhancement would remain significant if $1/Ro$ were sufficiently large.

\subsection{The Reynolds stress tensor and its anisotropy}

\begin{figure}
  \centering
  \includegraphics[width = 6.5cm]{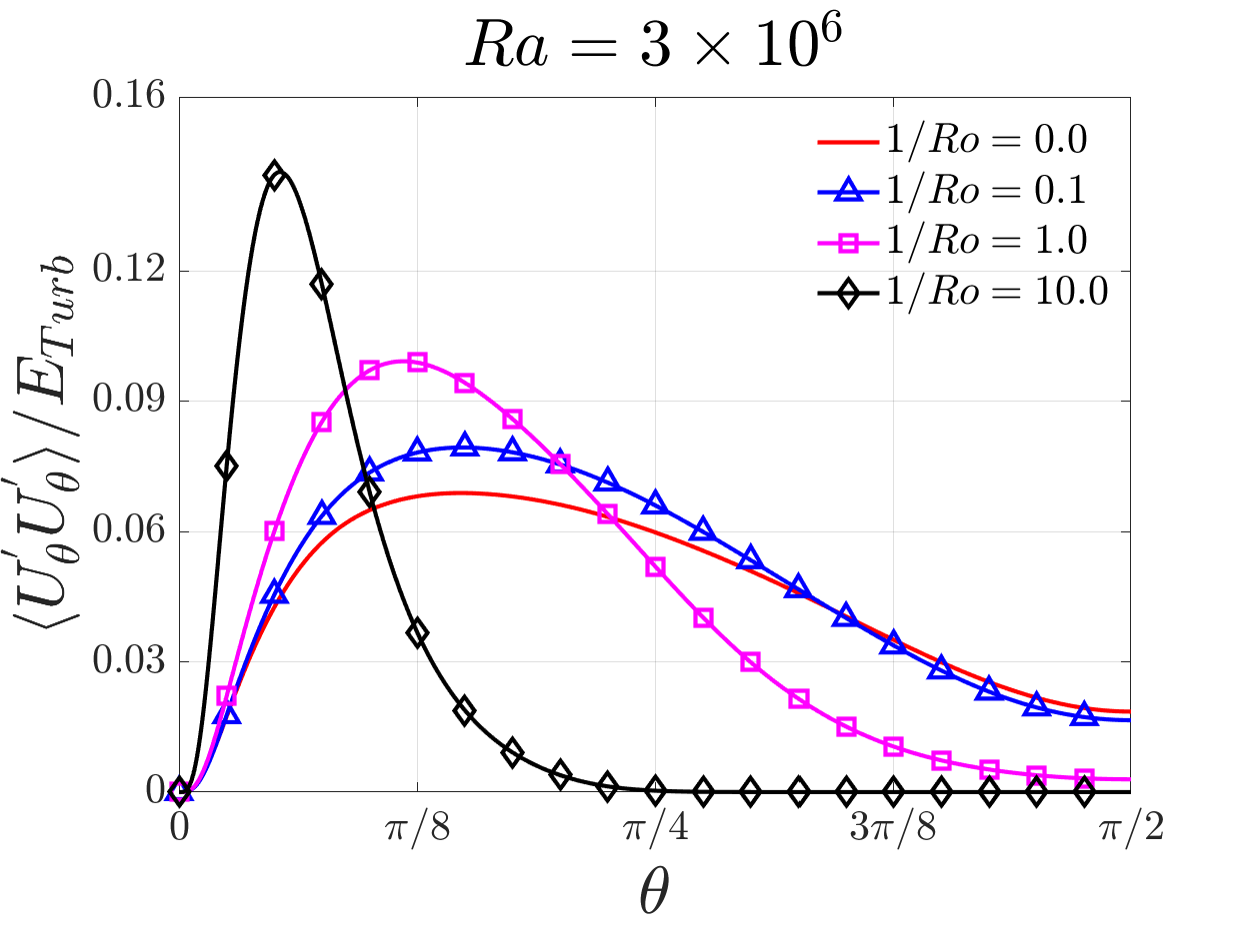}
    \includegraphics[width = 6.5cm]{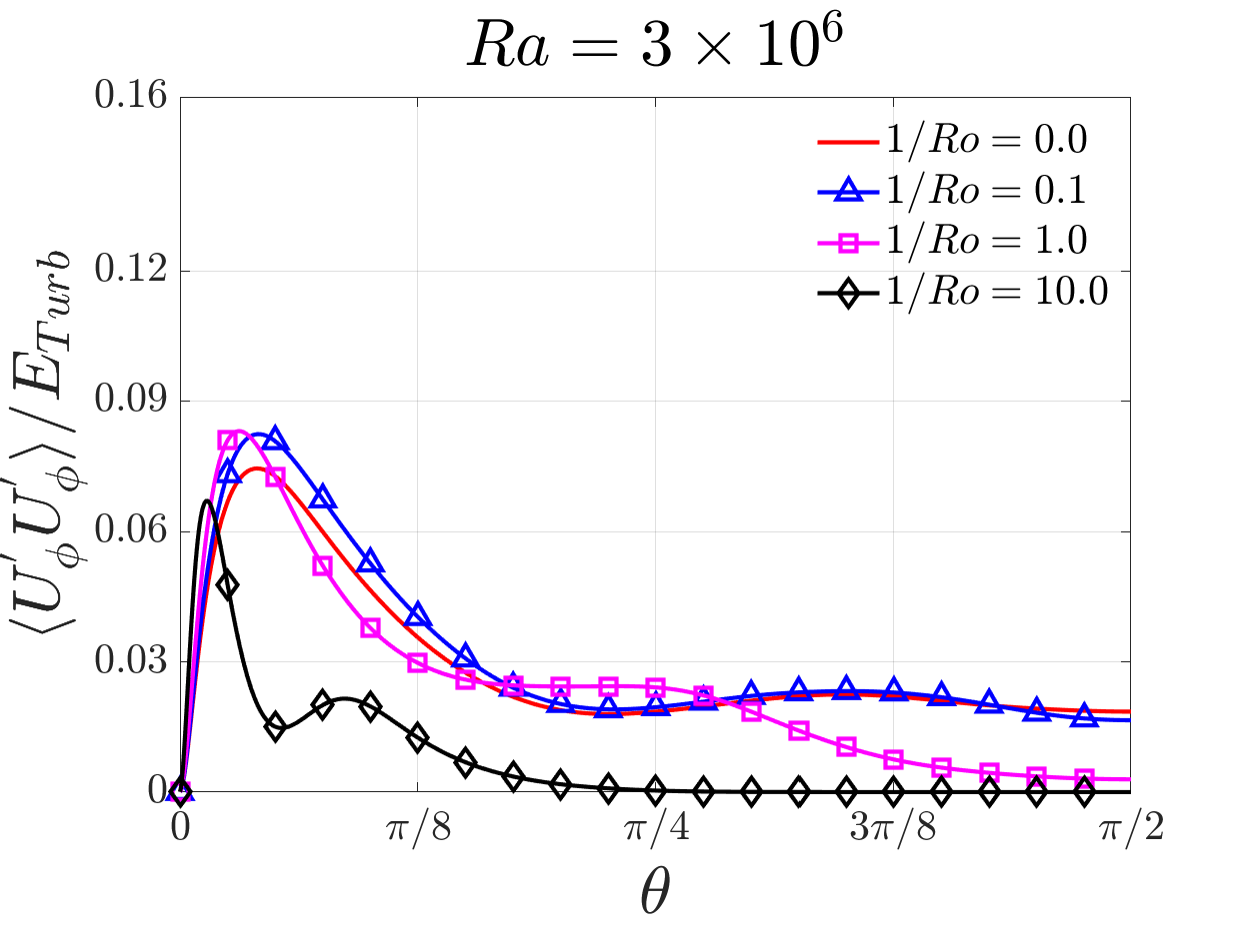}
    \includegraphics[width = 6.5cm]{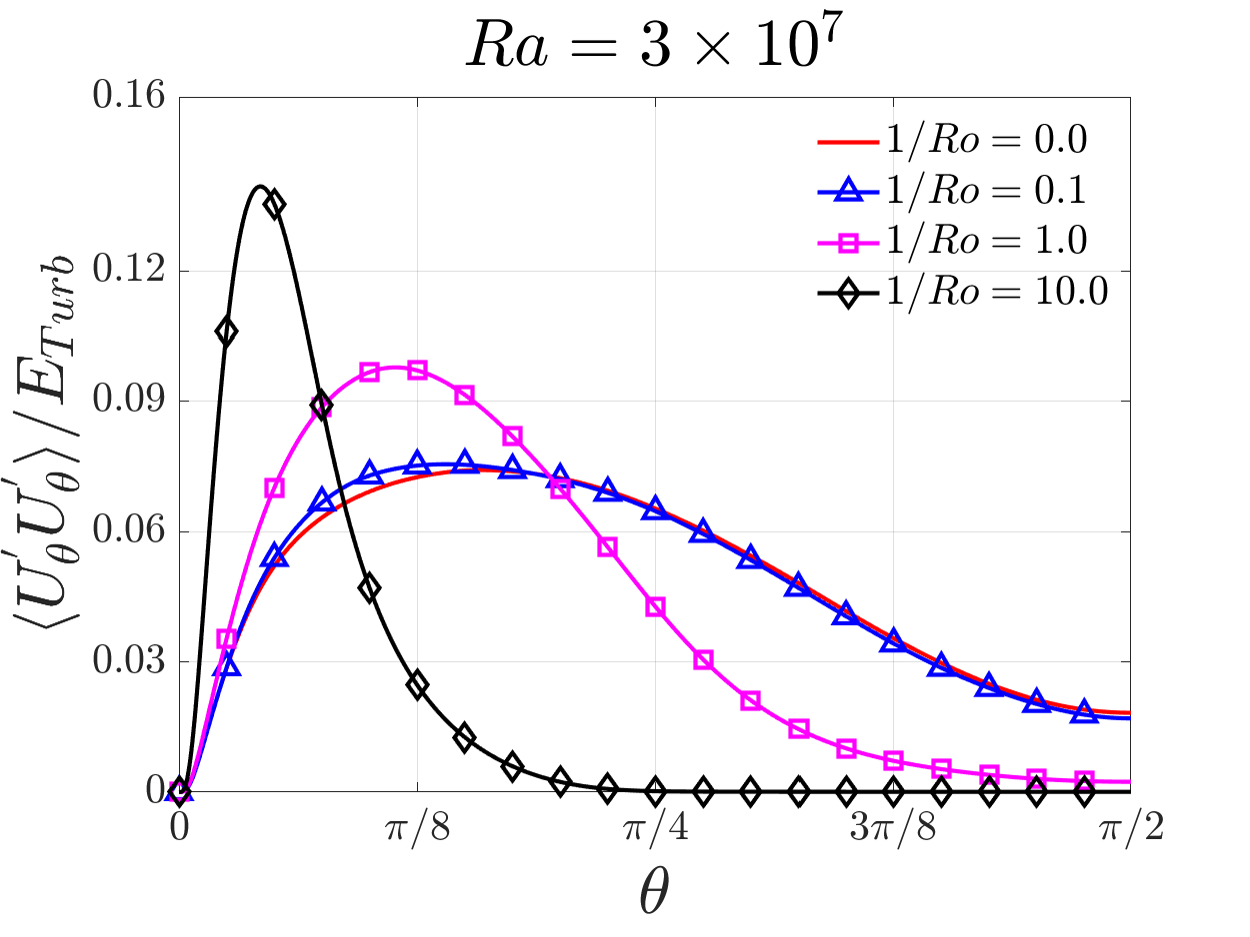}
        \includegraphics[width = 6.5cm]{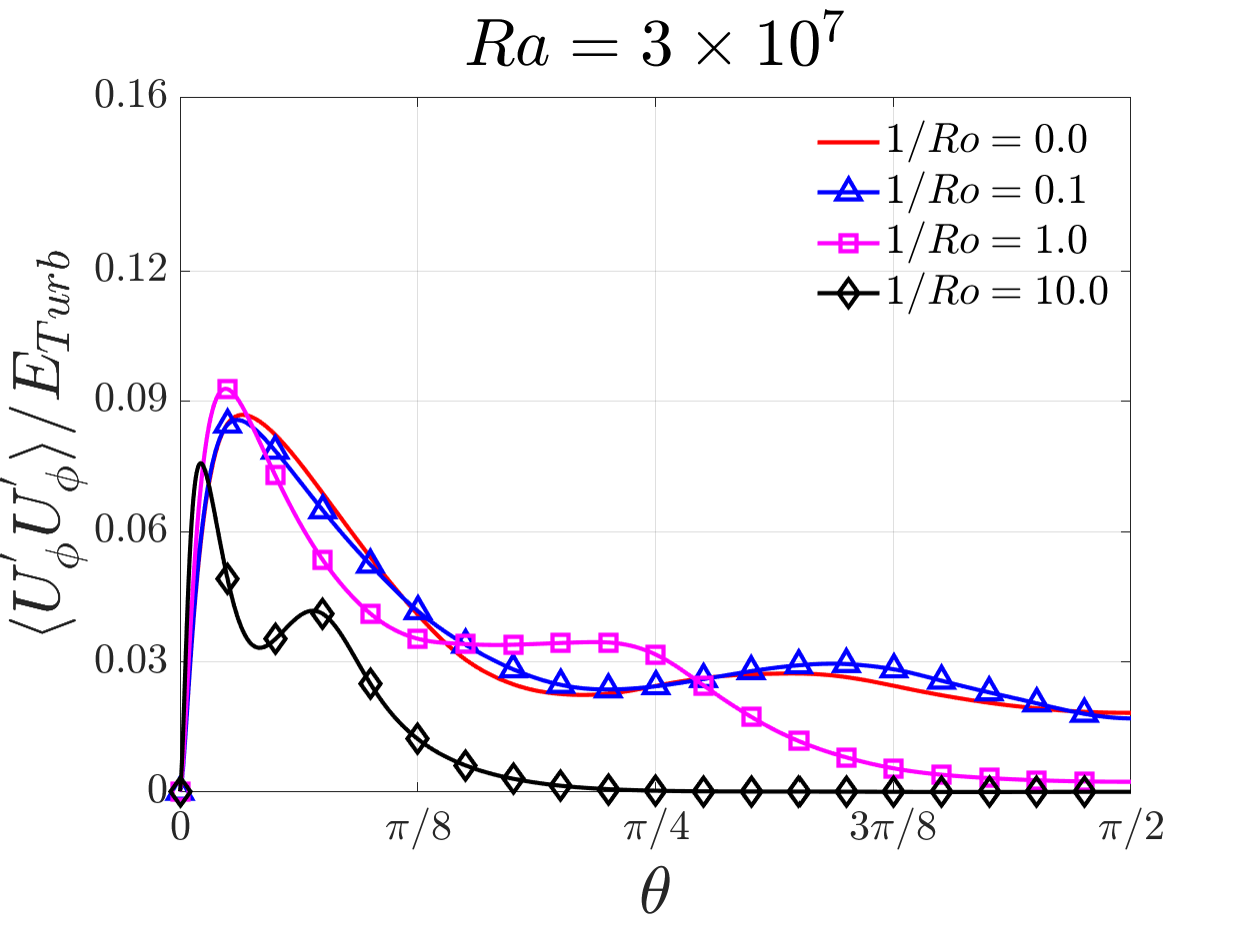}
      \includegraphics[width = 6.5cm]{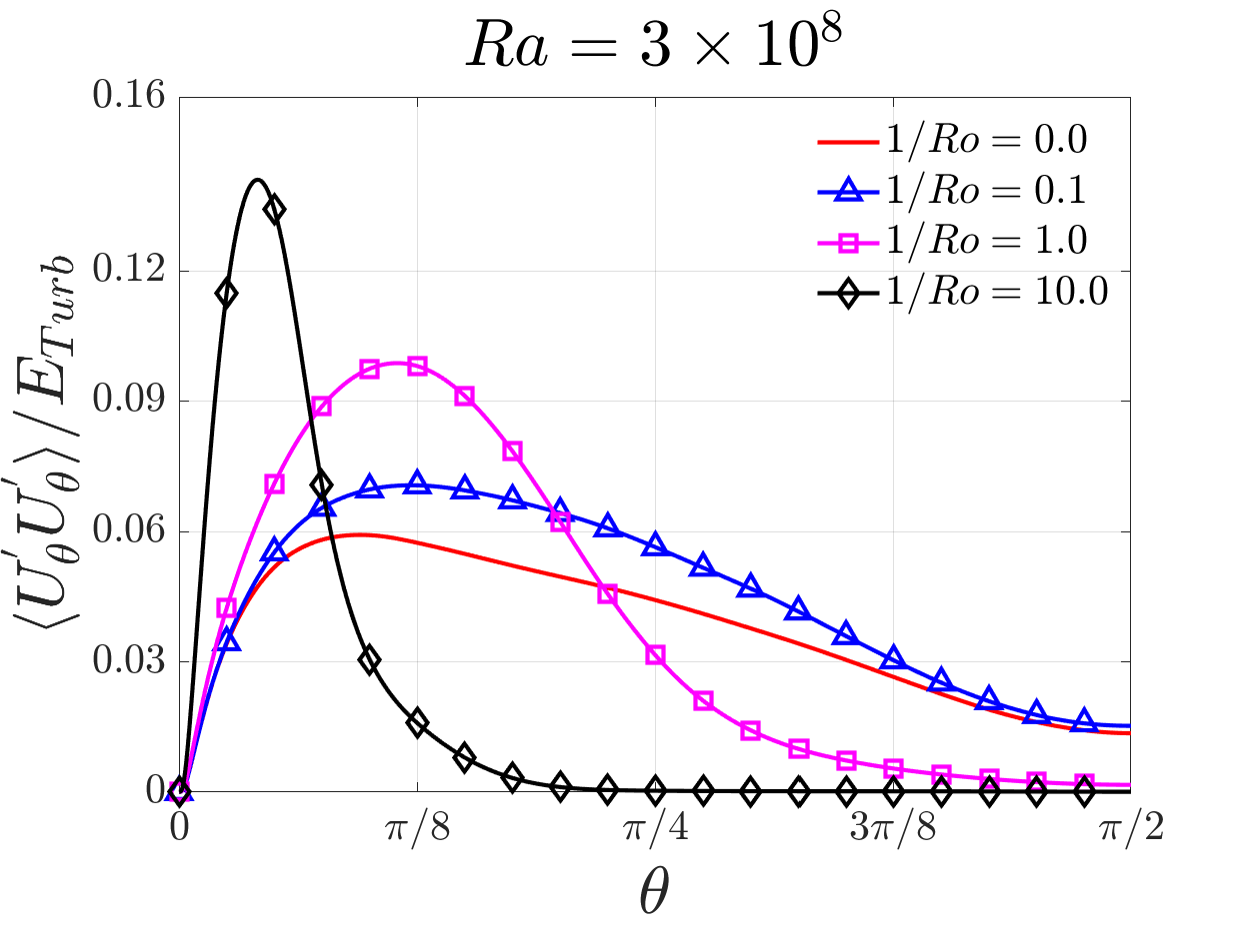}
          \includegraphics[width = 6.5cm]{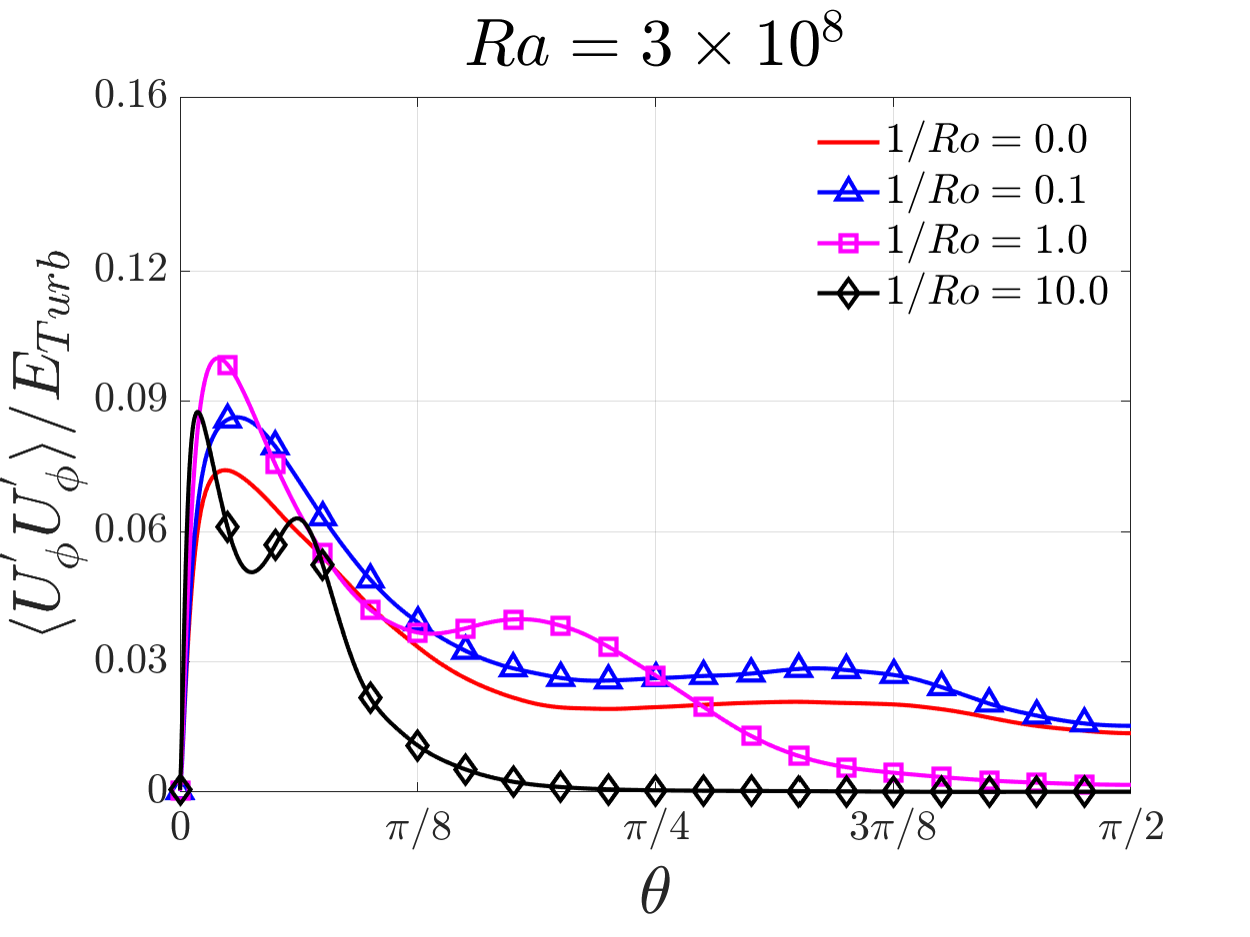}
        \includegraphics[width = 6.5cm]{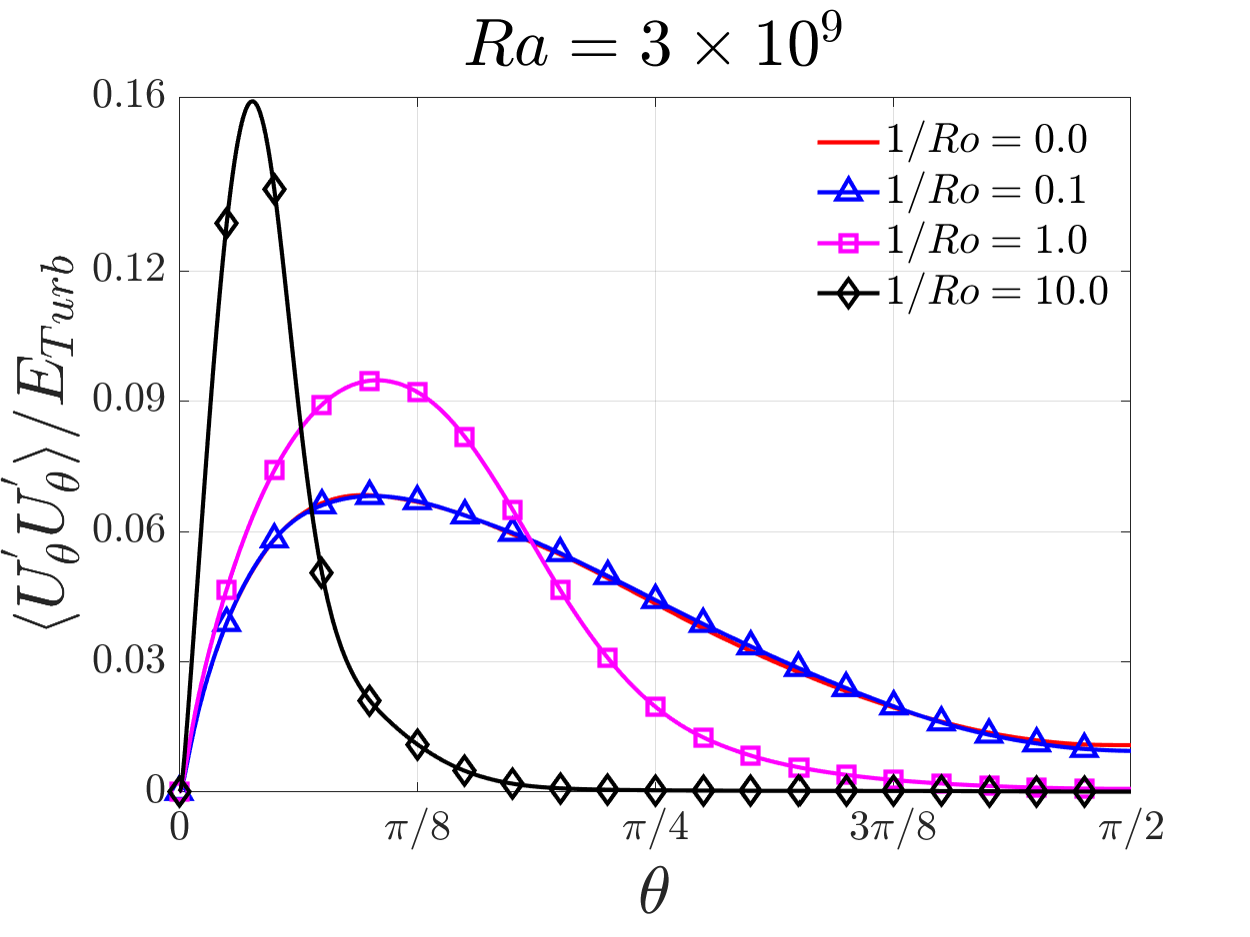}
            \includegraphics[width = 6.5cm]{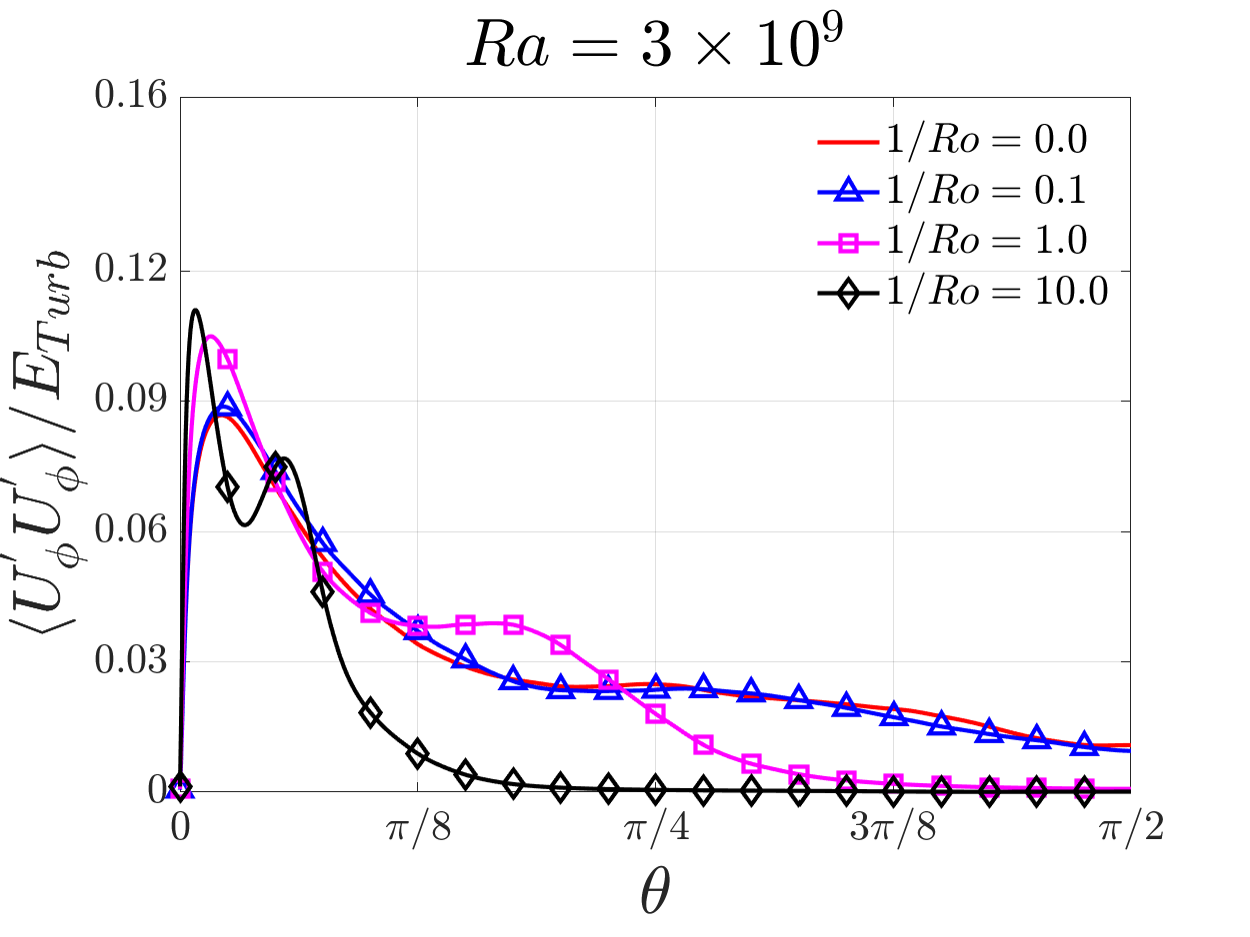}
  \caption{The diagonal Reynolds stress components $\sigma_{\phi\phi}$ and $\sigma_{\theta\theta}$, normalized by $E_{turb}$, as a function of latitude, and for different $Ra, 1/Ro$. }
  \label{fig:reStress11}
\end{figure}

\begin{figure}
  \centering
  \includegraphics[width = 6.5cm]{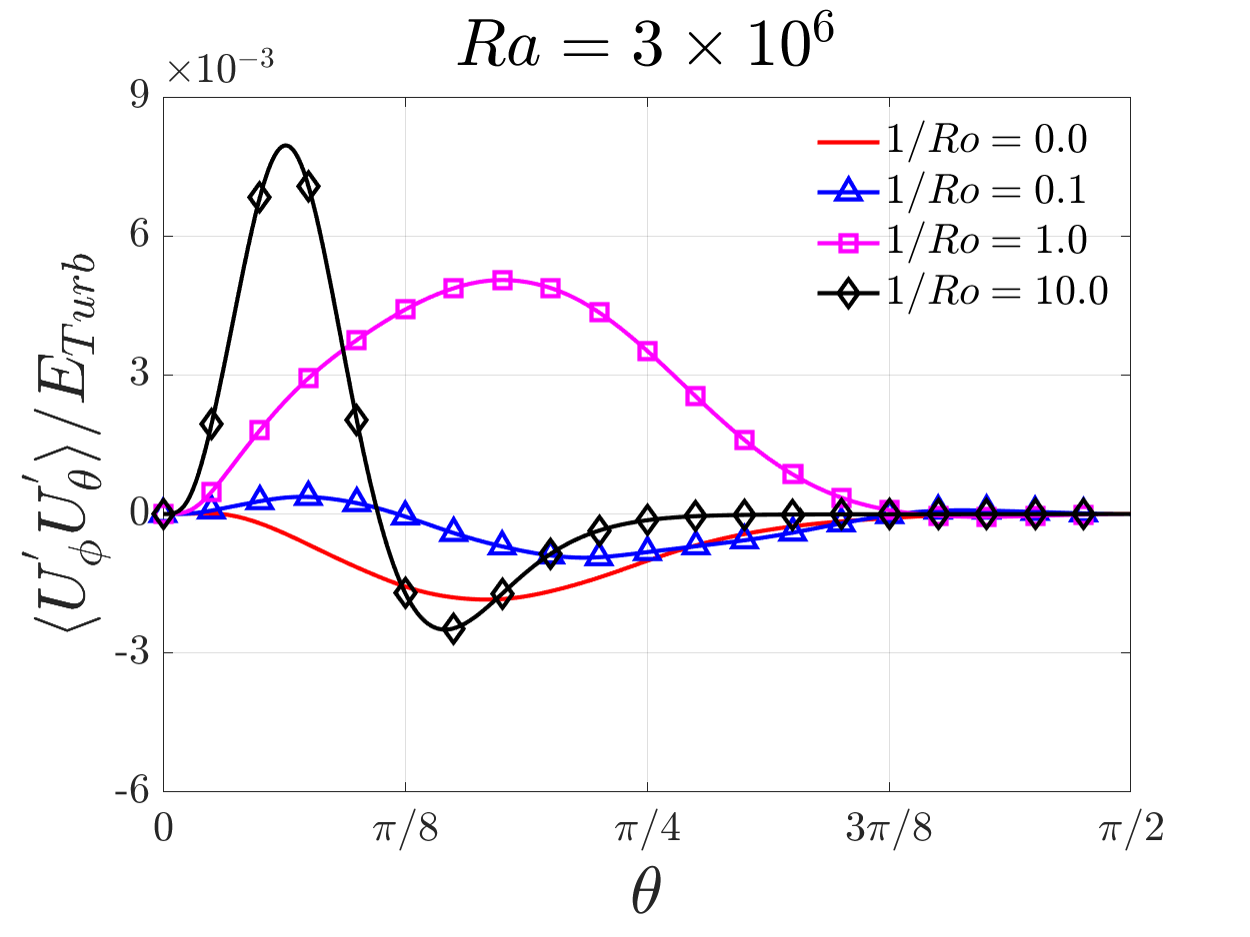}
    \includegraphics[width = 6.5cm]{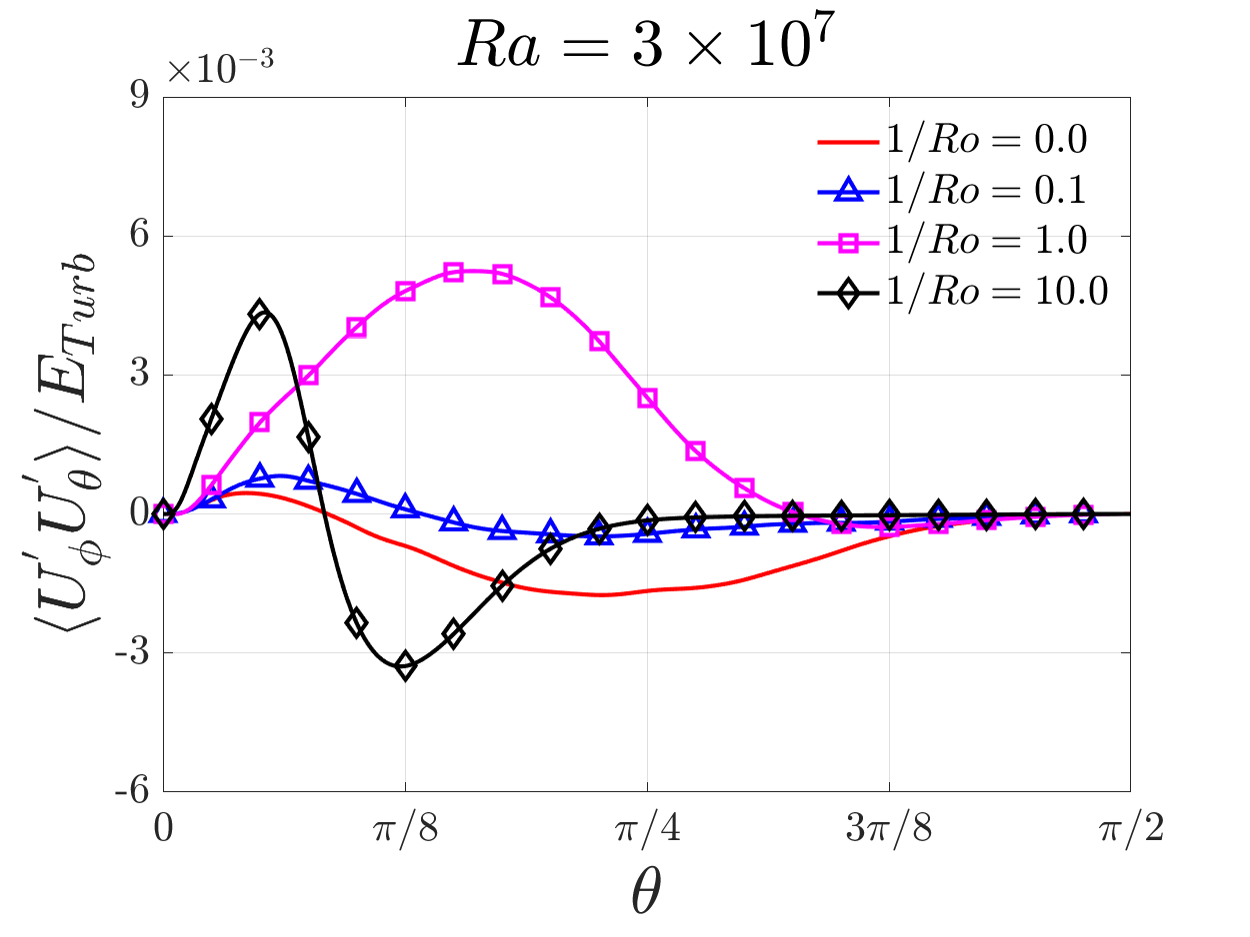}
      \includegraphics[width = 6.5cm]{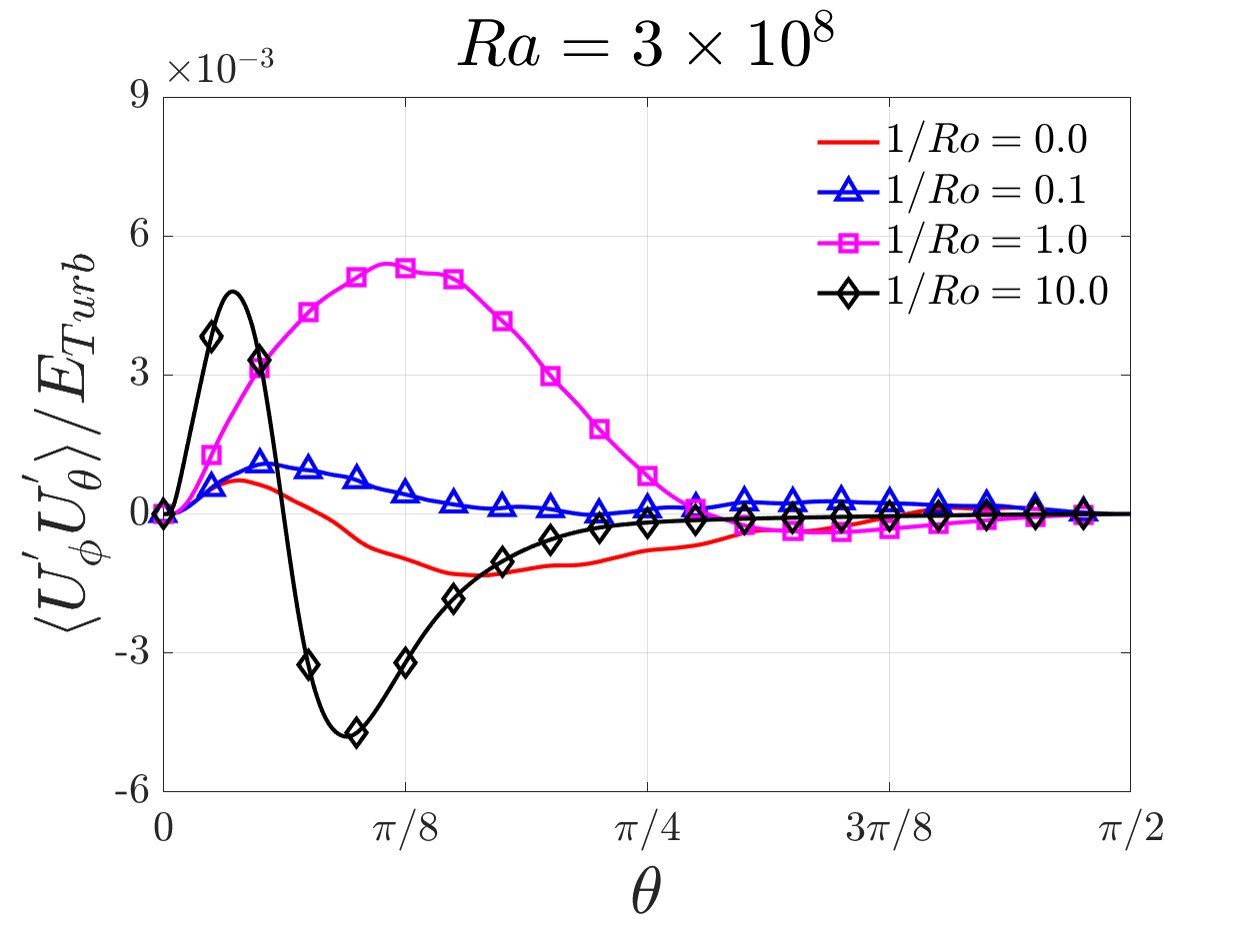}
        \includegraphics[width = 6.5cm]{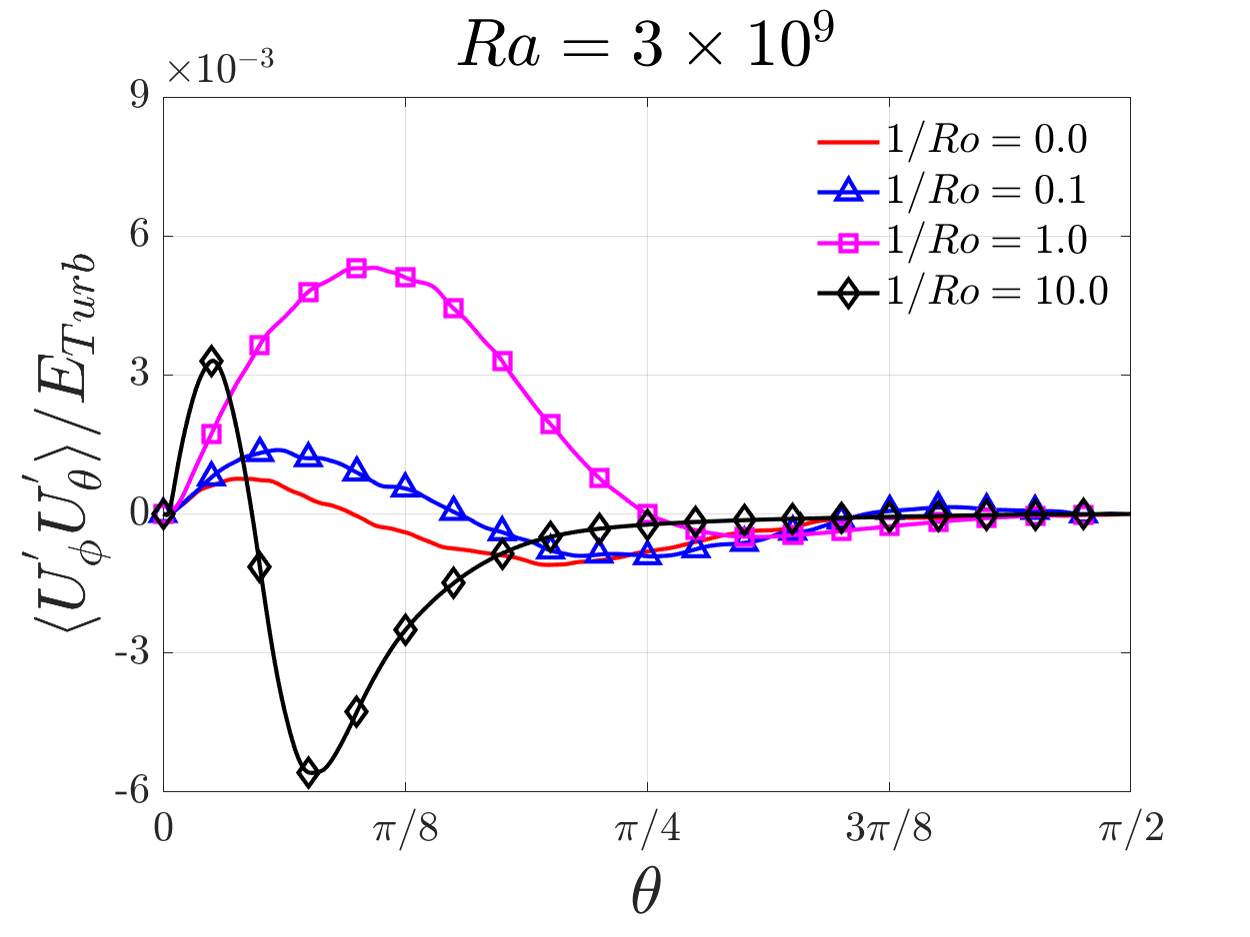}
  \caption{The averaged Reynolds shear stress $\sigma_{\phi\theta}$ along different latitudes of the soap bubble. }
  \label{fig:reStress12}
\end{figure}

We now turn to characterize the turbulence using the Reynolds stress tensor $\boldsymbol{\sigma}\equiv \langle(\boldsymbol{U}-\langle\boldsymbol{U}\rangle)^2\rangle$, and for the bubble flow, the components of this tensor may be written in matrix form as
\begin{align}
\boldsymbol{\sigma}=\begin{pmatrix}
\langle U'_\theta U'_\theta\rangle  & \langle U'_\phi U'_\theta\rangle \\
\langle U'_\phi U'_\theta\rangle & \langle U'_\phi U'_\phi\rangle
\end{pmatrix},\label{RS}
\end{align}
and whose components we denote by $\sigma_{ij}$, with $i,j$ either $\theta$ or $\phi$. We begin by considering the diagonal components of $\sigma_{\phi\phi}, \sigma_{\theta\theta}$ that contribute to the TKE, and the results are shown in figure~\ref{fig:reStress11}. The results show strong anisotropy in the flow, with $\sigma_{\theta\theta}$ generally significantly larger than $\sigma_{\phi\phi}$, except at lower latitudes where $\sigma_{\phi\phi}$ is typically larger (the region of $\theta$ for which this occurs depends upon $1/Ro$). The buoyancy term in the Navier-Stokes equation makes a contribution $\boldsymbol{B}+\boldsymbol{B}^\top$ to the Reynolds stress equation, where $\boldsymbol{B}\equiv \beta g_\theta\boldsymbol{e_\theta}\langle T' \boldsymbol{U}'\rangle$, and in matrix form
\begin{align}
\boldsymbol{B}+\boldsymbol{B}^\top=\beta g_\theta\begin{pmatrix}
2\langle T' U'_\theta \rangle  & \langle T'  U'_\phi\rangle \\
\langle T'  U'_\phi\rangle & 0
\end{pmatrix}.
\end{align}
From this it is clear that buoyancy does not make a direct contribution to $\sigma_{\phi\phi}$. Instead, buoyancy acts as a source for $\sigma_{\theta\theta}$, and some of this fluctuating energy is transferred to $\sigma_{\phi\phi}$ via redistribution mechanisms, such as the pressure-strain term \citep{pope}. While this explains why $\sigma_{\theta\theta}$ is greater than $\sigma_{\phi\phi}$ over much of the domain, it is not consistent with the fact that $\sigma_{\phi\phi}$ is larger than $\sigma_{\theta\theta}$ for low latitudes. One way to understand this is that since the only finite contribution to $\boldsymbol{\nabla}\langle\boldsymbol{U}\rangle$ in the flow comes from the component involving $\partial_{\theta}\langle U_\phi\rangle$, then the transport equation for $\sigma_{\phi\phi}$ will involve a finite shear-production term, whereas $\sigma_{\theta\theta}$ does not, and the data for $\sigma_{\theta\phi}$ (below) indicates that this term is positive in the region where $\sigma_{\phi\phi}$ is larger than $\sigma_{\theta\theta}$ at lower latitudes.

The Coriolis term also contributes to the redistribution of kinetic energy among the flow components. For example, the Coriolis acceleration projected along $\boldsymbol{e_{\phi}}$ is
\begin{align}
-2\boldsymbol{e_{\phi}}\boldsymbol{\cdot}(\boldsymbol{\Omega}\times\boldsymbol{U})= -2\Omega U_{\theta}\boldsymbol{e_{\phi}}\boldsymbol{\cdot}(\boldsymbol{e_z}\times\boldsymbol{e_{\theta}}),\label{CRe}
\end{align}
such the flow in the latitudinal direction can produce an acceleration in the longitudinal direction, modifying the Reynolds stress associated with $ U'_\phi$. The net effect on both components is however zero, since $(\boldsymbol{\Omega}\times\boldsymbol{U})\boldsymbol{\cdot U}=0$. Therefore, like the pressure-strain term, the Coriolis term produces a redistributive effect on the flow energy, only in this case, the effect arises from purely linear processes.

The results in figure~\ref{fig:reStress11} show oscillations in $\sigma_{\phi\phi}$ as $\theta$ is increased, which are not present for $\sigma_{\theta\theta}$. These oscillations may be due to the presence of LSC, whose influence on $\langle U_\phi\rangle$ was discussed earlier. As $1/Ro$ is increased, the effect on $\sigma_{\phi\phi}, \sigma_{\theta\theta}$ is non-monotonic, increasing or else decreasing their values for different $\theta$. As $\theta$ is increased, the buoyancy force tends to decrease, while the Coriolis term becomes dominant (see \S\ref{GEDNS}). The Taylor-Proudman effect hinders the transport of fluctuations towards larger $\theta$, and since the Coriolis term does not produce TKE, this then implies that both $\sigma_{\theta\theta}$ and $\sigma_{\phi\phi}$ should decay as $\theta$ is increased. The results in figure~\ref{fig:reStress11} are consistent with this behavior, and show that $\sigma_{\theta\theta}$ and $\sigma_{\phi\phi}$ reduce to almost zero for $\theta\geq \pi/4$. We note that for $\theta\to 0$, the effect of rotation seems to vanish (which is also apparent for several of the other results). This is most likely due to the fact that the Rossby number based on the boundary layer thickness is $>O(1)$, i.e. the Coriolis force does not strongly affect the small scales of motion in the boundary layer.

The results for the shear-stress term $\sigma_{\theta\phi}$, are shown in figure \ref{fig:reStress12}. The results indicate that the maximum value of these stresses are an order of magnitude smaller than the diagonal Reynolds stresses. For the higher $Ra$ cases and for $1/Ro=0$, $\sigma_{\theta\phi}$ is positive near the equator, but becomes negative at higher latitudes. As $1/Ro$ is increased, this feature still persists, however the location of the transition from positive to negative values, as well as the magnitudes of $\sigma_{\theta\phi}$ change significantly. The region of $\theta$ over which $\sigma_{\theta\phi}$ is finite decreases with increasing $1/Ro$ since in the regime $1/Ro\gg 1$, the Coriolis term suppresses $\sigma_{\theta\phi}$. In particular, while \eqref{CRe} implies that the Coriolis term can produce coupling between the components of $\boldsymbol{U}'$, the Coriolis term suppresses both components of $\boldsymbol{U}'$ so that \eqref{CRe} is also suppressed. Nearer to the equator this behavior does not emerge since buoyancy forces are important there. However, as for the diagonal components of $\boldsymbol{\sigma}$, the dependence on $1/Ro$ is strongly non-monotonic, with the maximum value for $\sigma_{\theta\phi}$ on the bubble emerging for the $1/Ro=1$ case. However, the enhancement as $1/Ro$ is increased from zero to one is significantly stronger than for either $\sigma_{\theta\theta}$ or $\sigma_{\phi\phi}$.

\section{Results \& Discussion: scale-dependent information}\label{SDI}
Having explored the behavior of the flow using one-point quantities that mainly characterize the large scales of the flow, we now turn to consider quantities that describe the properties of the flow at different scales. In this section we focus on the data for $Ra=3\times 10^{-6}$ for which there is the greatest scale separation in the flow, allowing us to most clearly explore the behavior at dynamically distinct scales in the flow.
\subsection{Kinetic energy and temperature spectrums}

For flows with strong buoyancy, there may exist a range of scales where buoyancy and inertial forces are both important, with viscous forces negligible. Assuming these forces balance, for stably stratified turbulence Bolgiano and Obukhov derived the well-known Bolgiano-Obukhov scaling laws (BO59, \cite{Bolgiano59,Obukhov59}). In recent decades, there has been significant interest in exploring whether BO59 also applies in other flows such Rayleigh-B\'enard convection \citep{LohseXia29} and Rayleigh-Taylor turbulence \citep{boffetta17}. Using Bolgiano-Obukhov scaling, then for two-dimensional turbulence with an inverse energy cascade the following scaling is predicted \citep{boffetta17}

\begin{align}
	E_U(k)&\varpropto k^{-11/5},\\
	E_T(k)&\varpropto k^{-7/5},
\end{align}
where $E_U(k)$ is the kinetic energy spectrum, and $E_T(k)$ is the thermal energy (half of the squared temperature).

The BO59 phenomenology is only expected to apply in regions of the flow sufficiently far from boundaries, and therefore will likely not apply at low latitudes in our flow. However, even for sufficiently high latitudes BO59 scaling may still not apply for at least two reasons. First, our flow takes place on a curved, not flat surface, and buoyancy in our flow is a function of latitude due to the variation of $\boldsymbol{g\cdot}\boldsymbol{e}_\theta$ with $\theta$, as discussed earlier. However, one could argue that BO59 scaling might nevertheless hold in a local sense  at scales small enough to be only weakly affected by the surface curvature and spatial variation of $\boldsymbol{g\cdot}\boldsymbol{e}_\theta$. Second, our flow is affected by rotation, and at scales where $1/Ro_\ell>1$ BO59 will no longer apply due to the role of the Coriolis force at these scales that is not included in BO59. The scaling that will emerge will then be determined by the relative roles of buoyancy, Coriolis and inertial forces in the flow at different length scales and different scalings may emerge in different scale ranges where different forces in the flow balance. However, at scales where $1/Ro_\ell<1$, BO59 should still apply since at these scales the role of the Coriolis force is weak.
\begin{figure}
	\centering
	\includegraphics[width = 6.5cm]{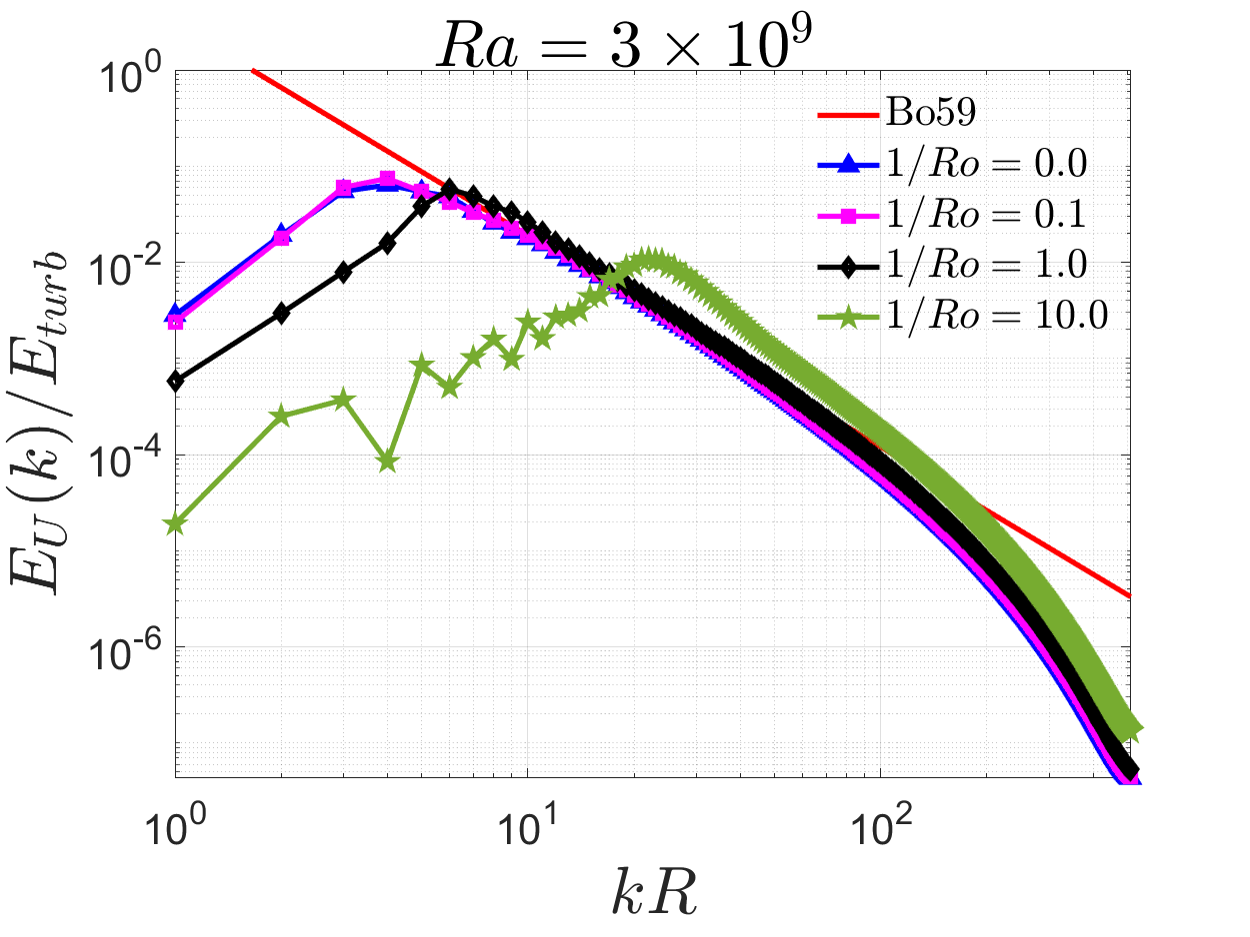}
	\includegraphics[width = 6.5cm]{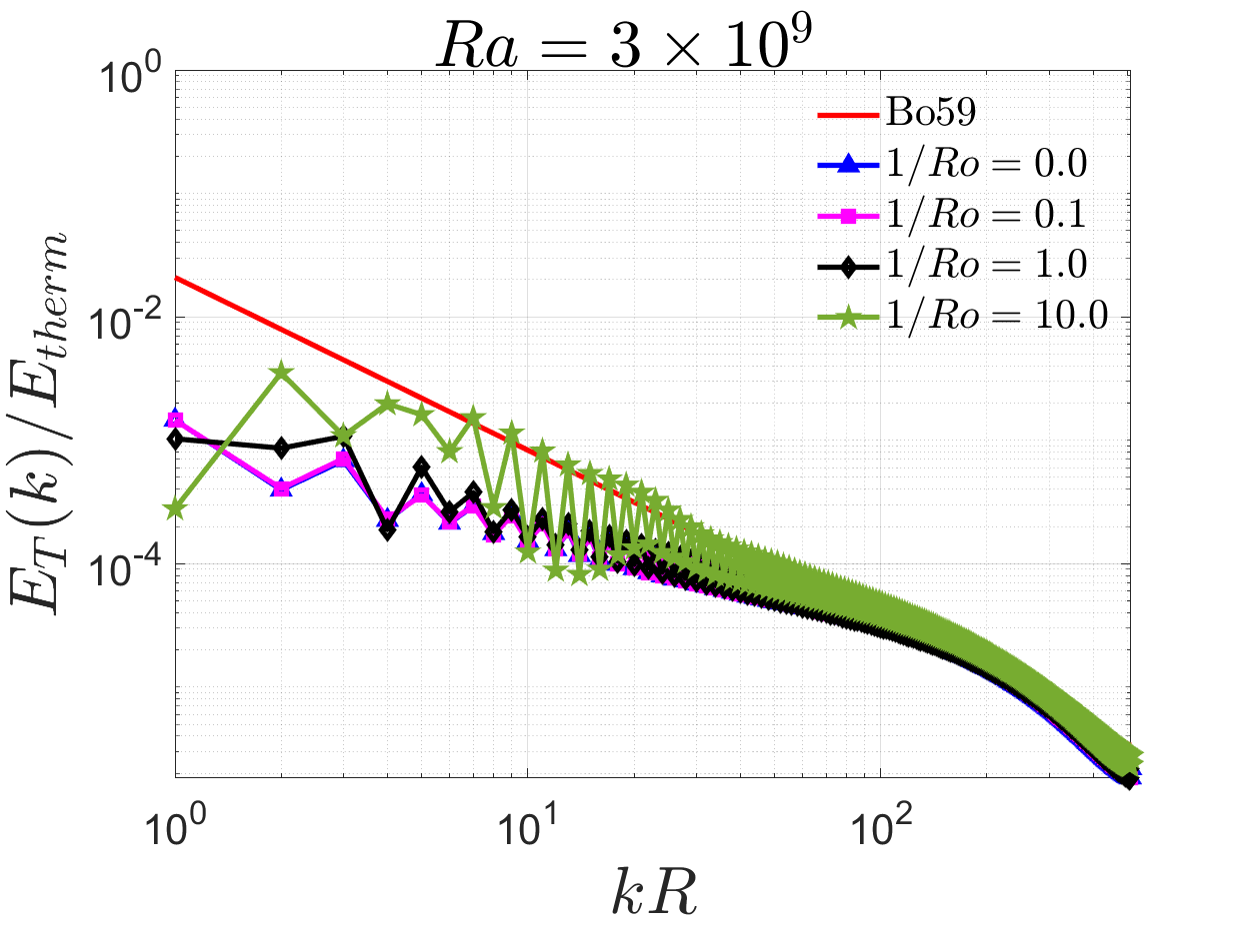}
	\caption{Kinetic energy and thermal energy spectrums for varying $1/Ro$ and for $Ra=3\times 10^9$. }
	\label{fig:Spectrums}
\end{figure}
In figure \ref{fig:Spectrums} we plot our results for $E_U(k)$ and $E_T(k)$. These are computed using spherical harmonics decompositions, with averaging over all $\theta,\phi$. Details on the method of computation may be found in \citet{BruneauFischer19}.

Our results show that for $1/Ro\leq 1$, a significant range of $kR$ exists over which the BO59 scaling accurately predicts the scaling behavior of $E_U(k)$. For $1/Ro>1$, however, BO59 does not apply since the effect of rotation strongly affects the range of $kR$ where BO59 would otherwise emerge. The effect of rotation causes $E_U(k)$ to decay faster than $k^{-11/5}$ as $k$ increases. This may be understood by notating that if the Coriolis force dominated the behavior of the spectrum, then the dynamically relevant parameter would be $\Omega$ and the scaling that would emerge is $E_U(k)\propto k^{-3}$.

In contrast, our results show that even for $1/Ro=0$, BO59 does not accurately describe the scaling behavior of $E_T(k)$ over any significant range of $kR$. In particular, $E_T(k)$ seems to decay more slowly than $E_T(k)\varpropto k^{-7/5}$ even for $1/Ro=0$. The most likely explanation for why BO59 does not describe $E_T(k)$ well, even though it does describe $E_U(k)$ well is that unlike the velocity field, the temperature field is maximum at the equator and quickly reduces as $\theta$ is increased. Hence, since $E_T(k)$ is computed as an integral over the surface of the whole bubble, it will be strongly affected by the thermal boundary layer, for which BO59 does not apply.

\subsection{Spectral fluxes of kinetic energy, temperature, and enstrophy}
\begin{figure}
	\centering
	\includegraphics[width = 6.5cm]{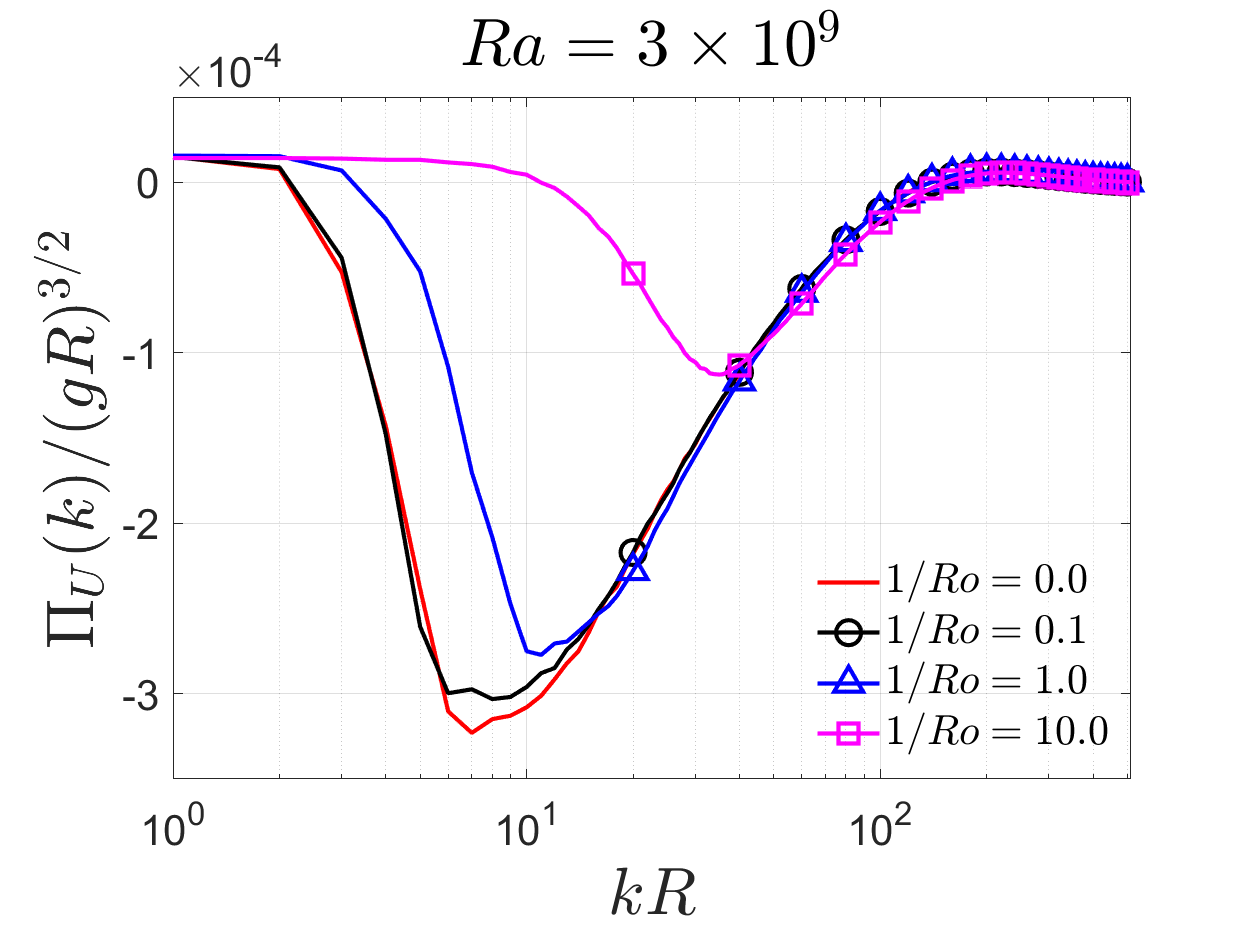}
		\includegraphics[width = 6.5cm]{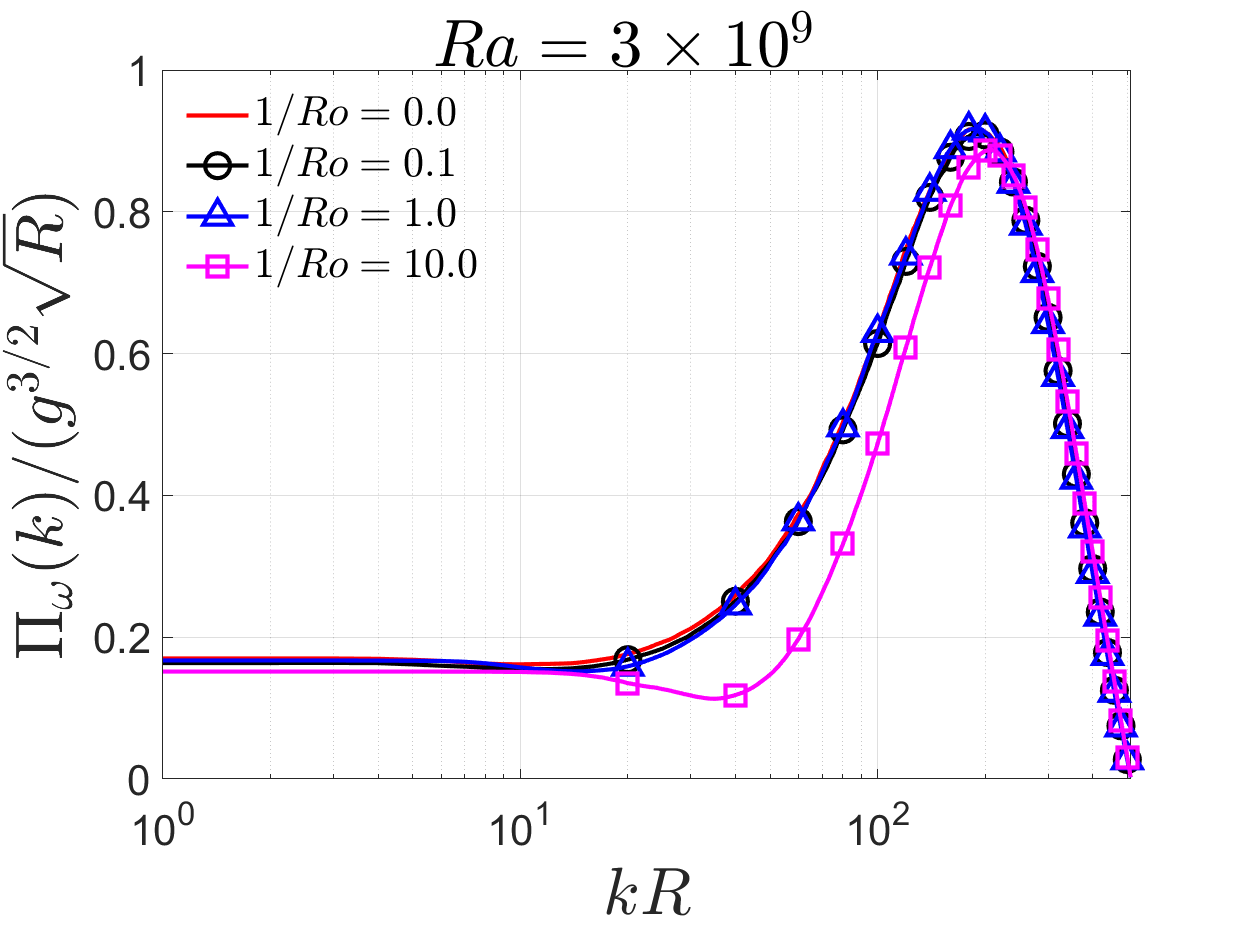}\\
			\includegraphics[width = 6.5cm]{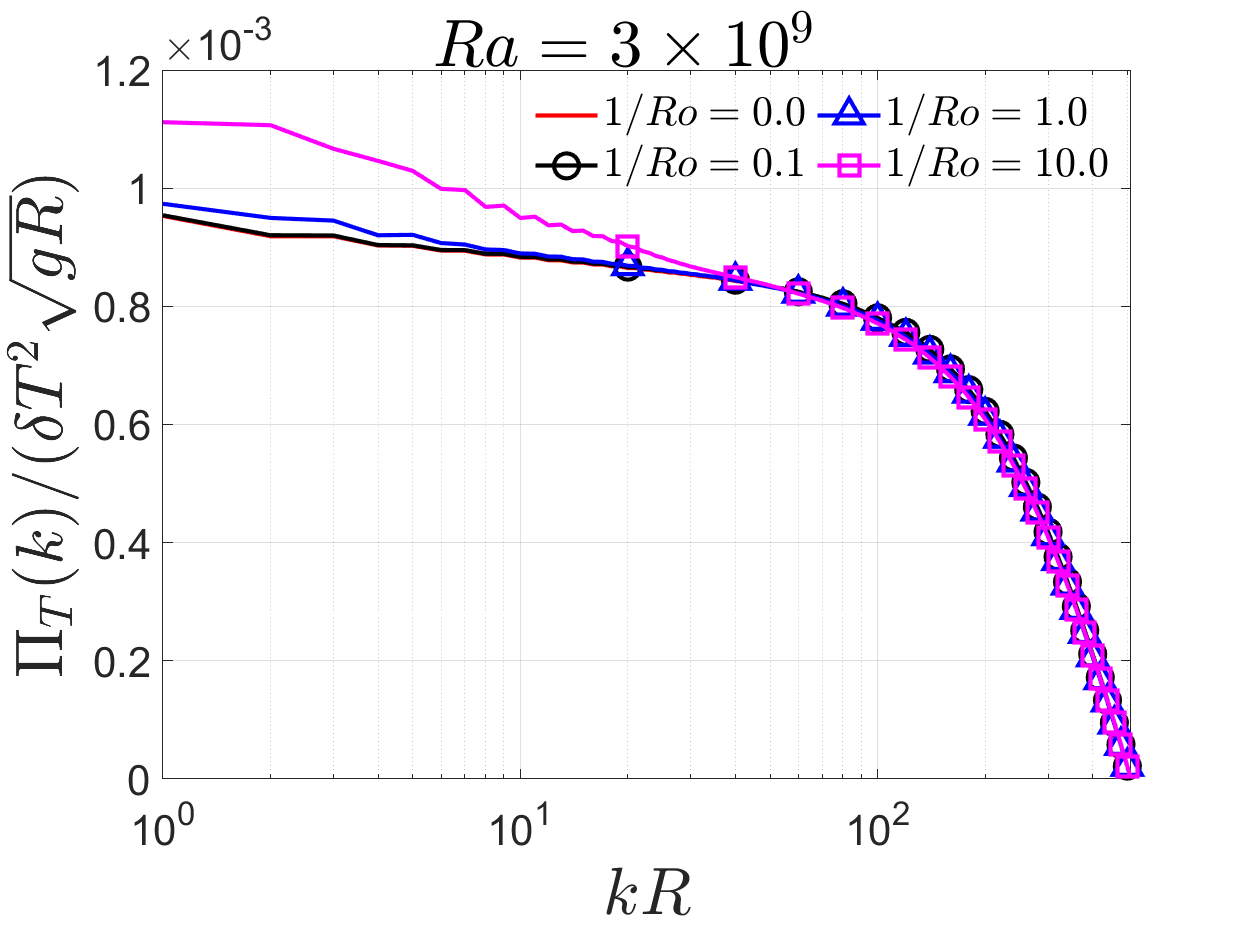}
	\caption{The kinetic energy, enstrophy, and thermal energy spectral fluxes. }
	\label{fig:Fluxes}
\end{figure}
The behavior of $E_U(k)$ and $E_T(k)$ are determined by the behavior of fluxes in spectral space that determine the energy and temperature fluctuations at different scales. In two dimensional turbulence in the absence of buoyancy, there are two inviscid integrals of motion, namely kinetic energy and enstrophy. According to standard arguments, at larger wavenumbers there is an upscale cascade of kinetic energy, and at smaller scales there is a downscale cascade of enstrophy. For our flow where buoyancy and Coriolis forces also play a role, the behavior of these cascades may differ, except at scales small enough for inertial forces to dominate over buoyancy and Coriolis forces.

In figure \ref{fig:Fluxes} we compute the kinetic energy flux $\Pi_U(k)$, the enstrophy flux $\Pi_{\omega}(k)$, and thermal energy flux $\Pi_{T}(k)$, as a function of $kR$. The results indicate that except at the highest wavenumbers where $\Pi_U$ becomes slightly positive, $\Pi_U$ is predominantly negative, indicating an inverse energy flux driving kinetic energy to larger scales in the flow.  However, the Reynolds number of our flow is too small to observe a cascade regime associated with a constant-flux. Up until a certain value of $kR$, the magnitude of $\Pi_U$ reduces as $1/Ro$ is increases. This is simply understood in view of the fact that in the limit $1/Ro\to \infty$, the dynamical equations are linear, and there is no energy transfer among scales. We also note, however, that above a critical $kR$ (that increases with increasing $1/Ro$), $\Pi_U$ is converges to $\Pi_U$ for the non-rotating case $1/Ro=0$. This can be understood in terms of the scale-dependent Rossby number $Ro_\ell$ introduced in \S\ref{GEDNS}, namely, that for any given rotation-rate $\Omega$ there is a scale below which $1/Ro_\ell<1$ indicating that the effects of rotation are subleading at these scales. As $\Omega$ is increased, one has to go to a smaller scale before this regime is observed. The results in figure~\ref{fig:Fluxes} are consistent with this, showing that $\Pi_U$ approaches the $1/Ro=0$ behavior at larger $kR$ (i.e. smaller scale) as $1/Ro$ is increased.

The results for the enstrophy flux $\Pi_{\omega}(k)$ as a function of wavenumber $k$, and for different $1/Ro$ are also shown in figure \ref{fig:Fluxes}. As with two-dimensional turbulence on a flat surface, the results show that there is a downscale flux of enstrophy, but the Reynolds number of our flow is too small to observe a constant-flux cascade regime. As $1/Ro$ is increased, $\Pi_{\omega}(k)$ tends to be reduced, as the role of nonlinearity in the flow is reduced. However, we again observe that for sufficiently high $kR$, the effect of $1/Ro$ on $\Pi_{\omega}(k)$ disappears, corresponding to scales of the flow where $/Ro_\ell$. Overall, $\Pi_{\omega}(k)$ is much less sensitive to $1/Ro$ than $\Pi_U(k)$, which is because enstrophy is a predominantly small scale quantity, whereas the velocity field is dominated (away from boundaries) by the large scales, for which the effect of rotation is the strongest.

The results for the thermal energy flux $\Pi_{T}(k)$ shown in figure \ref{fig:Fluxes} show very weak sensitivity to $1/Ro$, except for $kR\lesssim 20$. This may seem surprising given that the temperature field, like the velocity field (but unlike the vorticity), is dominated (away from boundaries) by the largest scales of the flow, and therefore should be strongly susceptible to the effects of rotation. However, as noted earlier, the strongest contributions to the temperature field come from the thermal boundary layer, and the Rossby number based on the boundary layer thickness is very small. This then explains the weak effect of $1/Ro$ on $\Pi_{T}(k)$.
\begin{figure}
	\centering
	\includegraphics[width = 6.5cm]{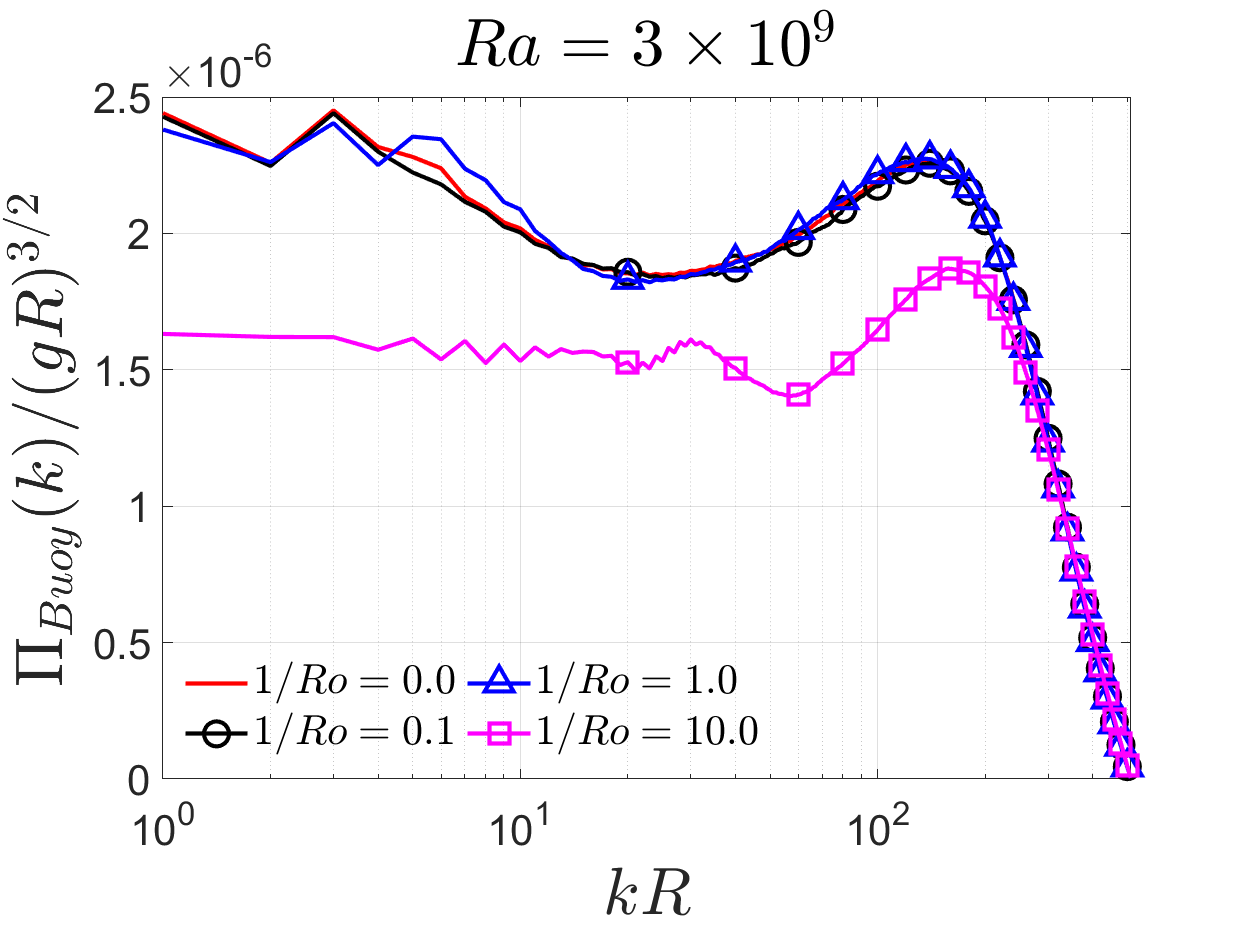}
	\caption{Buoyancy flux spectrum for varying $1/Ro$ and for $Ra=3\times 10^9$. }
	\label{fig:Buoyflux}
\end{figure}

In figure~\ref{fig:Buoyflux} we show the results for the buoyancy flux spectrum $\Pi_{Buoy}$ which is of utmost importance since buoyancy is the mechanism generating the turbulent flow on the bubble. Indeed, $\Pi_{Buoy}>0$ for all $kR$, indicating that buoyancy injects TKE at all scales of the flow. The results for $1/Ro=0$ show that at the lowest $kR$, $\Pi_{Buoy}$ decreases with increasing $kR$, which is in part associated with the fact that the velocity and temperature fluctuations are largest at the large scales. However, as $kR$ continues to increase, $\Pi_{Buoy}$ increases to a local maximum before vanishing for $kR\to \infty$ (due to smoothness of the velocity and temperature fields). This indicates that there is a local, strong injection of TKE at the smaller scales in the flow. This must in fact be the case, since as already shown, the two-dimensional turbulent flow on the bubble exhibits an inverse kinetic energy flux, and this requires an injection of TKE at smaller scales in the flow. Comparison with the results in figure \ref{fig:Fluxes} shows that the injection of TKE at small scales due to buoyancy occurs at a smaller scale (larger $kR$) than that at which $\Pi_u$ becomes negative. Therefore, the inverse flux takes the TKE injected at small scales through $\Pi_{Buoy}$, and passes it to the larger scales in the flow, though not via a conservative cascade at these Reynolds numbers. The results in figure~\ref{fig:Buoyflux} show that as $1/Ro$ is increased, $\Pi_{Buoy}$ is suppressed except at the largest $kR$ where the effect of the Coriolis force is negligible. This then shows in scale-space, how rotation suppresses the mechanism of TKE injection into the flow, and thereby suppresses turbulence in the flow as $1/Ro$ is increased.


\subsection{Structure Functions of Temperature and Velocity}

Having considered the behavior in Fourier-space, we now consider the multiscale behavior of the flow using structure functions. Not only does this provide insight in physical space, but it also allows us to consider the behavior of fluctuations in the flow beyond the second order information captured by the spectrums.

The $N^{th}$ order structure functions of $T$ and $\boldsymbol{U}$ are defined as

\begin{align}
	S^{{U}}_N&\equiv\langle |[\boldsymbol{U}(\boldsymbol{x}+\boldsymbol{d},t) - \boldsymbol{U}(\boldsymbol{x},t)]\boldsymbol{\cdot \hat{d}}|^N \rangle,\\
	S^{T}_N&\equiv\langle |T(\boldsymbol{x}+\boldsymbol{d},t) - T(\boldsymbol{x},t)|^N \rangle,
\end{align}
where $\boldsymbol{\hat{d}}\equiv \boldsymbol{d}/\|\boldsymbol{d}\|$. Since the flow we are considering is homogeneous in $\phi$ but not in $\theta$, we choose $\boldsymbol{d}$ such that $\boldsymbol{d\cdot} \boldsymbol{e}_\theta=0$, i.e. the two points used in constructing the structure functions have the same latitude.

For a 2D turbulent flow, the BO59 scaling predictions are \citep{boffetta17}
\begin{align}
	S^{{U}}_N&\varpropto d^{3N/5},\\
	S^{T}_N &\varpropto d^{N/5}.
\end{align}
For sufficiently small scales where molecular effects become important, the temperature and velocity fields are smooth and these scalings must give way to the alternative scaling $S^{T}_N \propto S^{{U}}_N\propto d^N$ that may be demonstrated using a Taylor-series expansion.

One issue with computing the structure functions on the bubble is that the uniform grid used to solve the governing equations on the steriographic plane corresponds to a non-uniform grid on the curved bubble surface. This could then introduce a bias into the computation of the structure functions since only regions where the grid spacing is $\geq d$ can contribute to the computation of the structure functions at scale $d$. Therefore, in order to compute the structure functions we used a high-order method to interpolate $\boldsymbol{U}$ and $T$ onto a grid that corresponds to points uniformly spaced on the surface of the bubble. Furthermore, in order to reduce statistical noise, we computed the structure functions by averaging over $\theta$, as well as over $\phi$ and time, which is consistent with how the spectral quantities were computed. One difference, however, is that for the structure functions we only average over the region $\theta\in[\pi/18,\pi/2]$ (unlike like the earlier spectral results which had to be computed by averaging over all $\theta$ due to the method of computation in terms of spherical harmonics). This is because close to the equator at $\theta=0$, the boundary layers, which are not accounted for in BO59, strongly influence the results (especially for the temperature). By only averaging over the region $\theta\in[\pi/18,\pi/2]$, the effect of these boundary layers on the computed structure functions is greatly reduced.

\begin{figure}
	\centering
	\includegraphics[width = 6.5cm]{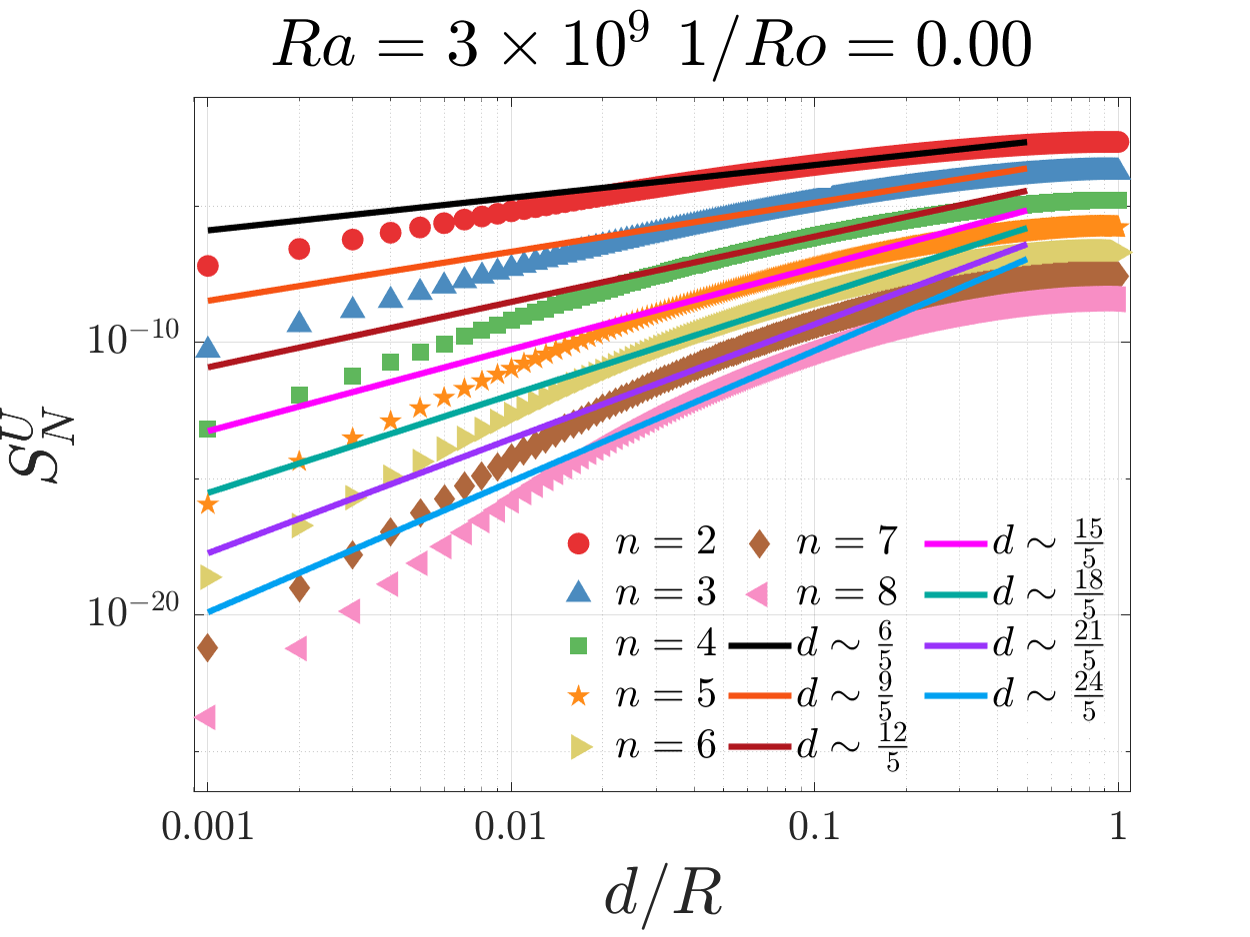}
		\includegraphics[width = 6.5cm]{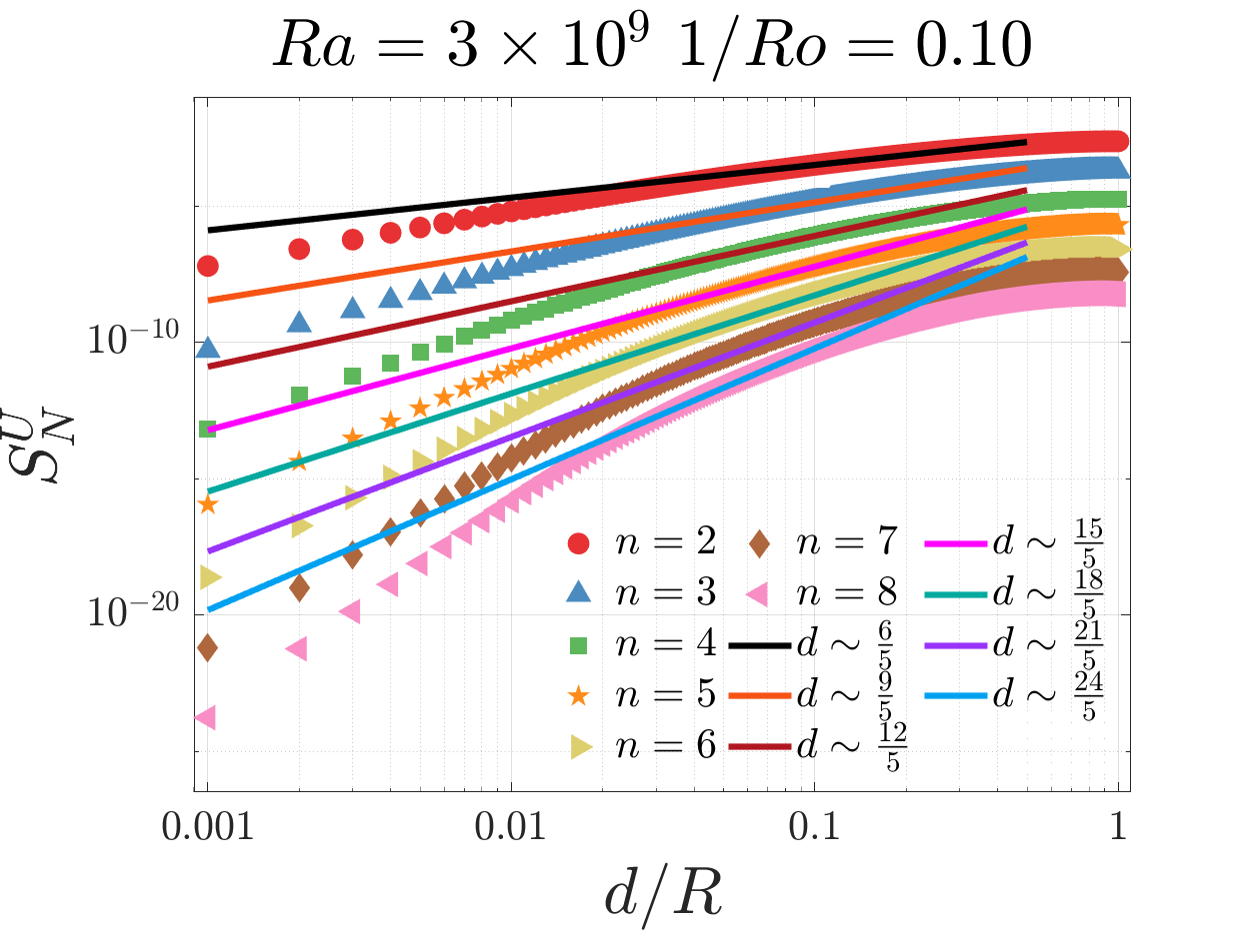}\\
			\includegraphics[width = 6.5cm]{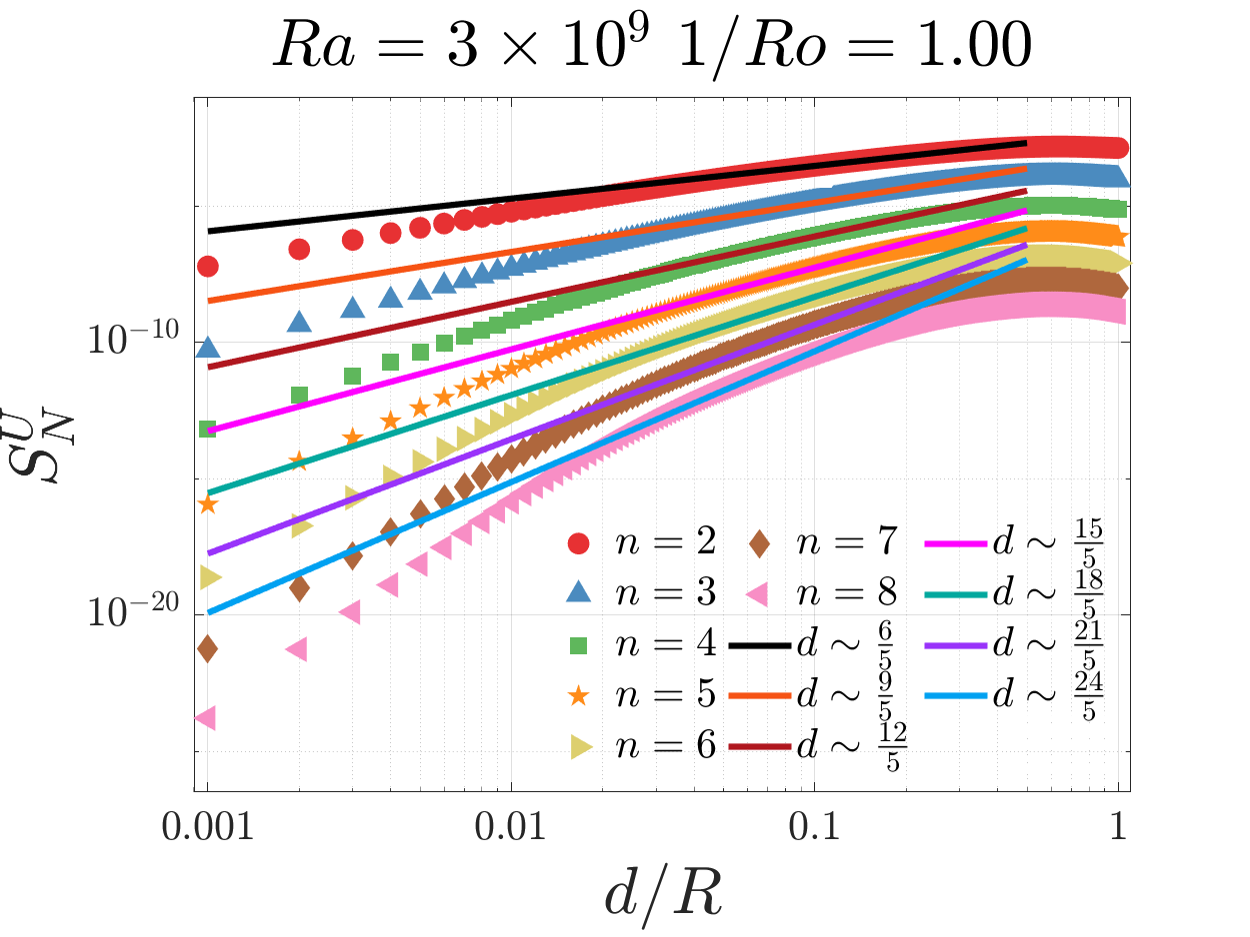}
						\includegraphics[width = 6.5cm]{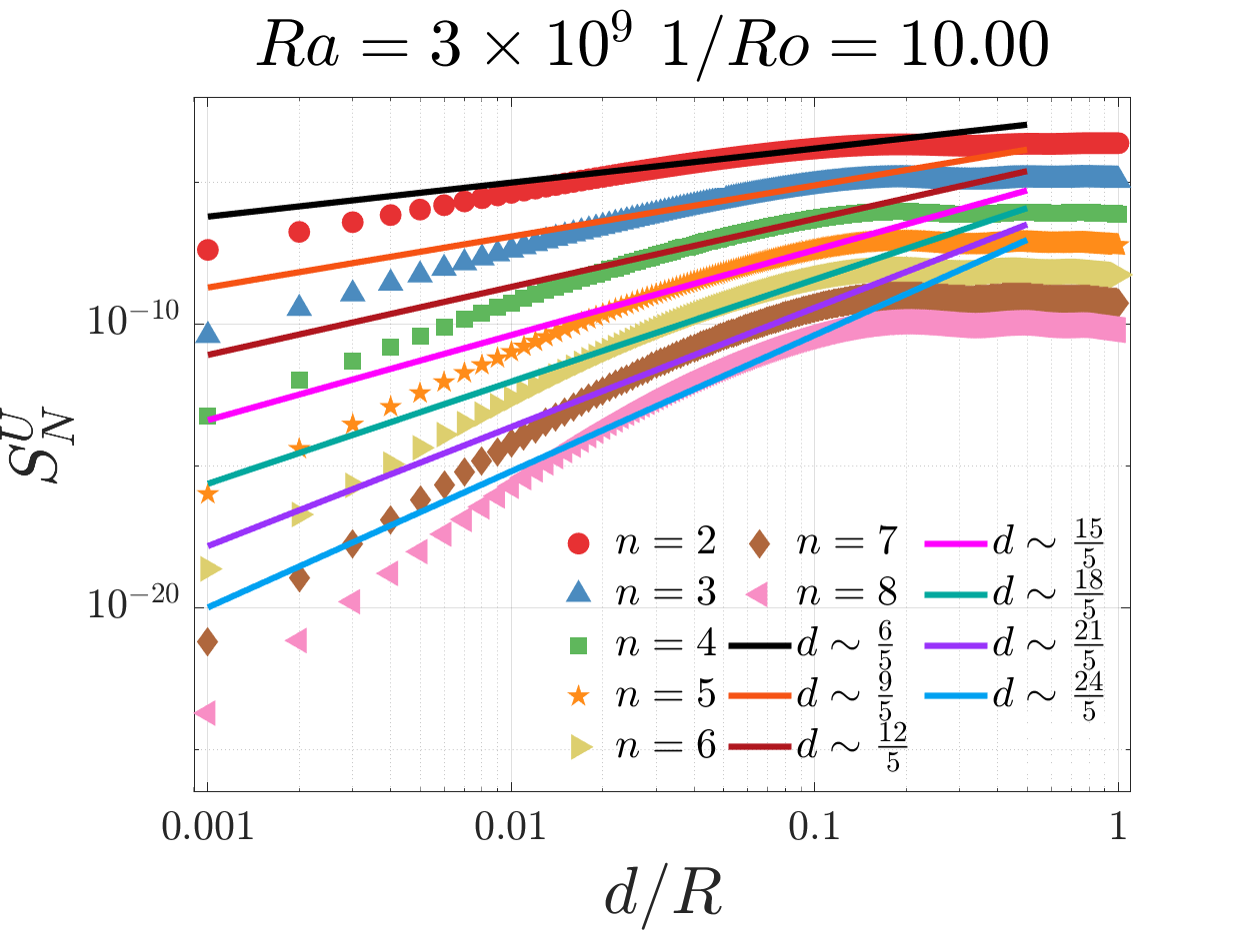}
	\caption{Velocity structure functions for $Ra=3\times 10^9$ and for different $1/Ro$.}
	\label{fig:VSF}
\end{figure}

The results for $S^{{U}}_N$ are shown in figure \ref{fig:VSF} for $N=1$ to $N=8$ and for different $1/Ro$. For $d/R\leq O(0.01)$, the smooth scaling $S^{{U}}_N\propto d^N$ emerges, but above this and for $1/Ro=0$, there is a clear range where BO59 seems to describe the behavior well for each $N$ considered. This indicates the absence of intermittency in the velocity field, as is expected for two-dimensional turbulent flows \cite{Boffetta12}, including those driven by buoyancy \cite{boffetta17}. As $1/Ro$ is increased, the BO59 scaling is still observed but over a region that becomes smaller as $1/Ro$ is increased. This is again simply due to the fact that for fixed $Ra$, as $1/Ro$ is increased, the Coriolis force affects increasingly smaller scales in the flow, and BO59 does not apply to scales significantly affected by the Coriolis force. As $1/Ro$ is reduced, the values of $S^{{U}}_N$ at the larger scales decrease. This can be understood by noting that for sufficiently large $d/R$, $S^{{U}}_N$ is related to the Reynolds stress component $\sigma_{\phi\phi}$, and as shown in figure \ref{fig:reStress11} and discussed earlier, this significantly reduces going from $1/Ro=0$ to $1/Ro=10$.

\begin{figure}
	\centering
	\includegraphics[width = 6.5cm]{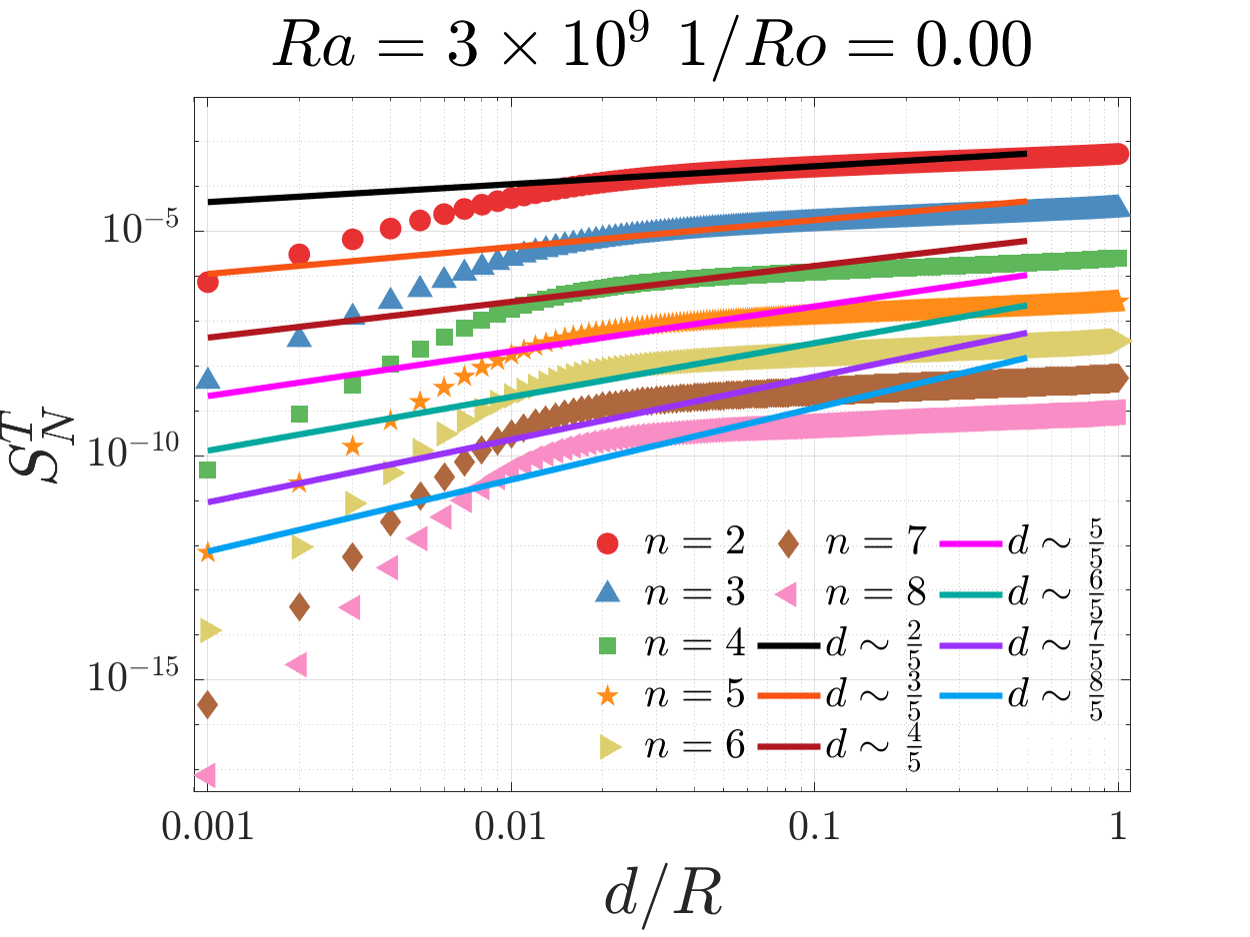}
		\includegraphics[width = 6.5cm]{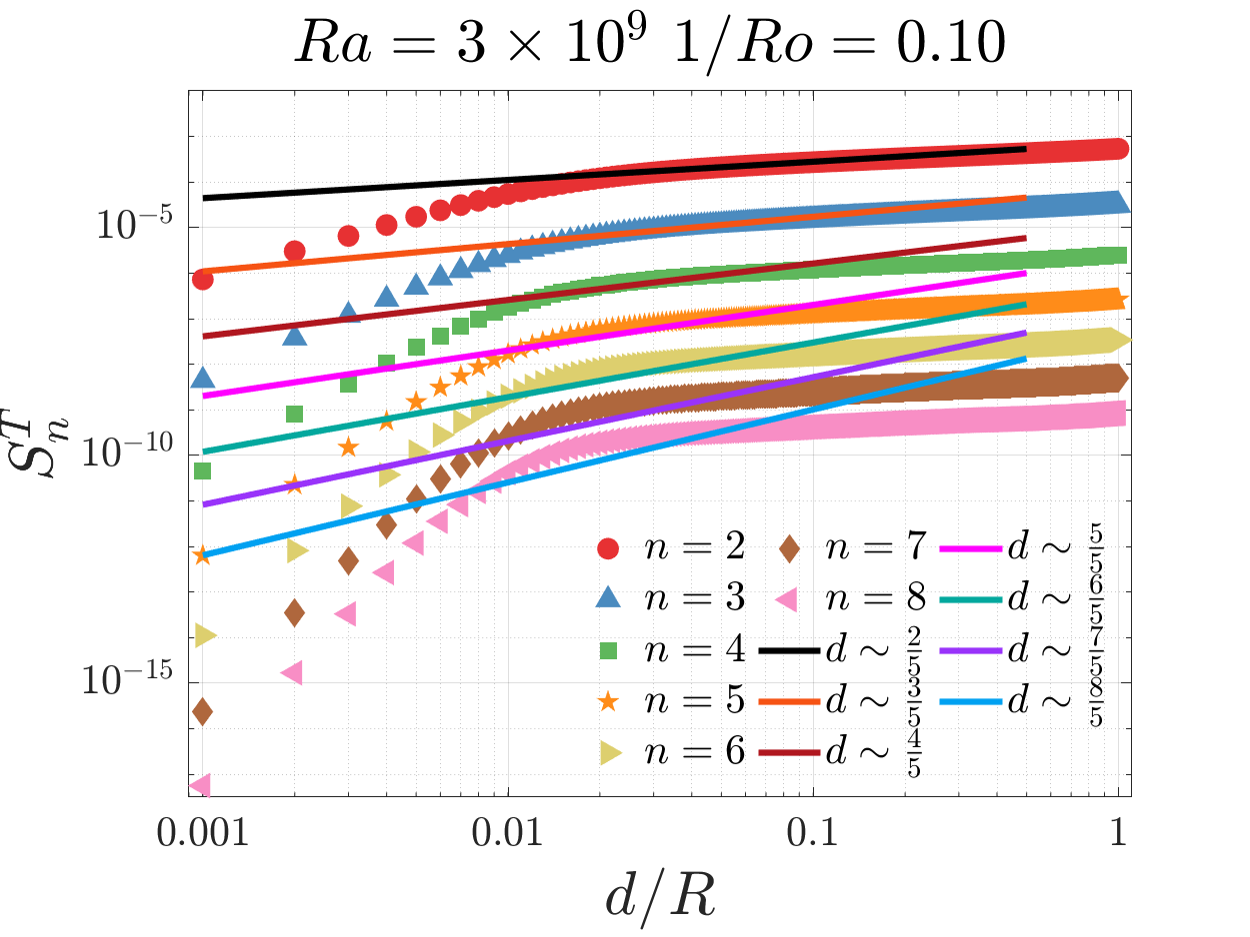}\\
			\includegraphics[width = 6.5cm]{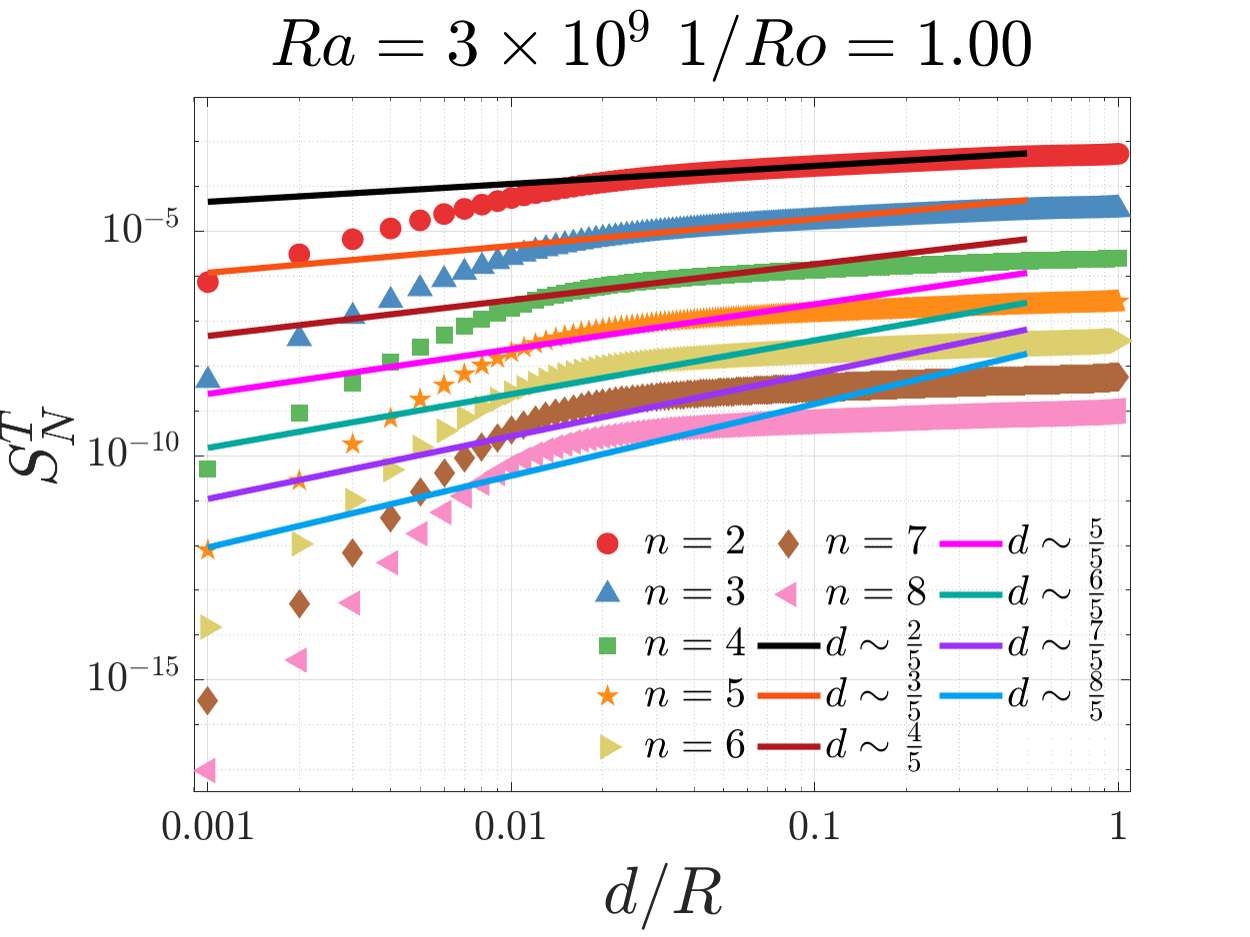}
						\includegraphics[width = 6.5cm]{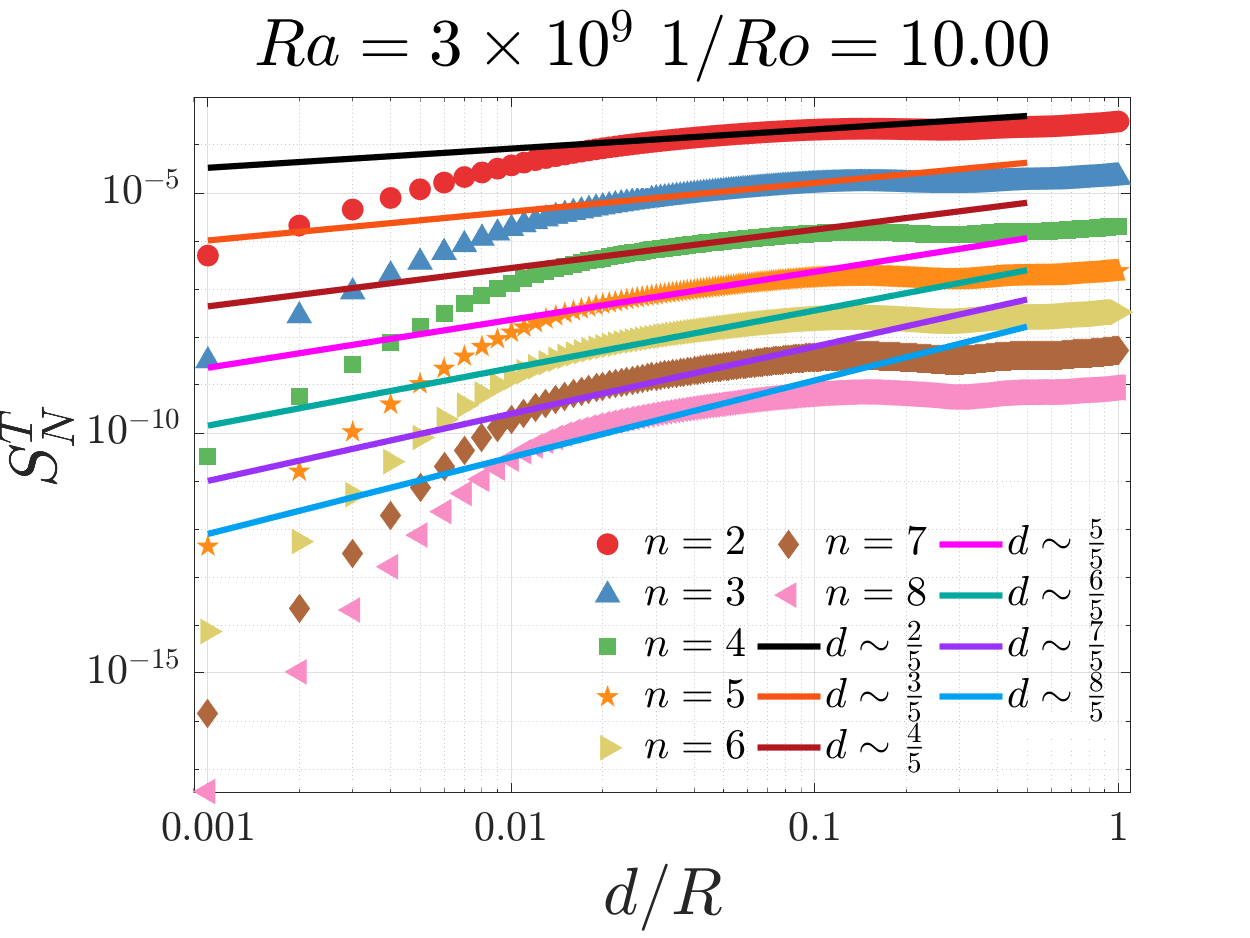}
	\caption{Temperature structure functions for $Ra=3\times 10^9$ and for different $1/Ro$. }
	\label{fig:TSF}
\end{figure}

In figure \ref{fig:TSF} we show the results for $S^T_N\propto d^N$. For $d/R\leq O(0.01)$, the smooth scaling $S^{{T}}_N\propto d^N$ emerges. Above this and for $1/Ro=0$, there is a clear range where BO59 seems to describe the behavior well for smaller $N$, but significant departures are observed for larger $N$. While some of these deviations could be due to the particularities of the rotating bubble flow we are considering, the study in \cite{Celani06} found similar behavior in a non-rotating, two-dimensional Rayleigh-Taylor flow with flat geometry. Therefore, as was also concluded in \cite{Celani06}, the deviations we observe from BO59 scaling for $S^T_N$ are most likely due to intermittency, something that is not captured in a mean-field theory like BO59. As $1/Ro$ is increased, the behavior remains the same, except that the range of $d/R$ over which BO59 is accurate for small $N$ reduces due to the Coriolis force affecting the larger scales of the flow, which is not accounted for in BO59. For sufficiently large $1/Ro$, the Coriolis force would make a leading order contribution at all scales in the flow, and BO59 scaling would not be observed at any scale or for any $N$.

\section{Conclusions}\label{conc}

We have used Direct Numerical Simulations to study the two-dimensional flow of a rotating, half soap bubble that is heated at its equator. This setup mimics the experimental study of \cite{MeuelCoudert11}, but the DNS enables us to consider the flow in greater detail. The heating at the equator of the bubble produces buoyancy, while rotation generates a Coriolis forces in the fluid. However, due to the curved surface of the bubble, the buoyancy and Coriolis forces vary with latitude on the bubble. This yields a flow that is strongly inhomogeneous and anisotropic, with rich flow behavior.

We began by exploring the single-point properties of the flow, including the Reynolds $Re$ and Nusselt $Nu$ numbers, mean fields, and Reynolds stresses, all as a function of latitude. Increasing the Rayleigh number $Ra$ increases $Re$ and $Nu$, associated with an increasingly strong production of turbulence due to convection in the flow. For a given $Ra$, we observe a non-monotonic dependence of the flow on the Rossby number $Ro$ for a range of different flow quantities. Moreover, the large scale mean circulations that appear due to convection are found to be strongly influenced by rotation, with the mean circulation becoming increasingly strong as $Ro$ decreases.

We then considered flow quantities that characterize the multiscale nature of the flow, including spectrums and spectral fluxes of kinetic and thermal energy, and enstrophy, and structure functions of velocity and temperature. The fluxes show that just a for non-buoyant two-dimensional turbulence on a flat surface, there is an upscale flux of kinetic energy at larger scales, and a downscale flux of enstrophy at smaller scales. The kinetic energy spectrum and velocity structure functions are well described by Bolgiano-Obukhov (BO59) scaling at scales where the effects of rotation are weak. The thermal energy spectrum and temperature structure functions are sensitive to contributions from the thermal boundary layer where most of the thermal fluctuations are contained. Provided the temperature statistics are computed away from this boundary layer they are found to satisfy BO59 scaling quite well for low-order structure functions, but deviations are strong at higher-orders due to intermittency. This is unlike the velocity structure functions, which satisfy BO59 scaling at all orders due to the absence of intermittency in the velocity field, associated with the inverse energy flux in the flow.

One interesting direction for future work would be to perform the DNS at much larger $Ra$ for which the scale separation in the flow will be large. This will allow to explore scales of the flow where, for example, Coriolis and buoyancy approximately balance or else Coriolis and inertial forces approximately balance (depending on the flow parameters) as well as scales much smaller than the buoyancy scale where the role of temperature fluctuations become dynamically passive and inertially dominated dynamical ranges may emerge.

\section{Acknowledgement}
This work was funded by the National Natural Science Foundation of China grant number 11872187, 12072125.  The authors thank SCTS/CGCL HPCC of HUST for providing computing resources and technical support. Y. L. Xiong acknowledges Cyclobulle Collaboration and the ANR grant for supporting his postdoctoral research opportunity at Institut de Math\'ematiques de Bordeaux from 2011 to 2013, which initialized the present study.

\section{Declaration of Interests} 
The authors report no conflict of interest

\appendix

\bibliographystyle{jfm}
\bibliography{newref}
\nocite{*}
\end{document}